\RequirePackage{fix-cm}
\documentclass[twocolumn]{svjour3mod}          
\smartqed  
\usepackage{graphicx}
\usepackage{algorithmic,algorithm}
\usepackage{amsmath}%
\usepackage{amsfonts}%
\usepackage{amssymb}%
\usepackage{graphicx}
\usepackage{subfigure}
\usepackage{multirow}
\usepackage[pagebackref=true]{hyperref}
\usepackage[binary-units=true,]{siunitx}
\DeclareSIUnit{\bph}{\textup{bph}}
\usepackage{caption}
\usepackage{datetime}
\usepackage{xspace}
\usepackage{cite}
\usepackage{textcomp}
\usepackage{multirow}

\newcommand{\vm}{VM\xspace}
\newcommand{\vms}{VMs\xspace}

\newcommand{\ms}[1]{mesh\#$#1$}
\newcommand{\Ms}[1]{Mesh\#$#1$}
\newcommand{\actn}{\textsc{Actnum}\xspace}
\newcommand{\VMS}{volume meshes\xspace}
\newcommand{\hs}{Hexa\-Shrink\xspace}
\newcommand{\jtwoktd}{J2K-3D\xspace}

\let\OldTexttrademark\texttrademark
\renewcommand{\texttrademark}{\OldTexttrademark\xspace}%

\newcommand{\etal}{\emph{et al.}\xspace}

\graphicspath{{./}}
\journalname{Computational Geosciences}

\begin{document}

\title{\hs,  an exact scalable framework for hexahedral meshes with  attributes and discontinuities:  multiresolution rendering and storage of  geoscience models\\\thanks{This work was partly presented in \cite{Peyrot_J_2016_p-icip_hexashrink_mclshmdg}. This is a post-peer-review, pre-copyedit version of an article published in Computational Geosciences. The final authenticated version is available online at: http://dx.doi.org/10.1007/s10596-019-9816-2}}

\titlerunning{\hs:  multiresolution reversible  compression of hexahedral meshes for geoscience simulation models}        

\author{Jean-Luc Peyrot \and Laurent Duval  \and Fr\'ed\'eric Payan \and Lauriane Bouard \and L\'ena\"ic Chizat \and S\'ebastien Schneider   \and Marc Antonini}

\authorrunning{Peyrot, Duval,  Payan,  Bouard, Chizat, Schneider, Antonini}

\institute{ Laurent Duval and L\'ena\"ic Chizat                                \at IFP Energies nouvelles \\
                                                                1 et 4 avenue de Bois-Pr\'eau \\
                                                                F-92852 Rueil-Malmaison \\
                                                                Tel.: +33 (0) 1 47 52 61 02 \\
                                                                \email{laurent.duval@ifpen.fr,lenaic.chizat@inria.fr}
                                                            \and
Laurent Duval \at University Paris-Est, LIGM \\
ESIEE Paris \\
F-93162 Noisy-le-Grand
\and
Jean-Luc Peyrot, S\'ebastien Schneider and Lauriane Bouard      \at IFP Energies nouvelles \\
                                                                Rond-Point de l'\'echangeur de Solaize, BP3 \\
                                                                F-69360 Solaize \\
                                                                \email{lauriane.bouard@ifpen.fr,sebastien.schneider@holomake.fr}
                                                            \and
            Fr\'ed\'eric Payan and Marc Antonini            \at Universit\'e C\^ote d'Azur, CNRS, I3S \\ 
                                                                2000, route des Lucioles\\
                                                                F-06900 Sophia Antipolis \\
                                                                Tel.: +33(0) 4 89 15 43 22 / +33(0) 4 92 94 27 18 \\
                                                                \email{fpayan@i3s.unice.fr,am@i3s.unice.fr}
																																\and
L. Chizat is now with INRIA, ENS, PSL Research University Paris, France. 	S\'ebastien Schneider is 	now with HoloMake.																											
}

\date{Last compiled: \today, \currenttime}


\maketitle
\begin{abstract}
With  huge  data acquisition progresses realized in the past decades and acquisition systems  now able to produce high resolution grids and point clouds, the digitization of physical terrains becomes increasingly more precise. Such extreme quantities of generated and modeled data greatly impact computational performances on many levels of high-performance computing (HPC):  storage media, memory requirements, transfer capability, and finally  simulation interactivity, necessary to exploit this instance of big data. Efficient representations and storage are thus becoming ``enabling technologies'' in HPC experimental and simulation science \cite{Foster_I_2017_incoll_computing_jwynodares}. We propose \hs, an original decomposition scheme for structured hexahedral \VMS. The latter are used for instance in biomedical engineering,  materials science, or geosciences. \hs provides a comprehensive framework allowing efficient mesh visualization and storage.  Its exactly reversible multiresolution decomposition yields a hierarchy of meshes of increasing levels of details, in terms of either geometry, continuous or categorical properties of cells.

Starting with  an overview of volume meshes compression techniques, our contribution blends coherently different multiresolution wavelet schemes in different dimensions. It results in a global framework preserving discontinuities (faults) across scales, implemented as a fully reversible upscaling at different resolutions. Experimental results are provided on meshes of varying size and complexity. They emphasize the consistency of the proposed representation, in terms of visualization, attribute  downsampling and distribution at different resolutions. Finally, \hs yields  gains in storage space when combined to lossless compression techniques.


\end{abstract}



\section{Introduction\label{sec_introduction}}
Simulation sciences and scientific modelling in high-performance computing employ meshes with increasing precision and dynamics. Among them, hexahedral meshes are commonly handled in biomedical engineering \cite{Kober_C_2001_j-eng-comput_case_shmgshm}, computational materials science \cite{Owen_S_2017_j-procedia-eng_hexahedral_mgcmm} and in geosciences.  They are for instance used by geologists to study flow simulations for reservoir modelling \cite{Cannon_S_2018_book_reservoir_mpg}, and  benefit from an increasing interest for geologic model building \cite{Caumon_G_2013_j-ieee-tgrs_three-dimensional_ismbrsdtmtarmlpbnem,Lie_K_2017_j-computat-geosci_successful_ammrrse}. Huge progresses in data acquisition produce  increasingly more accurate  digitization of physical terrains.
The tremendous quantity of data thus generated prominently impacts computational resources and performances:
 memory size required for their storage and visualization, but also their transmission and transfer, and ultimately their processing. Consequently, it greatly affects the overall simulation interactivity. This trend affects the oil and gas sector at large  \cite{Perrons_R_2015_j-energy-pol_data_awogsloiabd}.

We propose \hs, an efficient multiscale representation dedicated to hexahedral meshes with attributes and discontinuities. \hs combines four wavelet-like decompositions to adapt to the heterogeneous nature of geoscience meshes. Geometrical, continuous and categorical properties are consistently  downsampled (upscaled in geoscience terms \cite[pp. 181--204,  Simulation and Upscaling]{Cannon_S_2018_book_reservoir_mpg}) in an exactly reversible manner. In addition to  regular structures, \hs also strives to manage mesh externalities like boundaries and borders   tagged as inactive cells for simulation purposes. It produces a hierarchy of meshes at dyadic resolutions maintaining geometrical coherency over scales, consistent with geomodeler/simulator upscaling operations. It finally lends itself to efficient lossless storage, in combination with state-of-the-art  compression algorithms.

The paper is structured as follows:
Section \ref{sec_context} presents the specificities of structured hexahedral meshes and the discontinuities they may contain. Section \ref{sec_soa}  reviews  prior methods for volume mesh  representation or compression \cite{Dupont_F_2013_incoll_3d_mc}. We introduce the \hs structured mesh representation in Section \ref{sec_methodology}. We detail the four main multiscale wavelet-like schemes that entail an exactly invertible hierarchy of downsampled meshes, consistently with respect to geometry and properties. A special care is taken on the accurate representation of faults and the management of mesh borders.
Section \ref{sec_results} presents  visual results, and evaluates the quality of the \hs with respect to categorical property coherency across scales, and dyadic upscaling by a geomodeler. An exhaustive evaluation  obtained from combining \hs with different lossless compression algorithms, at different resolution levels, shows the interest of the proposed representation in terms of lossless compression for storage.
Finally, Section \ref{sec_perspectives} summarizes our contributions and proposes future works.

\section{Volume meshes: a primer\label{sec_context}}
\subsection{Generic definitions\label{sec_generic-definitions}}

\begin{figure}[htbp]
          \centering
          \includegraphics[width=.6\linewidth]{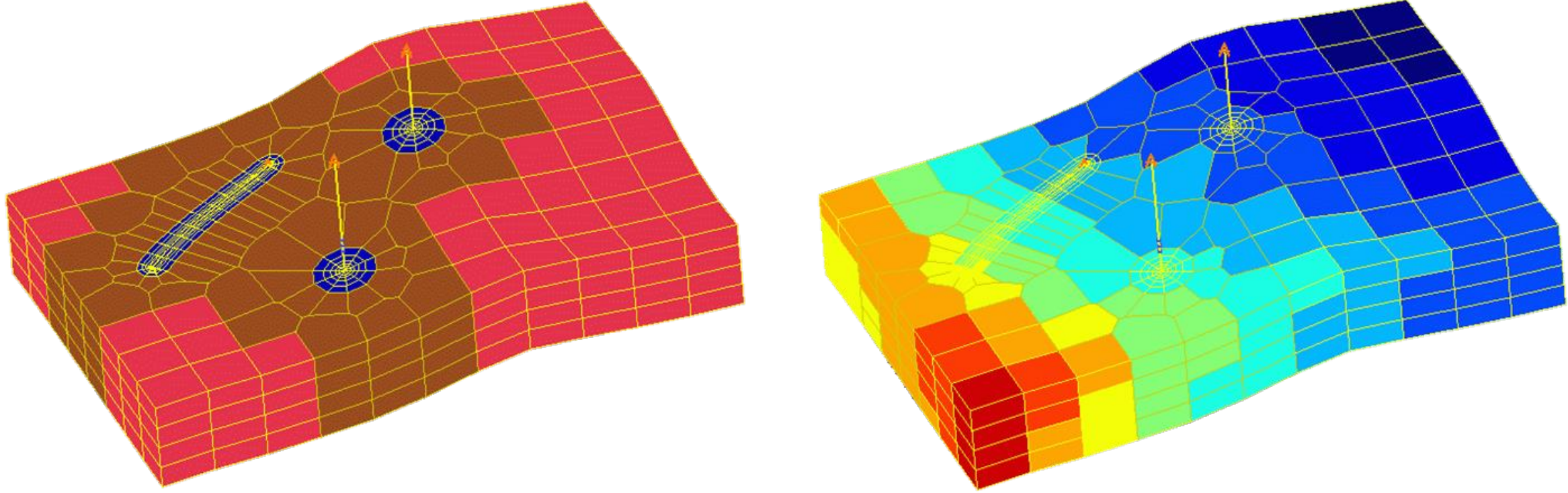}
          \caption{Example of a \vm used in geosciences (left); same mesh with the  associated porosity property (right). Note that this mesh is hybrid and unstructured, with both hexahedral and tetrahedral elements.\label{fig:hybrid_volume_mesh}}
         \end{figure}

Volumetric or \VMS (\vms) discretize the interior structure of $3D$ objects. They partition their inner space with a set of three-dimensional elements named cells (or zones). While pyramid and triangular prism partitions exist, most of the existing \vms are composed of \emph{tetrahedral} (4 faces) or \emph{hexahedral} (\num{6} faces) elements. They are called tets  or hexes (sometimes bricks), respectively. A \vm composed of different kinds of cells, tetrahedra and hexahedra for instance, is termed hybrid.
A \vm is described by the location of vertices in $3D$ space (\emph{geometry}) and the incidence information between cells, edges, and vertices (\emph{connectivity}).
In function of the application domain, \vms also contain \emph{physical properties} associated to vertices, edges, or cells. In geosciences, properties can be scalar (like a single porosity value, Figure \ref{fig:hybrid_volume_mesh}) or vectorial (a vector of compositional proportions in different rocks within a cell\ldots). We also distinguish  \emph{categorical} or \emph{nominal} variables (symbolic and discrete values describing the composition of rocks: sandstone (\num{0}), limestone  (\num{1}), shale  (\num{2})\ldots)  from \emph{continuous properties} (saturation, porosity, permeability, or a temperature taking values in a given range $[-T_\flat, T_\sharp]$).

\begin{figure}[htbp]
          \centering
          \includegraphics[width=0.3\linewidth]{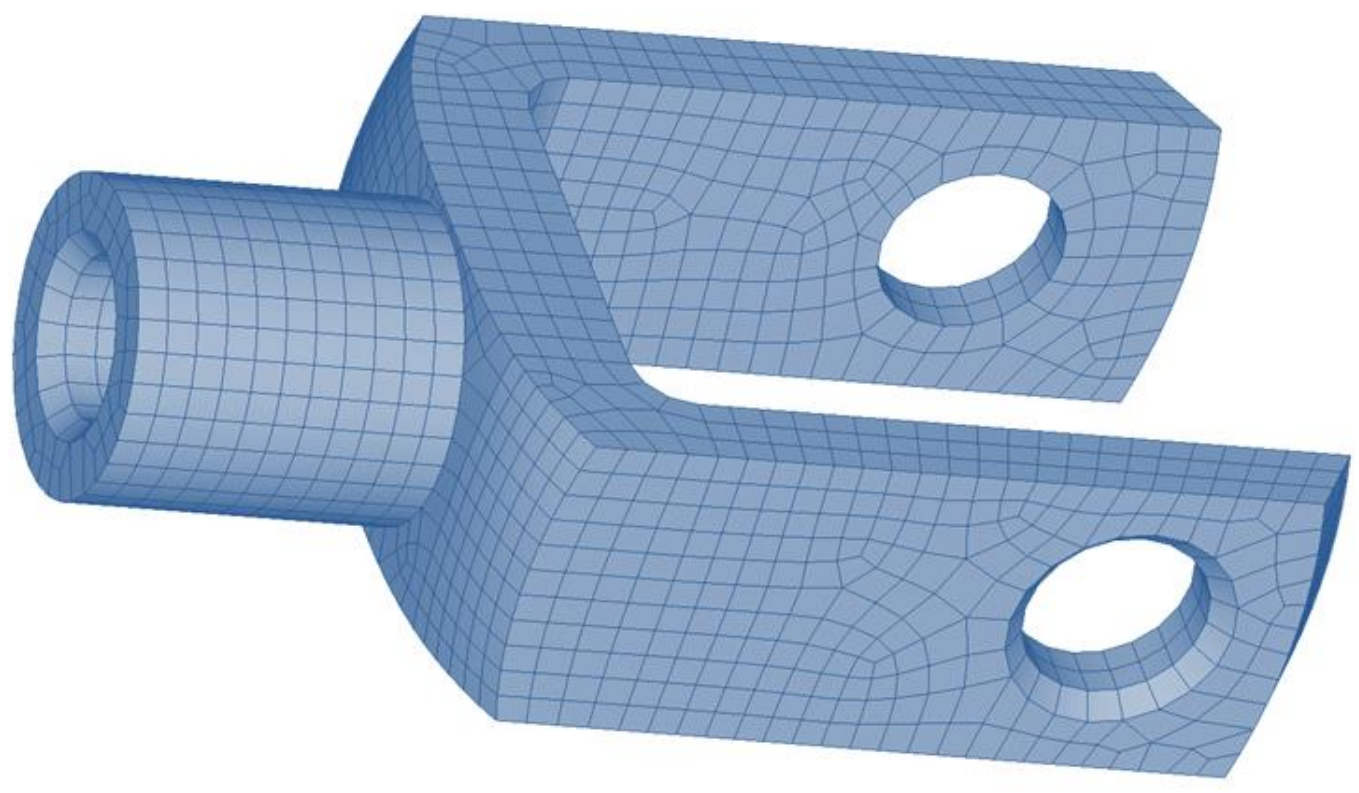}\\
          \caption{CAD model defined by an unstructured \vm.\label{fig:unstructured_volume_mesh}}
        \end{figure}

A non-degenerate hex has 6 \emph{faces} named quads, 12 \emph{edges}, and 8 \emph{vertices}.
Depending on incidence information between cells, edges and vertices, hexahedral meshes are either unstructured or structured. The degree of an edge is the number of adjacent faces.
An hexahedral mesh is \emph{unstructured} if cells are placed irregularly in the volume domain, \emph{i.e.}, if degrees are not the same for all edges of the same nature. Unstructured meshes have an important memory footprint, as all the connectivity information must be described explicitly. However, they are well-suited to model complex volumes,  Computer-aided design (CAD) models for instance, as shown in Figure \ref{fig:unstructured_volume_mesh}.

\begin{figure}[htbp]
  \centering
  \includegraphics[width=0.5\linewidth]{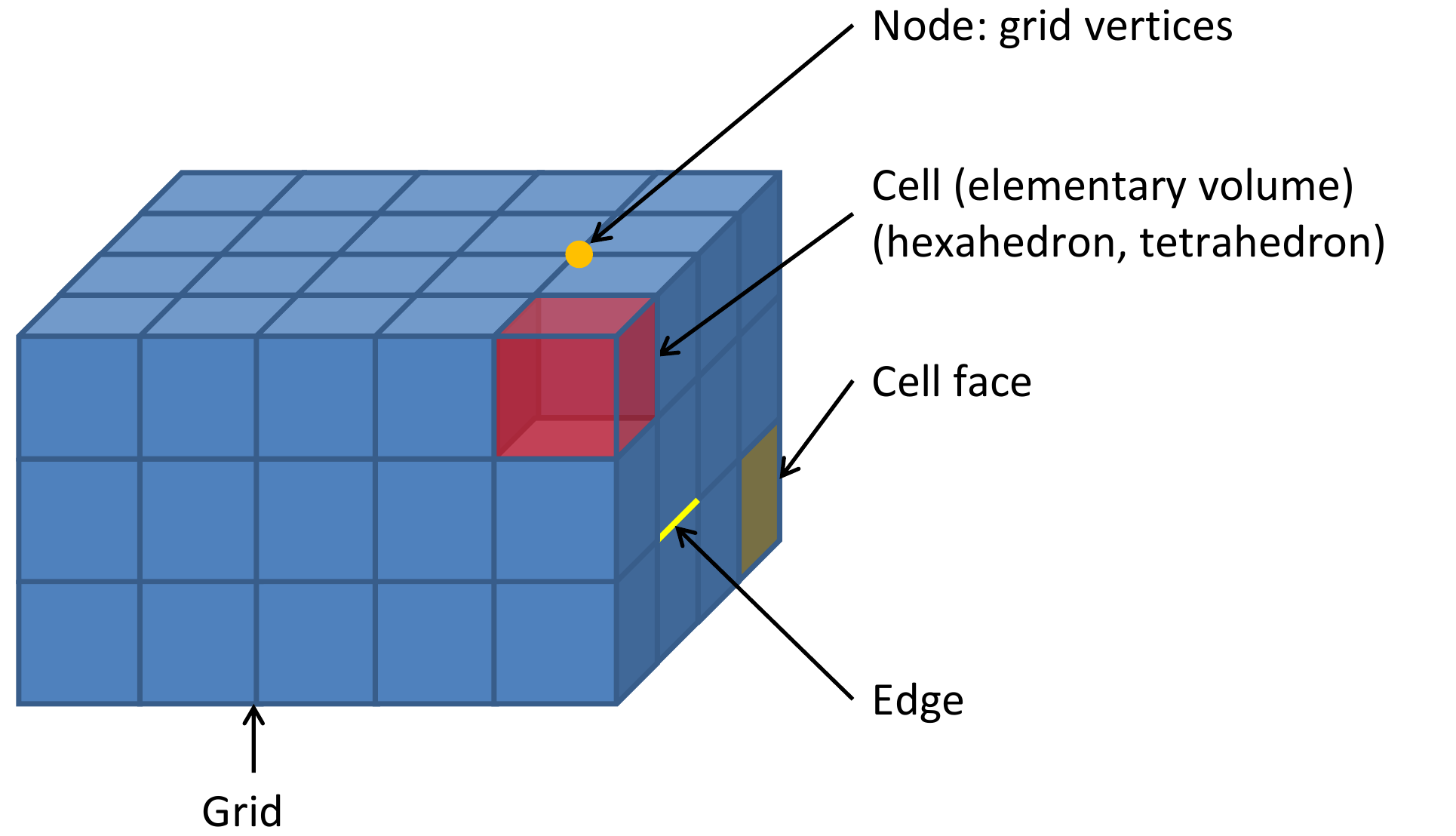}\\
  \caption{Structured hexahedral mesh composed of $\left(5 \times 4 \times 3 \right)$ cells.\label{fig:grid_presentation}}
\end{figure}

An hexahedral mesh is \emph{structured} if cells are regularly organized in the volume domain, \emph{i.e.}, if the degree is equal to four for \emph{interior} edges (inside the volume), and equal to two for \emph{border} edges (on a border of the volume). In that case, the set of hexahedral cells is topologically aligned on a $3D$ Cartesian grid (see Figure \ref{fig:grid_presentation}). Each vertex of the mesh can be associated to a node of the grid. Hence, each cell can be indexed by only one triplet $(i,j,k)$, and the connectivity information becomes implicit: only the position of the vertices is needed to model the mesh.

\subsection{Hexahedral meshes \& geometrical discontinuities\label{subsec:structured_meshes}}

\begin{figure}[h!]
\centering     
\subfigure[Free-fault area.]{\label{fig:fault_node_configuration_wo}\includegraphics[width=0.475\linewidth]{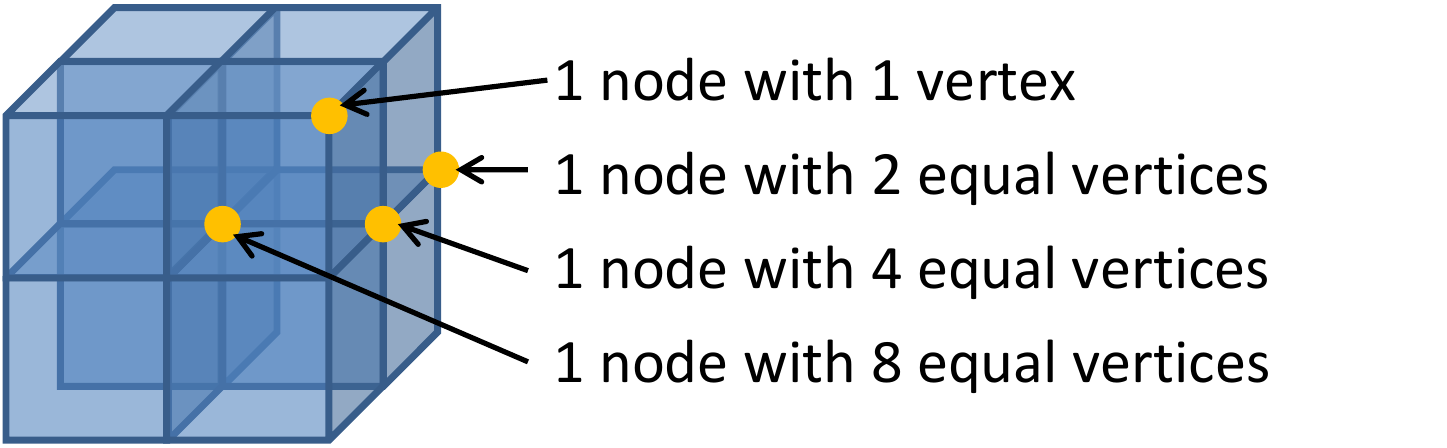}}
\subfigure[Area with a vertical fault.]{\label{fig:fault_node_configuration_w}\includegraphics[width=0.475\linewidth]{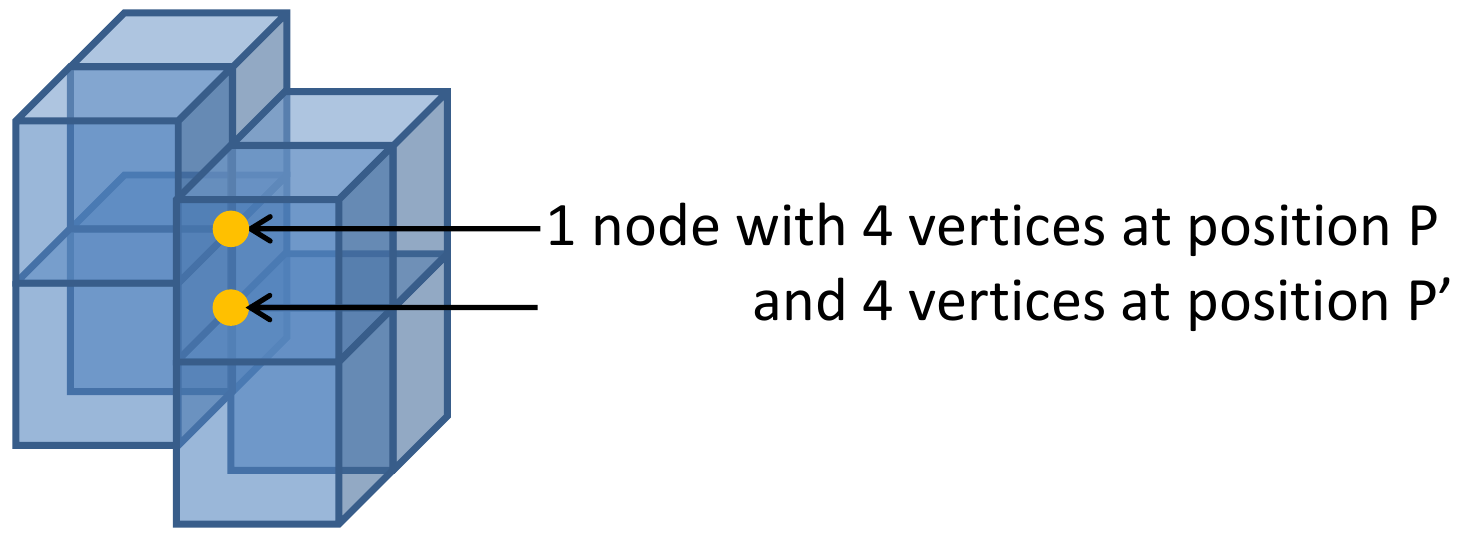}}
\caption{A fault-free and a fault area.\label{fig:fault_node_configuration}}
\end{figure}

Hexahedral meshes in geosciences are generally structured, and thus based on a Cartesian grid. But these meshes may contain \emph{geometrical discontinuities}. They correspond, for instance, to geological faults. It  induces a vertex disparity in space at the same node. The association of one node  of the Cartesian grid with \num{8} vertices (one for each adjacent cell) handles this specificity.
Figure \ref{fig:fault_node_configuration}-(a) provides an illustration of a fault-free volume. On Figure \ref{fig:fault_node_configuration}-(b), we see that this structure allows to describe for instance a vertical fault (by positioning vertices differently about the node), while preserving the Cartesian grid.

\begin{figure}[h!]
  \centering
  \includegraphics[width=0.75\linewidth]{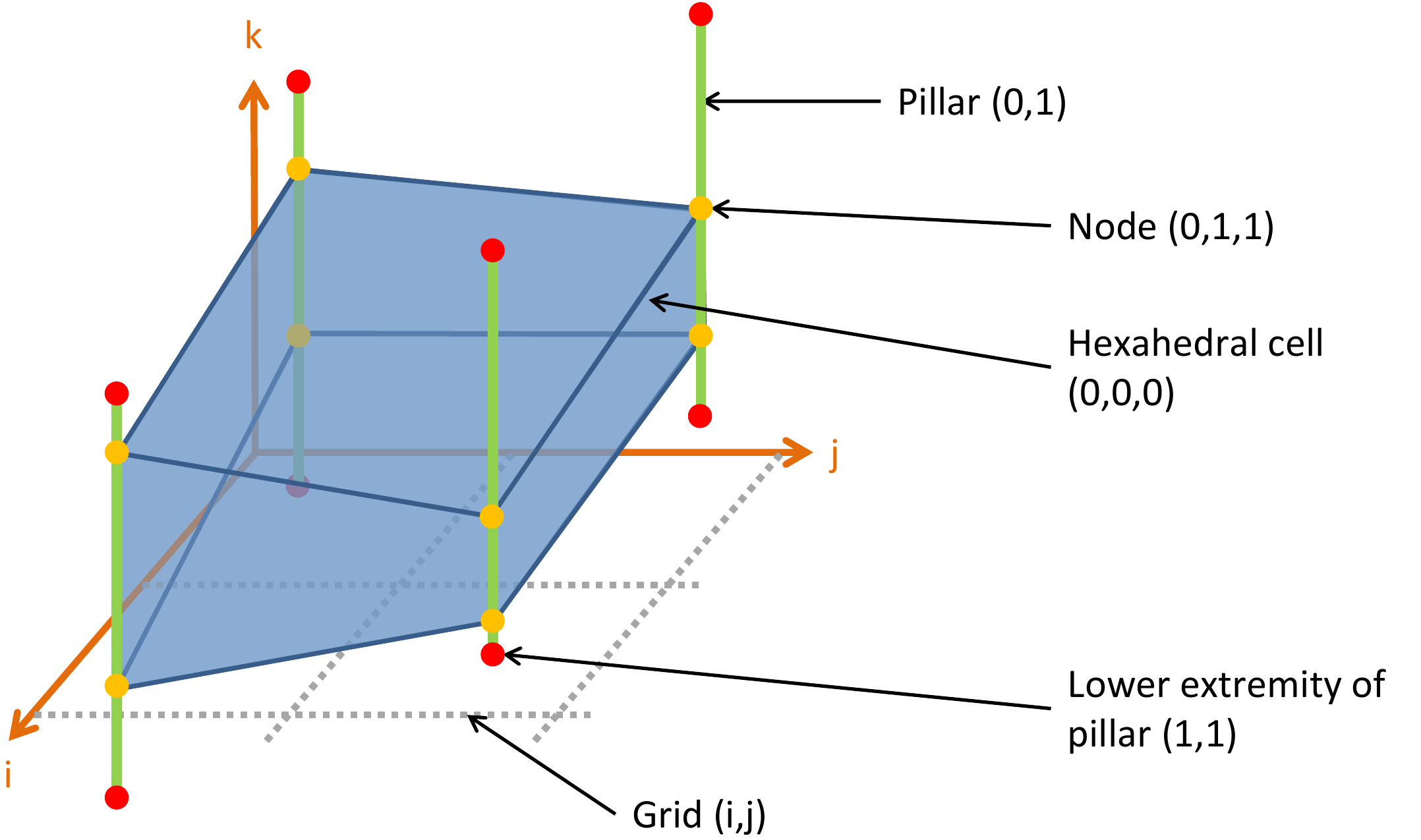}\\
  \caption{An hexahedron, according to the \emph{pillar grid} structure.}\label{fig:pillar_grid_description}
\end{figure}

The most popular data structure for structured hexahedral meshes with geometrical discontinuities is the \emph{Corner Point Grid} \cite{Roe_P_2016_j-computat-geosci_volume-conserving_rcfcpg,Lie_K_2016_book_introduction_rsumugmrstmrst} tessellation of an  Euclidean $3D$ volume. This structure is often termed \emph{pillar grid}. It is based on a set of vertical or inclined pillars running from the top to the bottom of the geological model. A cell is defined by its 8 adjacent vertices (2 on each adjacent pillar, see Figure \ref{fig:pillar_grid_description}), and the vertices of the adjacent cells are described independently, in order to model faults and gaps. Across the associated Cartesian grid, each cell can be indexed by a triplet $(i,j,k)$.

\begin{figure}[h!]
\centering
\subfigure[Single.]{\label{fig:degenerated_cells_case_0}\includegraphics[width=0.29\linewidth]{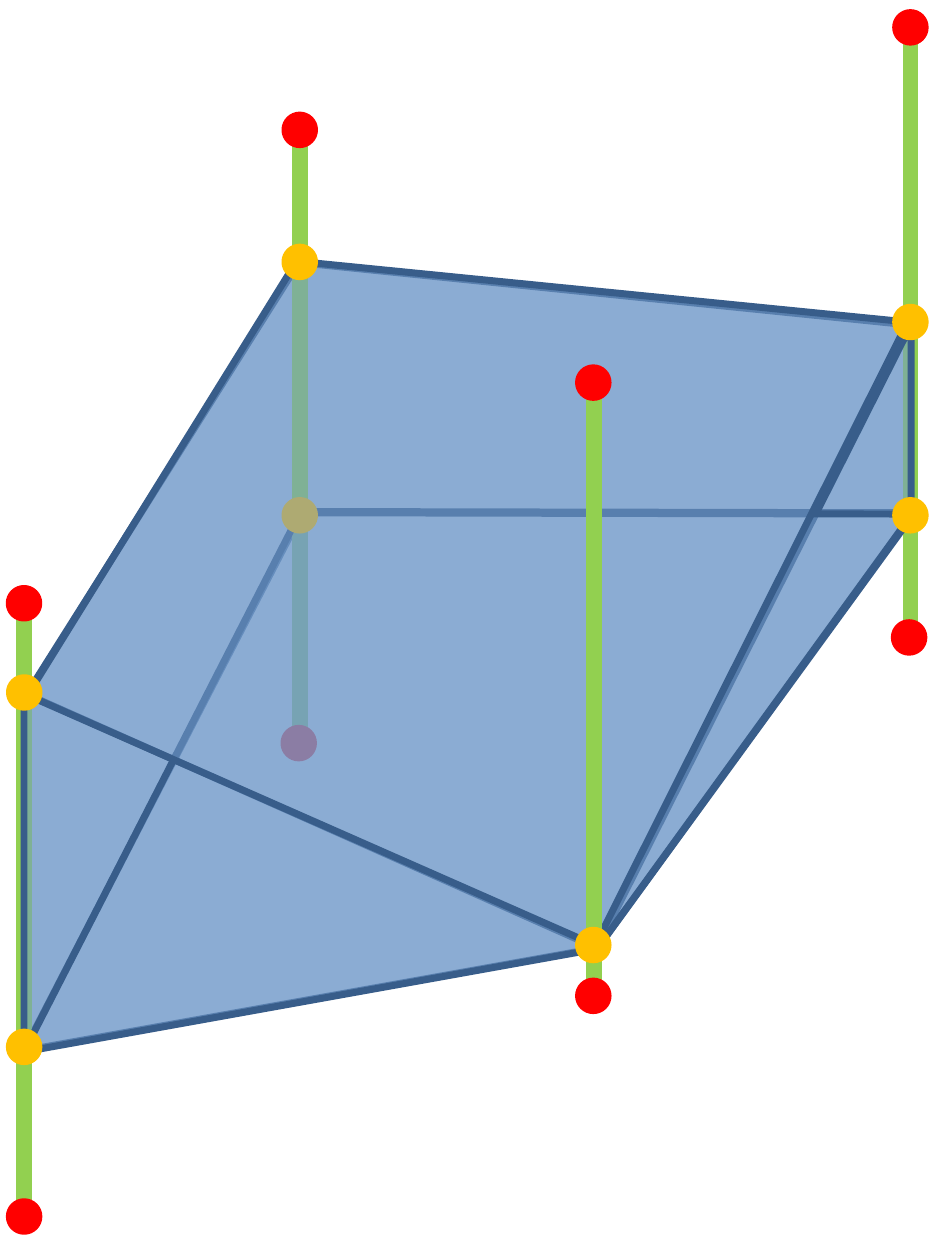}}
\subfigure[Two opposed.]{\label{fig:degenerated_cells_case_1}\includegraphics[width=0.29\linewidth]{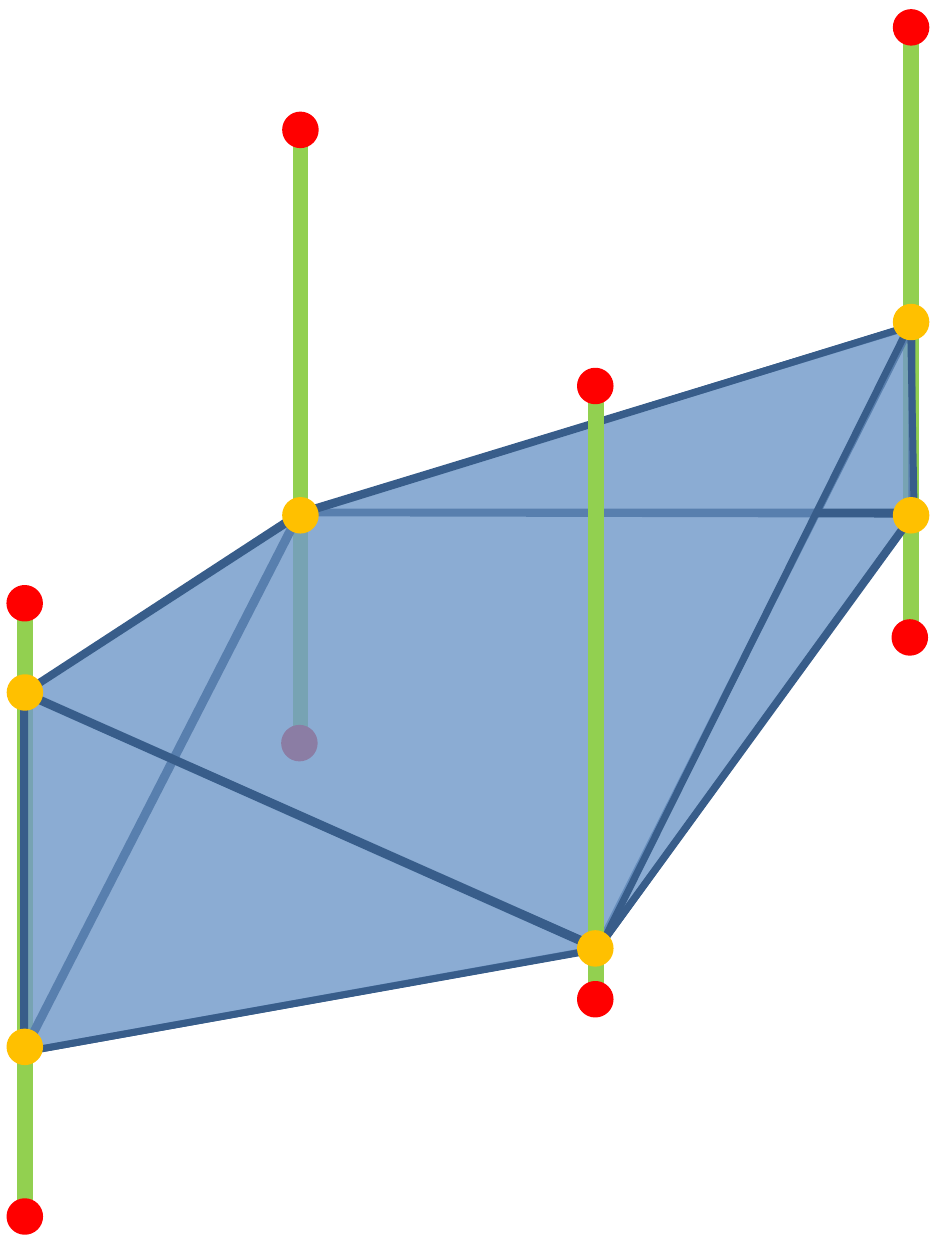}}
\subfigure[Two adjacent.]{\label{fig:degenerated_cells_case_2}\includegraphics[width=0.29\linewidth]{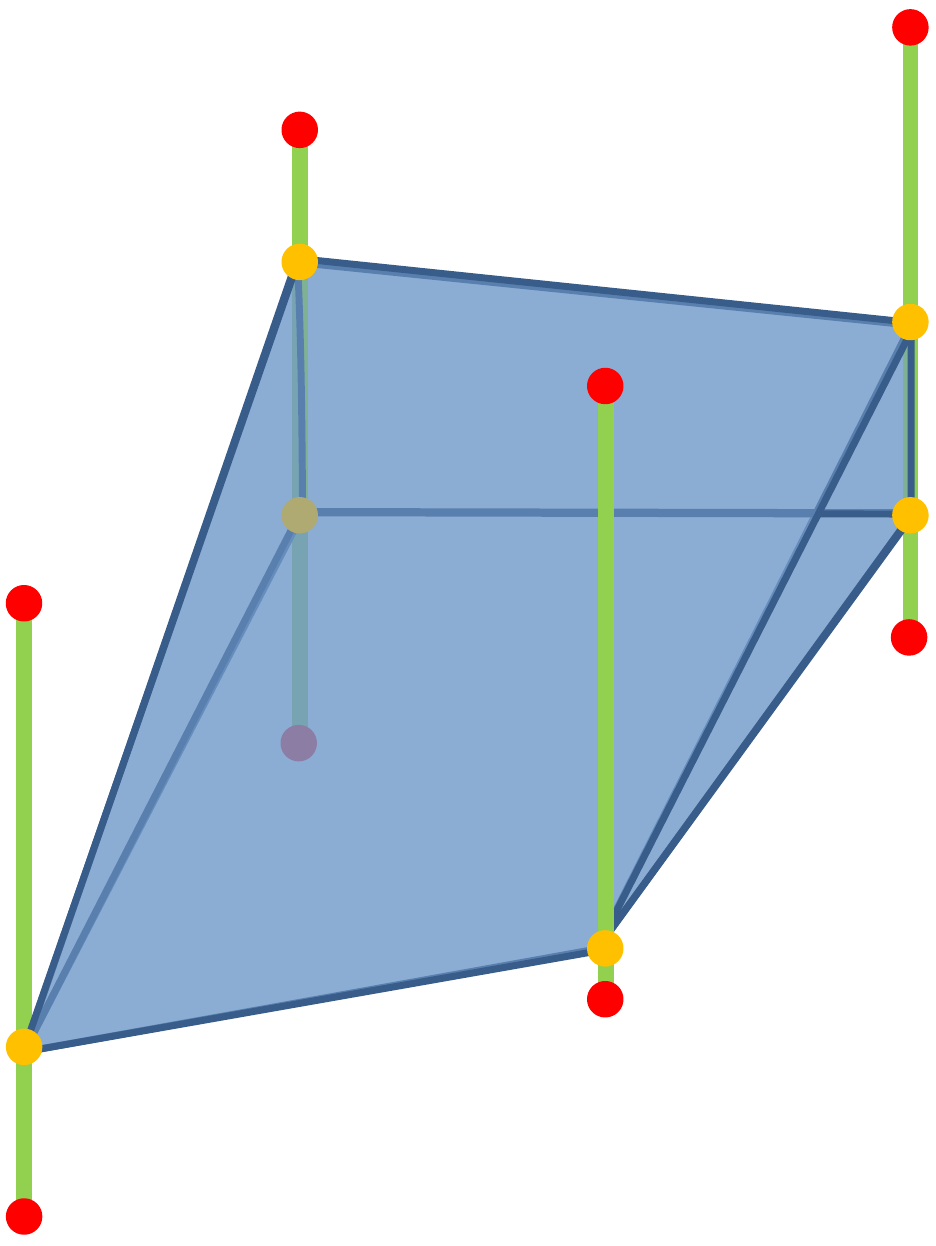}}
\caption{Degenerate hexahedral cells due to a single collapsed pillar (a) or two different collapsed pillar  locations (b), (c).\label{fig:degenerated_cells}}
\end{figure}

This pillar grid also allows to model geological collapses  (or erosion surfaces), by using \emph{degenerate} cells, {i.e.}, cells with (at least) two vertices on one pillar located at the same position (see Figure \ref{fig:degenerated_cells} for different degenerate configurations).

\section{Mesh compression: an overview\label{sec_soa}}

We deal in this section mostly with the ontological description of volume meshes, leaving aside the specificities related to actual data format. 
  \subsection{Basic techniques for mesh encoding}
  The most straightforward technique to encode a \vm is to use an indexed data structure: the list of all the vertex coordinates (three floating-point values, which amounts to $96$ bits par vertex), followed by their connectivity. The connectivity is defined cell after cell, each cell being defined by the set of indexes of the adjacent vertices (8 integers per hex). Physical property encoding depends on their nature: categorical/continuous, dimension, associated to cells or vertices. They thus only provide estimates of an actual compression performance.

To reduce the memory footprint or make the transmission of \vms faster, well-known techniques exist. The simplest tool for the geometry is the \emph{quantization} of  vertex coordinates. It consists of constraining the vertex coordinates to a  discrete and finite set of values. Hence, it becomes possible to encode each coordinate with an integer index, instead of a \num{32}-bit floating-point value. It is common to quantize the coordinates with  \num{12} or \num{16} bits, reducing the geometry information by a compression factor of \num{2.6} or \num{2} respectively. Quantization inevitably introduces an irreversible loss in accuracy. Visualization typically tolerates precision loss (as long as visual \emph{distortion} remains negligible), unlike some numerical simulations requiring more precise computations.

\emph{Prediction} (as well as related interpolation methods) further improves the geometry compactness. Predictive coding resorts to estimating the position of a vertex from already encoded neighbor vertices. \emph{Prediction errors} (differences between predicted and actual positions) are generally smaller in amplitude and sparser, which makes their entropy coding (which codes differently frequently occurring patterns)  efficient \cite[p. 63 sq.]{Salomon_D_2009_book_handbook_dc}.

Regarding connectivity, when meshes are unstructured, the most frequent technique performs a traversal of mesh elements, and  describes the incidence configurations with a reduced list of symbols. These symbols are then entropy coded. When meshes are structured, the connectivity is implicit, reducing its cost to zero. For such meshes, the only additional information to encode are geometrical discontinuities describing faults and gaps.

\subsection{Volume mesh compression: prior works\label{sec:mesh_compression}}

The basic tools previously presented can be implemented on the ontological structure of meshes, and improved in many different ways. Their combination, with the assistance of  advanced compression techniques, permits more efficient  tetrahedral or hexahedral mesh coding. Previously proposed algorithms are presented below, classified  into two categories:  \emph{single-rate} and \emph{progressive/multiresolution}.

\subsubsection{Single-rate mesh compression\label{subsubsec:single-rate_algorithms}}

They lead to a compact mesh representation, most of the time driven by efficient connectivity encoding.
The first method for tetrahedral meshes, Grow \& Fold, was presented by Szymczak and Rossignac \cite{Szymczak_A_2000_j-comput-aided-des_grow_fcctm} at the end of the nineties. It is an extension of \emph{EdgeBreaker} \cite{Rossignac_J_1999_j-ieee-trans-visual-comput-graph_edgebreaker_cctm} developed for triangle meshes. The method consists in building a tetrahedral spanning tree from a root tetrahedron. The traversal is arbitrary among the three neighboring tets  (Section \ref{sec_generic-definitions}) of the cell currently processed, and \num{3} bits are needed to encode each cell. The resulting spanning tree does not retain the same topology as the original mesh, because some vertices are replicated during the traversal. ``Fold" and ``glue" techniques are thus needed during encoding to restore the original mesh from the tetrahedron tree. The additional cost is \num{4} bits, leading to a total cost of \num{7} bits per tetrahedron.

The \emph{cut-border} initiated in \cite{Gumhold_S_1998_p-acm-siggraph-comput-graph_real_tctmc} was adapted to tetrahedral meshes \cite{Gumhold_S_1999_p-visualization_tetrahedral_mccbm}. It denotes the frontier between  tetrahedra already encoded and those to encode. At each iteration, either a triangle or an adjacent tetrahedron is added to the cut-border. In this case, if the added vertex is not already in the cut-border, this latter is included by a connect operation, and is given a local index. As the indexing is done locally, the integers to encode are very small, leading to a compact connectivity representation. In addition, two methods are proposed to encode geometry and associated properties, based on prediction and entropy coding. This method yields good bit rates (\num{2.4} bits per tetrahedron for connectivity) for usual meshes, handles non-manifold borders, but worst-cases severely impact bitrates and runtimes (which tend to be quadratic).

Isenburg and Alliez \cite{Isenburg_M_2003_j-graph-model_compressing_hvm} are the first to deal with hexahedral \vms. The connectivity is encoded as a sequence of edge degrees --- in a way similar to \cite{Touma_C_1998_p-gi_triangle_mc} for triangular meshes --- \emph{via} a region-growing process of a convex hull called \textit{hull surface}. It relies on the assumption that hexahedral meshes are often highly regular, which implies that the majority of vertices are shared by \num{8} cells. It involves an almost constant edge degree all over the mesh, which significantly decreases the entropy of the connectivity information. A context-based arithmetic coder \cite{Witten_I_1987_j-comm-acm_arithmetic_cdc} is then used to encode the connectivity at very low bit rates, between \num{0.18} and \num{1.55} bits  per hexahedron. Regarding geometry, a user-defined quantization first restricts the number of bits for coordinates, and then a predictive scheme based on the parallelogram rule encodes the position of vertices added during the region-growing process.

Krivograd \etal \cite{Krivograd_S_2008_j-comput-aided-des_hexahedral_mccvd} proposed a variant to \cite{Isenburg_M_2003_j-graph-model_compressing_hvm} that encodes the vertex degrees --- number of non-compressed hexahedra around a given vertex --- instead of the edge degrees. On the one hand, this variant achieves better compression performances than \cite{Isenburg_M_2003_j-graph-model_compressing_hvm} for dense meshes. On the other hand, it only deals with manifold meshes, and the algorithm is complex as  interior cells are encoded after  border cells (it involves many specific cases to process when encoding the connectivity).

Lindstrom and Isenburg proposed an original algorithm for unstructured meshes called Hexzip \cite{Lindstrom_P_2008_p-dcc_lossless_chm}. This algorithm is considered as fully lossless, because the initial indexing of vertices and hexahedra is preserved. For this purpose, connectivity is encoded directly in its indexed structure, by predicting the eight indices of an hexahedron from preceding ones. This technique is suitable because hexahedral meshes generally have regular strides between indices of subsequent hexahedra. A hash-table is then used to transform the index structure into a very redundant and byte-aligned list of symbols, that can be compressed efficiently with gzip (discussed in Section \ref{sec_results_lossless}). Concerning the geometry, spectral prediction \cite{Ibarria_L_2007_p-dcc_spectral_p} is used. This algorithm is faster and less memory intensive than \cite{Isenburg_M_2003_j-graph-model_compressing_hvm} as the connectivity is not modified. It handles non-manifold meshes and degenerate elements. On the other hand, bitrates are higher because of the lossless constraints.
Unlike methods presented above, Chen \etal \cite{Chen_D_2005_p-eusipco_geometry_ctmuop} focus on  geometry compression for tetrahedral meshes. The authors proposed a \emph{flipping} approach based on an extension of the \emph{parallelogram rule} (initially proposed for triangle meshes \cite{Touma_C_1998_p-gi_triangle_mc}) to tetrahedra. It consists in predicting the position of an outer vertex of two face-adjacent tetrahedra, with respect to the other vertices. To globally optimize the geometry compression, a Minimum Spanning Tree minimizing the global prediction error for the whole mesh is computed. This method is more efficient than prior flipping approaches whose traversal does not depend on the geometry, but solely on the connectivity.

Streaming compression is a subcategory of single rate compression, dedicated to huge data that cannot fit entirely in the core memory. A particular attention to I/O efficiency is thus required, to enable the encoding of huge meshes with a small memory footprint.
Isenburg and coworkers are the first to propose streaming compression for \vms (extended from his method for triangular meshes \cite{Isenburg_M_2005_p-eurographics-symp-geom-process_streaming_ctm}): for tetrahedral meshes  \cite{Isenburg_M_2006_p-graphics-interface_streaming_ctvm}, and then for hexahedral meshes \cite{Courbet_C_2010_j-vis-comput_streaming_chm}. In the latter, for instance, the compressor does not require the knowledge of the full list of vertices and cells before encoding. The compressor starts encoding the mesh as soon as the first hexahedron and its eight adjacent vertices have been read. For a given hexahedron: i) its face-adjacency is first encoded in function of its configuration with hexahedra already processed; ii) positions of vertices that are referenced for the first time are predicted (spectral prediction from adjacent cells); iii) prediction errors are encoded; iv) data structures relative to vertices, becoming useless (because their incidence has been entirely described) are finally removed from memory. Compared to other single rate techniques, streaming  tends to achieve similar compression performances for geometry, but poorer performances for connectivity.

\subsubsection{Progressive/multiresolution mesh compression}
\label{subsubsec:progressive_algorithms}
\emph{Progressive} algorithms (also called scalable or \emph{multiresolution}, see Section \ref{sec_background-mr} for details)  enable the original meshes to be represented and compressed at successive  LODs (levels of details). The main advantage is that it is not necessary to decompress a mesh entirely before vizualising it. A coarse approximation of the mesh (also known as its lowest resolution) is first decompressed and displayed. Then this coarse mesh is updated with the successive LOD (termed higher resolutions) that are decompressed progressively. While they cannot achieve yet  compression performance  of single-rate algorithms, progressive algorithms are popular because they enable LOD, and also adaptive transmission and displaying, in function of user constraints (network, bandwidth, screen resolution\ldots).

Pajarola \etal \cite{Pajarola_R_1999_p-visualization_implant_scptmc} are the first to propose in 1999 a progressive algorithm dedicated to \vm compression. This work is inspired by a simplification technique for tetrahedral meshes \cite{Staadt_O_1998_p-visualization_progressive_t}. It simplifies a given tetrahedral mesh progressively, by using successive \emph{edge collapses} \cite{Hoppe_H_1993_p-acm-siggraph-comput-graph_mesh_o}. Each time an edge is collapsed, its adjacent cells are removed, and all the information required to reverse this operation is stored: index of the vertex to split, and the set of incident faces to ``cut". Thus, during decompression, the LODs can be also recovered iteratively, by using the stored data describing \emph{vertex splits}. During coding, an edge is selected such as its collapse leads to the minimal error, with respect to specific cost functions. This algorithm gives a bitrate inferior to \num{6} bits per tetrahedron (for connectivity only).

In 2003, Danovaro \etal \cite{Danovaro_E_2002_p-smi_multiresolution_tmac} proposed two progressive representations based on a decomposition of a field domain into tetrahedral cells. The first is based on \emph{vertex splits}, as the previous method, the second is based on \emph{tetrahedron bisections}. This operation consists in subdividing a tetrahedron into two tetrahedra by adding a vertex in the middle of its longest edge. Unlike with vertex splits, the representation based on \emph{tetrahedron bisections} is obtained by following a coarse-to-fine approach, \emph{i.e.}, by applying successive bisections to an initial coarse mesh. Also, this representation only needs to encode the difference vectors between the vertices added by bisections and theirs real positions. This representation is thus more compact, as the mesh topology does not need to be encoded, but it only deals with structured meshes.

\vms multiresolution decomposition based on wavelets \cite{Jacques_L_2011_j-sp_panorama_mgrisdfs} was proposed by Boscard\'in \etal \cite{Boscardin_L_2006_j-istec-jcst_wavelets_bdot}  for tetrahedral meshes.
It is based on the tetrahedron subdivision scheme \cite{Bey_J_1995_j-computing_tetrahedral_gr} that transforms a tetrahedron into 8 sub-tetrahedra, by introducing 6 new vertices on each edge. After analysis, the input mesh is replaced by a base tetrahedral mesh, and several sets of detail wavelet coefficients. Although coefficients corresponding to differences between two resolutions could be encoded for mesh synthesis, this work does not provide an actual compression scheme.

\begin{figure}[htb]
  \centering
  \includegraphics[width=0.9\linewidth]{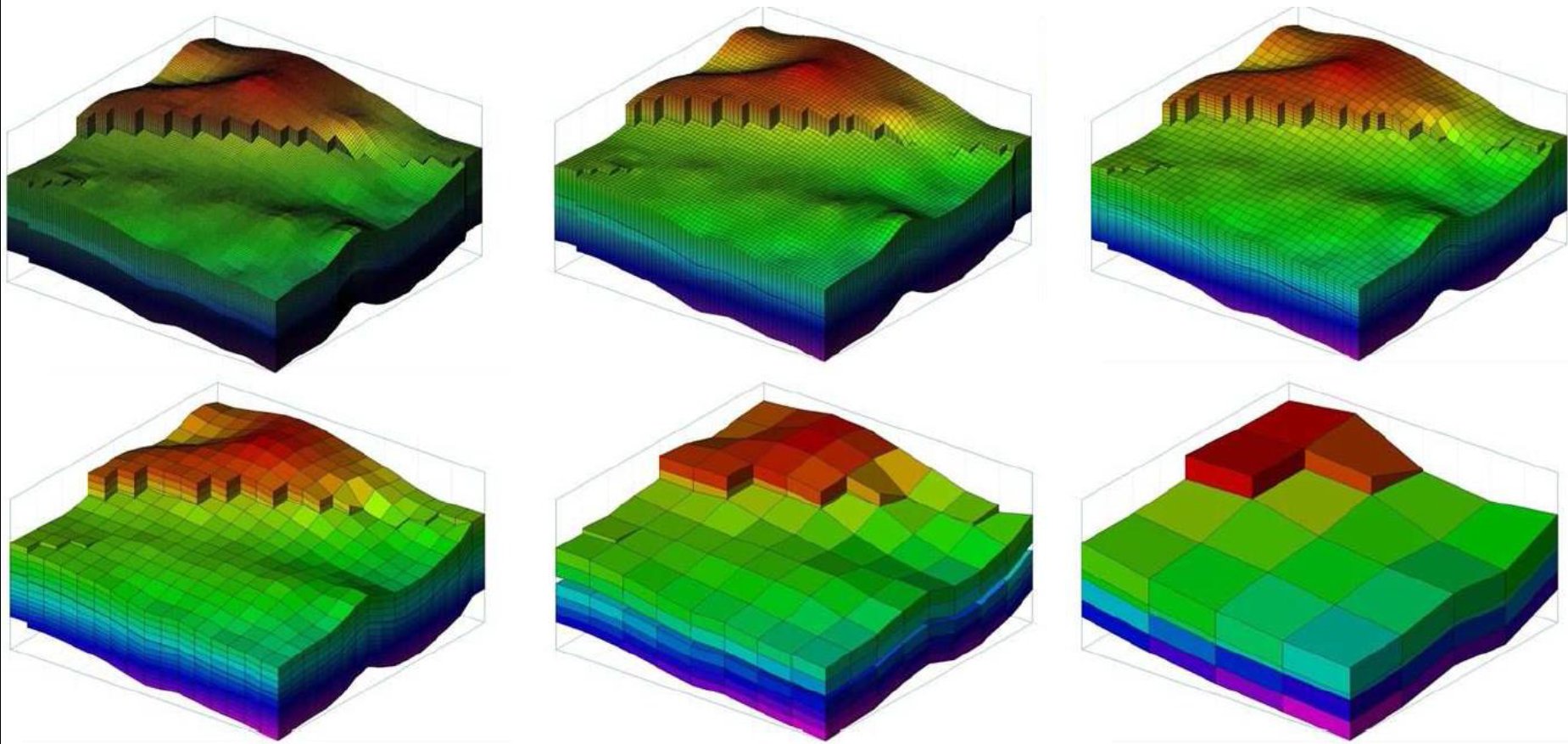}
  \caption{Dyadic non-separable multiresolution rendering  on a simple geologic mesh \cite{Chizat_L_2014_msc_multiresolution_scea}.\label{fig:Chizat_L_2014_msc_multiresolution_scea_results}}
\end{figure}

In \cite{Chizat_L_2014_msc_multiresolution_scea},  Chizat proposed a prototype for a multiresolution decomposition of geoscientific hexahedral meshes with the pillar grid structure (Section \ref{subsec:structured_meshes}). His main contribution resides in a multiresolution analysis (MRA) that partially manages  geometrical discontinuities representing the faults. It can be achieved by using a morphological wavelet transform (Section \ref{sec_2d-morphological-wavelet}). This non-separable transform enables the preservation of some fault shapes at different resolutions, as depicted in Figure \ref{fig:Chizat_L_2014_msc_multiresolution_scea_results}.

The latter work is a seed for the upcoming description of \hs.

\section{Global \hs algorithmic workflow\label{sec_methodology}}
\subsection{Multiresolution analysis: background\label{sec_background-mr}}

MRA or  multiscale \emph{approximation} can be interpreted as a decomposition of data at different resolutions, LOD or  scales, through a recursive \emph{analysis} process. It is called exact, reversible or invertible when a \emph{synthesis} scheme can retrieve the original data. Inter-scale relationships \cite{Chaux_C_2007_j-ieee-tit_noise_cpdtwd,Chaux_C_2008_j-ieee-tsp_nonlinear_sbemid} often yield  \emph{sparsification} or increased  \emph{compressibility} on sufficiently regular datasets. In discrete domains, each \emph{analysis} stage transforms a set of values (continuous or categorical, in one or several dimensions), denoted by $S^{0}$. The resulting representation consists in one  subset $S^{-1}$  of approximation coefficients at a lower resolution identified by a negative  index $S^{-1}$ that approximates the original signal, \emph{plus} one subset of details $D^{-1}$ or a combination thereof. The latter  represents information missing in the approximation $S^{-1}$. Depending on the MRA scheme, the lower \emph{resolution} $S^{-1}$ may represent a coarsening or  ``low frequencies'' of the original samples, or an upscaling in geosciences (cf. Section \ref{sec_methodology}). The subset $D^{-1}$ represents refinements, fast variations or  ``high-frequency'' details removed from $S^{0}$. We consider here exact systems, allowing the perfect recovery of $S^{0}$ from a combination of subsets $S^{-1}$ and $D^{-1}$. Hence, a similar  analysis stage can be applied iteratively, and perfectly again, to the lower resolution $S^{-1}$, in a so-called pyramid scheme. Thus, with the non-positive extremum decomposition level $L$, and indices $0\ge l\ge L$, after an $|L|$-level multiresolution decomposition, the input set $S^{0}$ is now decomposed and represented by the subset $S^{L}$ --- a (very) coarse approximation of $S$ --- and $|L|$ subsets of details $D^{L},\ldots, D^{l},\ldots, D^{-1}$, representing  information missing between each two consecutive approximations.

We consider in the following four different MRA flavors, all called wavelets for simplicity. They stem from iterated, (rounded) linear or non-linear combinations of coefficients, as well as separable (applied separately in $1D$ on each direction) or non-separable ones. Without going into  technicalities here (cf. Section \ref{sec_rounded-wavelet}), computations are performed using the \emph{lifting scheme}. It suffices to mention that lifting uses complementary interleaved grids of values, often indexed with odd and even indices. Values on one grid are usually  predicted (approximations) and updated (details) from the others. The main interests reside in reduced computational load, in-place computations  and the possibility to maintain exact integer precision, using for instance only dyadic-rational coefficients (written as $m/2^n$, $(m,n)\in \mathbb{Z}\times\mathbb{N}$) and rounding. We refer to \cite[sections 2.3, 3.2 and 4.3]{Jacques_L_2011_j-sp_panorama_mgrisdfs} for a concise account on both  non-separable and non-linear wavelet MRAs, and to \cite{Sweldens_W_1996_j-acha_lifing_scdcbw,Bruekers_F_1992_j-ieee-sel-areas-com_new_npipr,Rao_R_1998_book_wavelet_tita,Kovacevic_J_2012_book_signal_pfwr} for  more comprehensive visions of wavelets and their lifting implementations. A recent use in geological model upscaling is given in \cite{Rezapour_A_2019_j-transp-porous-med_upscaling_gmorugulbgwt}.

\begin{figure}[htb]
\centering
\includegraphics[width=\linewidth]{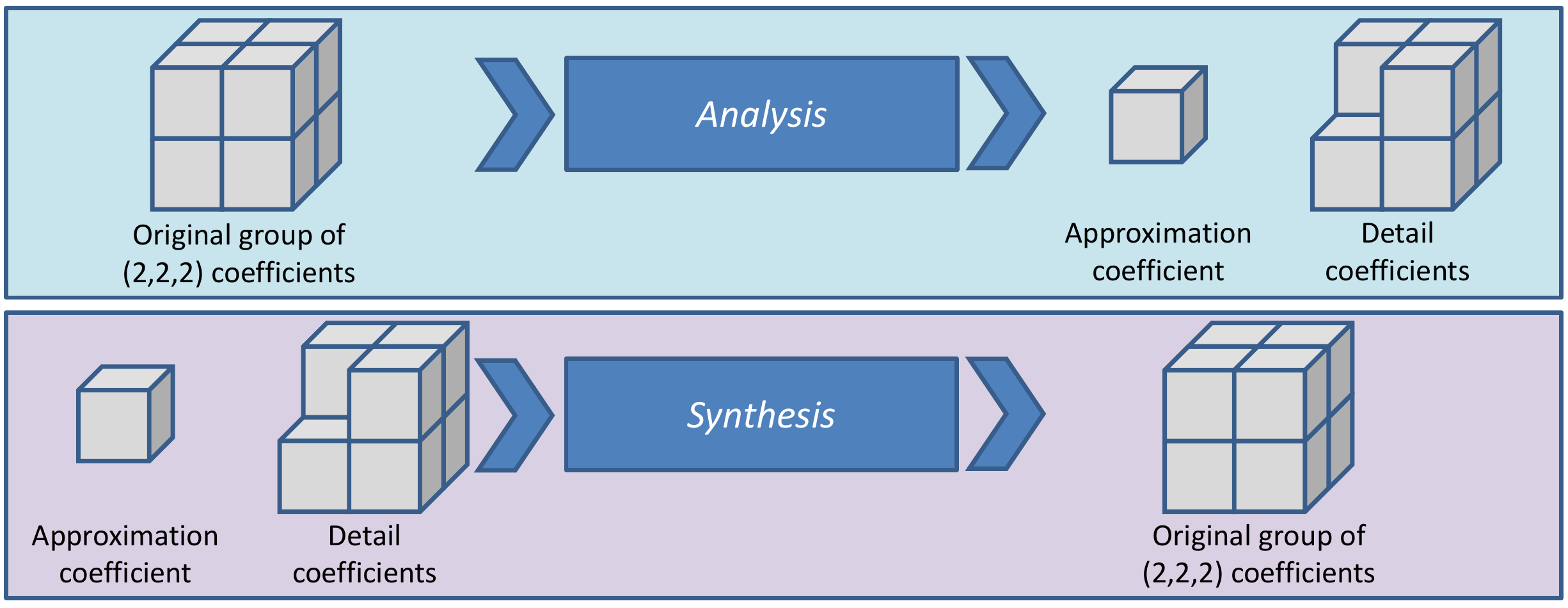}
\caption{Analysis and synthesis stages for \vms.\label{fig:Wavelet_transform}}
\end{figure}

More simply put, for our hexahedral \vms, the dyadic analysis stage transforms each cell block $\mathcal{C}^l$ of values around $2^3=8$ contiguous cells (possibly borrowing  values from a limited cell neighborhood) at resolution ${l}$. They are turned into one \emph{approximating cell} (lower resolution ($S^{l-1}$), and a subset of $2^3-1=7$ detail cells  $D^{l-1}$, as depicted in Figure \ref{fig:Wavelet_transform} along with the reverse synthesis stage.
Hence, if a \vm at resolution $l$ is composed of $\left(\mathcal{C}_{i}^l \times \mathcal{C}_{j}^l \times \mathcal{C}_{k}^l \right)$ cells, $\mathcal{C}_{i}^l$ being the number of cells in  direction $i$, the \vm of lower resolution will be of dimension $\left\lceil \frac{\mathcal{C}_i^l}{2} \right\rceil \times \left\lceil \frac{\mathcal{C}_j^l}{2} \right\rceil \times \left\lceil \frac{\mathcal{C}_k^l}{2} \right\rceil $, to take into account non-power-of-two sized grids.
As several digital attributes are associated to each cell (geometry, continuous or categorical properties), different types of MRA are  performed separately on the different variables defining these properties, as explained in the following sections.

\subsection{Multiresolution scheme for geometry\label{sec_geometry_compression}}
\begin{figure}[htb]
		\centering
		\subfigure[A node and its $8$ surrounding cells.]{\label{fig:Cube_naming_convention_a}\includegraphics[width=0.4\linewidth]{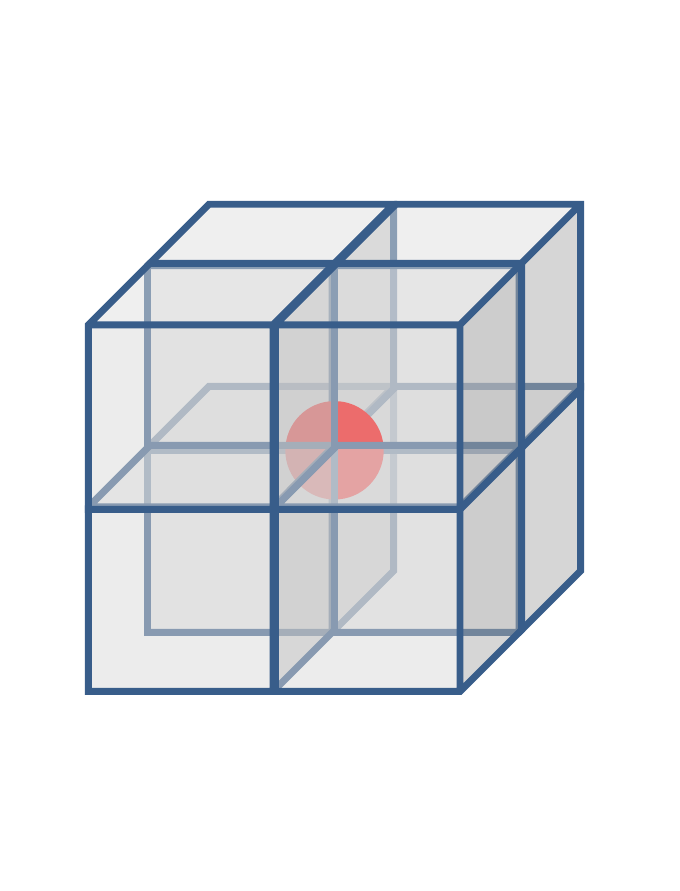}}
		\ \ \ \subfigure[Splitting view of the node into its $8$ vertices.]{\label{fig:Cube_naming_convention_b}\includegraphics[width=0.4\linewidth]{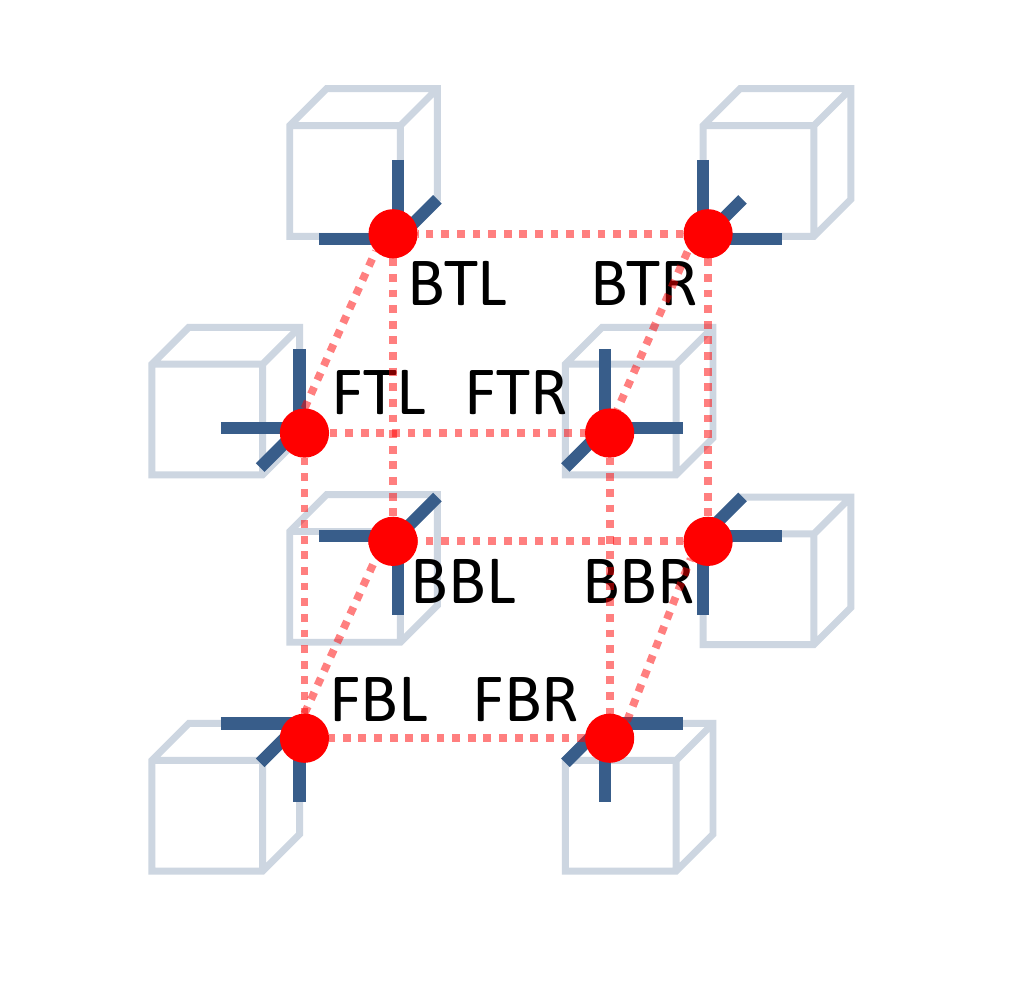}}
		\caption{Vertex naming  with the Back (B)/Front (F), Bottom (B)/Top (T), Left (L)/Right (R) convention.\label{fig:Cube_naming_convention}}
		\end{figure}

Standard linear MRA schemes rely on smoothing or averaging and difference filters for approximations and details, respectively. To preserve coherency of  representation of geometrical discontinuities --- whatever the resolution --- a special care is taken to avoid excessive smoothing, while at the same time  allowing the reverse synthesis.
 As the pillar grid format is used (see Section \ref{subsec:structured_meshes}), vertices are inevitably positioned along pillars. So, our multiresolution scheme for geometry data only focuses on:
 \begin{itemize}
   \item the $z$ coordinates of the 8 vertices associated to each node. According to the naming convention presented in Figure \ref{fig:Cube_naming_convention}, those $8$ vertices can be differentiated according to their relative positions [Back (B)/Front (F), Bottom (B)/Top (T), Left (L)/Right (R)];
   \item  the $x$ and $y$ coordinates of the nodes describing the low (bottom) and high (top) extremities of all the pillars (the $x$ and $y$ coordinates of  intermediary nodes being implicit). The nodes are called hereinafter the \emph{floor} and \emph{ceil} nodes, respectively.
 \end{itemize}

\begin{figure}[htb]
\centering
\includegraphics[width=0.7\linewidth]{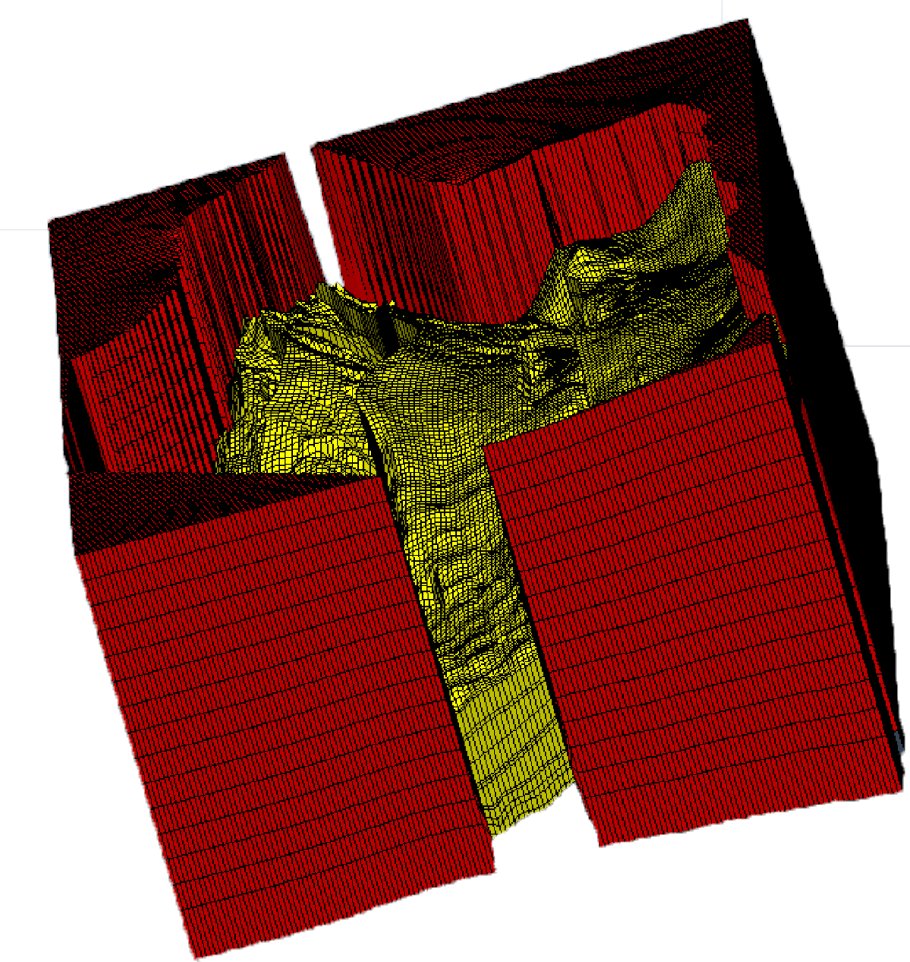}
\caption{\Ms{5} (in yellow) has inactive cells (in red) to describe its boundaries using the \actn field.\label{fig:fullfield_actnum_0}}
\end{figure}

Actual $3D$ meshes can exhibit very irregular boundaries. Hence, a Boolean field called \actn may be associated to each cell to inactivate its display (and its influence during simulations as well). It enables the description of either mesh boundaries (Figure \ref{fig:fullfield_actnum_0}), or caves/overhangs. Resultantly, this \actn field must be carefully considered  during the multiresolution analysis of the geometry data, to avoid artifacts at lower resolutions on frontiers between active and inactive cells (see Section \ref{sec-externalities} and Figure \ref{fig:fullfield_bad_good}).

By construction, most geological \vms have no horizontal fault, as there is no vertical gap between any two adjacent layers of cells. For every node, each of the four top vertices has the same $z$ coordinate as its counterpart bottom vertex. Therefore, from now on, our geometry multiscale representation method only deals with the $z$ coordinates
of the bottom vertices BBL, BBR, FBL, and FBR of each node.\\

\begin{figure}[htb]
		\centering
	\includegraphics[width=\linewidth]{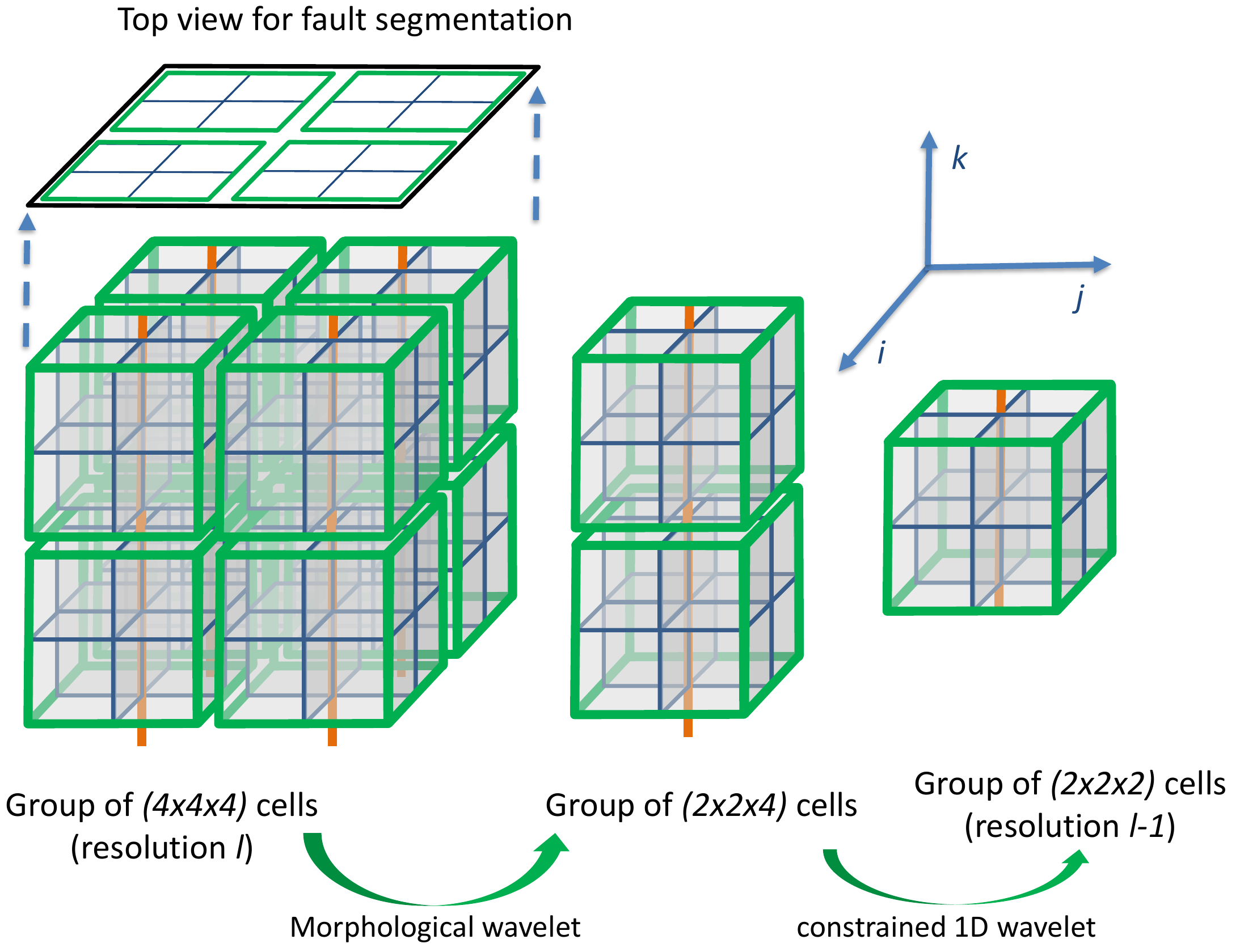}
		\caption{\hs multiresolution scheme for geometry: (left) input grid and its top view; (middle) output from the  non-separable, non-linear $2D$ morphological wavelet based on a fault segmentation (based on the top view);  (right) non-linear $1D$ wavelet transform along pillars (orange lines). \label{fig:principle_Z_wavelet_fred}}
		\end{figure}

An instance of the decomposition shown in Figure \ref{fig:Wavelet_transform} can be implemented with the proposed two-step technique depicted by arrows in Figure  \ref{fig:principle_Z_wavelet_fred}:
 \begin{itemize}
   \item A non-linear and non-separable \textbf{$2D$ morphological wavelet transform} applied on the nodes, in order to detect the faults in the input \vm, and then to preserve their coherency in the lower resolutions. This step relies on a \textbf{fault segmentation} within the input \vm obtained by studying all possible fault configurations for the top view of the \vm (see Figure \ref{fig:Fault_node_configurations});
   \item A non-linear \textbf{$1D$ wavelet transform} applied on the output of the above first step to analyze \textbf{the $z$ coordinates of the vertices} along each pillar. The same $1D$ wavelet transform is also applied on the sets of $x$ and $y$ coordinates of the \emph{floor} and \emph{ceil} nodes, to complete the ``horizontal'' decomposition.
 \end{itemize}

\subsubsection{Fault segmentation}
\begin{figure}[ht]
      \centering
      \includegraphics[width=0.8\linewidth]{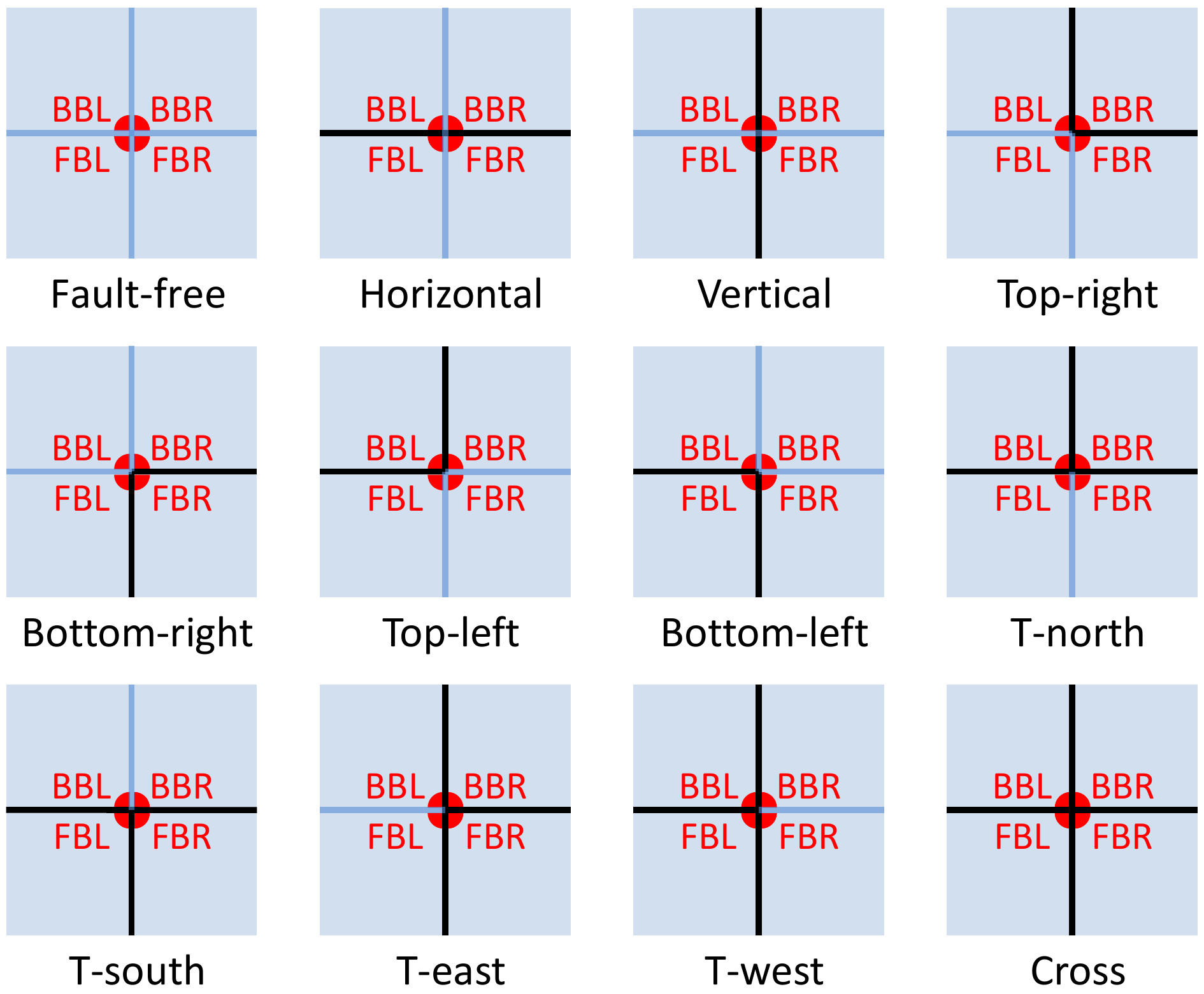}
      \caption{The $12$ possible fault configurations (in black lines) at a given node. \label{fig:Fault_node_configurations}}
      \end{figure}

This stage detects the faults in the original mesh, in order to preserve them during the morphological wavelet analysis. For each node, a dozen of fault configurations, depending on BBL, BBR, FBL, and FBR, is possible: fault-free (1), straight (2), corner (4), T-oriented (4) or cross (1), as illustrated in Figure \ref{fig:Fault_node_configurations}.

\begin{figure}[htb]
		\centering
		\includegraphics[width=0.8\linewidth]{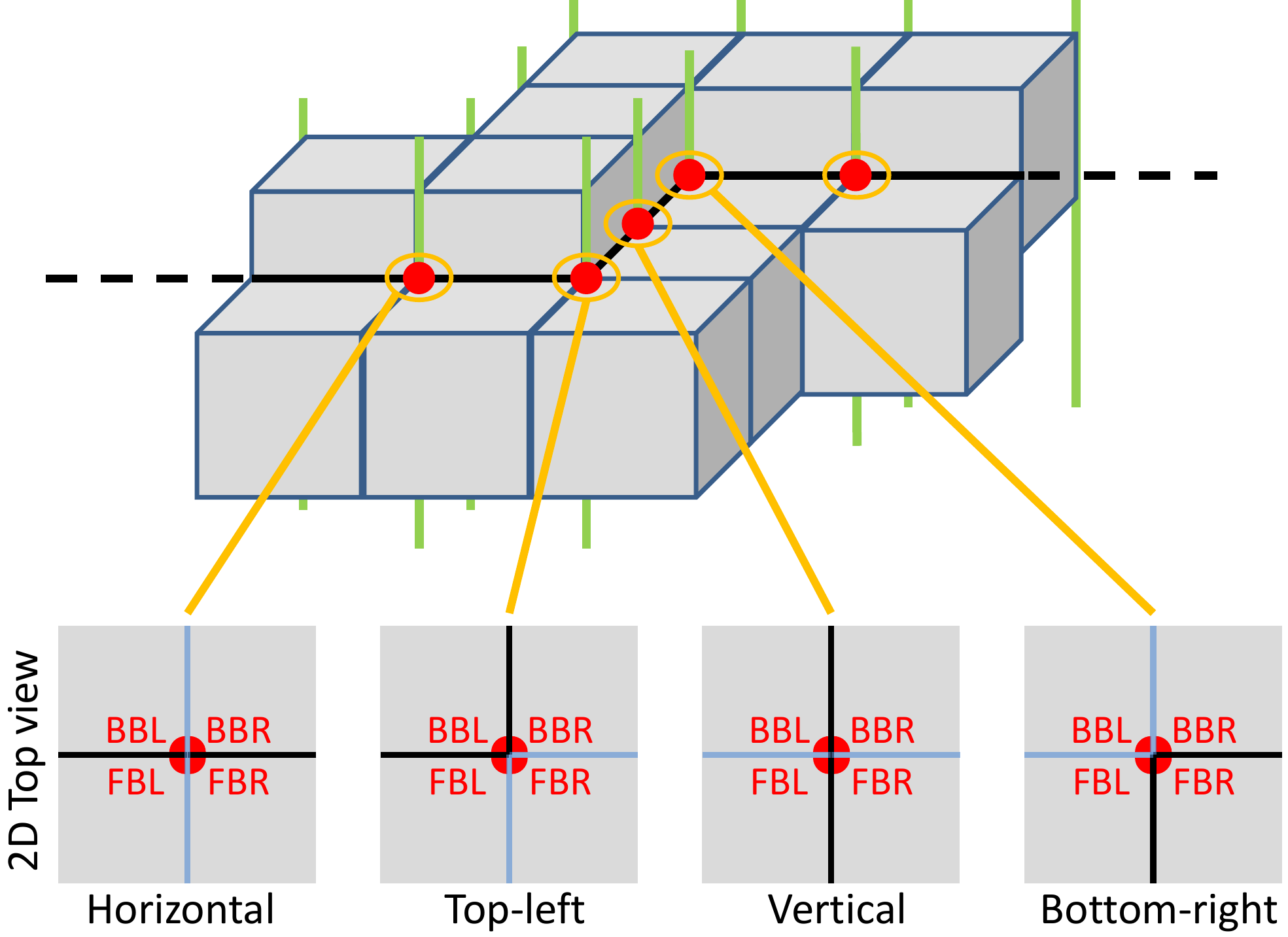}
		\caption{Fault segmentation within the original mesh.\label{fig:Fault_segmentation}}
		\end{figure}

Each configuration depends on the four orientations of the cardinal axes (north, south, east and west), which are either active or inactive. For instance, the T-north configuration has its south axis inactive, while the three remaining ones are active.
Assuming that a fault configuration is $z$-invariant, meaning that the nodes belonging to the same pillar present the same fault configuration, a single $2D$ configuration map is sufficient to represent the fault configuration of the whole mesh, as illustrated by Figure \ref{fig:Fault_segmentation}.

\subsubsection{Horizontal $2D$  morphological wavelet transform\label{sec_2d-morphological-wavelet}}
The fault segmentation guides the multiresolution analysis to preserve faults, as much as possible, all over the decomposition process.
The fault configuration of  \num{4} associated nodes at resolution $l$ is used to predict the extension of the downsampled fault structure at resolution $l-1$.

\begin{figure}[htbp]
		\centering \includegraphics[width=0.6\linewidth]{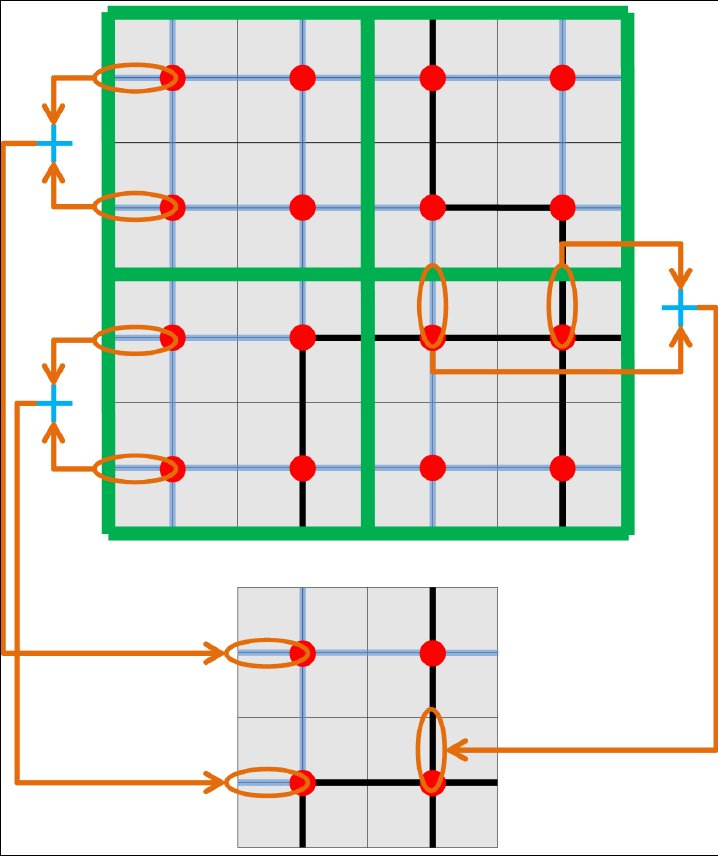}
		\caption{Prediction of a fault node  at resolution $l-1$ from the four parents' configuration at resolution $l$, orange ovals denoting $\lor$ operands. \label{fig:Fault_configuration_prediction}}
\end{figure}

This horizontal prediction is based on the logical function OR ($\lor$), computed on each side of each group of \num{4} nodes. For instance, a resulting fault node configuration contains a west axis if the fault configurations of the \num{2} left nodes contain at least \num{1} west axis, as illustrated in Figure \ref{fig:Fault_configuration_prediction}. By repeating the procedure for the north, south and east axes of each resulting node, fault node configurations at lower resolutions are fully predicted. This  non-linear and peculiar choice is meant to maintain a directional flavor of orientated faults for  flows; other choices could be devised, depending on physical rules and geological intuitions.

Finally, from this prediction, the node whose configuration minimizes its distance with the predicted one, corresponds to the aforementioned approximation coefficient, which will be part of the novel $Z$ matrix at  lower resolution $l-1$. The same procedure  can be applied recursively until the wanted resolution.

\subsubsection{Rounded linear $1D$  wavelet transform\label{sec_rounded-wavelet}}

\begin{figure}[tbh]
		\centering
		\includegraphics[width=\linewidth]{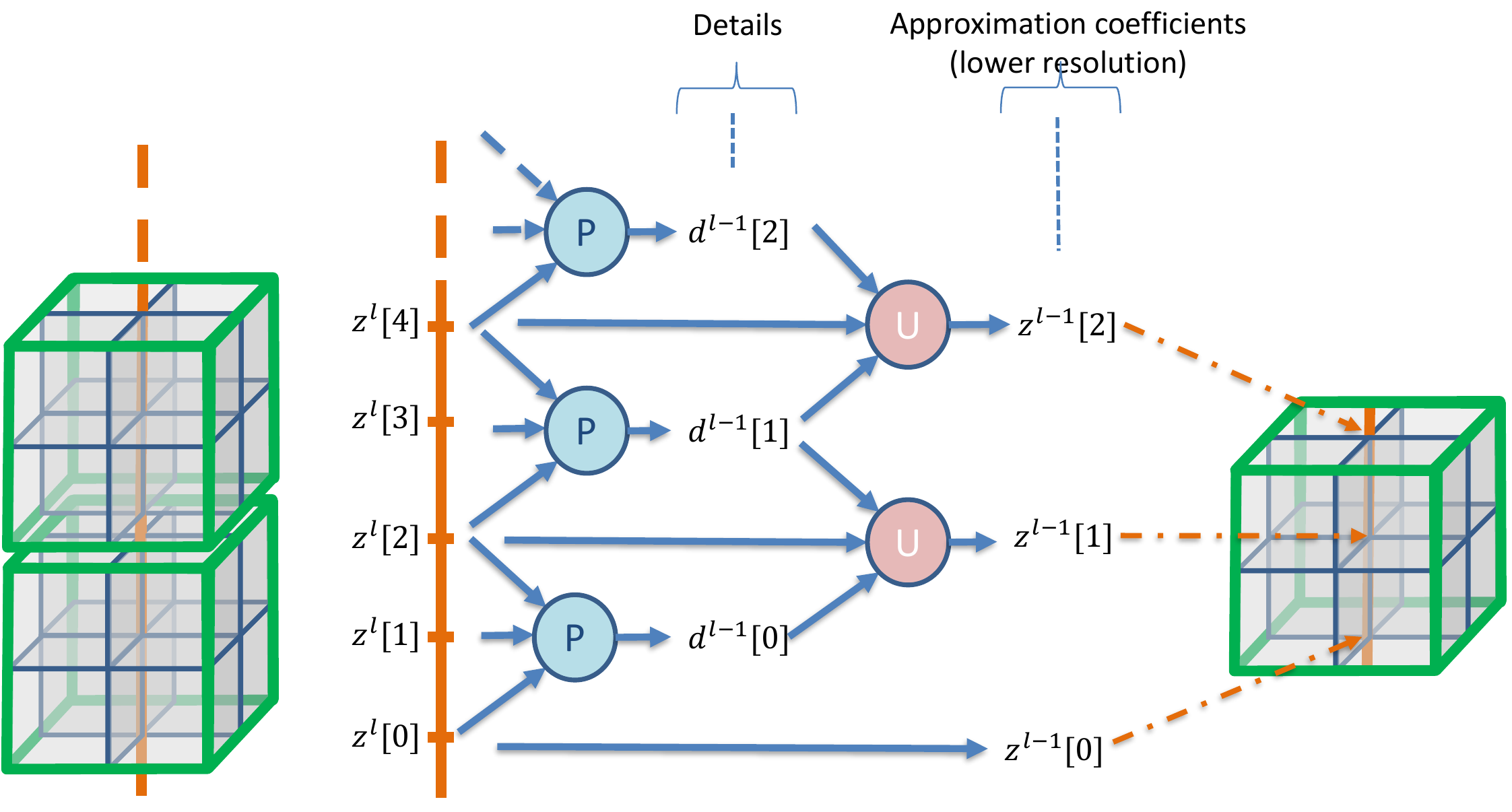}
		\caption{Principle of the lifting scheme (Prediction and Update) for the rounded linear $1D$  wavelet from \eqref{eq:ForwardDetailFilterUsualBior22}-\eqref{eq:ForwardApproximationFilterUsualBior22}, to analyze  $z$ coordinates of  vertices BDR, FDR, BDL and FDL  along each pillar, as well as $x$ and $y$ coordinates of the \emph{floor} and \emph{ceil} nodes.\label{fig:principle_Z_wavelet_fred2}}
		\end{figure}

 This  $1D$ wavelet transform is applied on the output of the above horizontal $2D$ morphological wavelet, to analyze the $z$ coordinates of the \num{4} sets of vertices BDR, FDR, BDL and FDL separately, along each selected pillar. 
Hence,  \hs here decomposes at each scale $z^{l}$ into a subsampled pillar coordinate $z^{l-1}$ and its associated detail $d^{l-1}$.  By geomodel construction, coordinate behavior along the pillars is expected to be relatively smooth.  This  entails the use of  a modified, longer spline wavelet. The latter can be termed LeGall \cite{LeGall_D_1988_p-icassp_sub-band_cdiusskfact}, or CDF $5/3$ (after Cohen, Daubechies and Feauveau \cite{Cohen_A_1992_j-comm-acm_biorthogonal_bcsw}), or biorthogonal $2.2$ from its vanishing moments.

The lifting analysis operations \emph{Prediction} and \emph{Update} are depicted by Figure \ref{fig:principle_Z_wavelet_fred2}. To retrieve respectively the sets of details $ d^l $ and the $z^l$ coordinates at resolution $l-1$ from scale $l$, the following equations are used ($\forall n \in \mathbb{N}$):
\begin{align}
 d^{l-1}[n] &= z^{l}[2n+1] - \left\lfloor \frac{z^{l}[2n] + z^{l}[2n+2]}{2} \right\rfloor \label{eq:ForwardDetailFilterUsualBior22}\,,\\
 z^{l-1}[n] &= z^{l}[2n+0] + \left\lfloor \frac{d^{l-1}[n-1] + d^{l-1}[n]}{4} \right\rfloor\,, \label{eq:ForwardApproximationFilterUsualBior22}
\end{align}
where both dyadic integers and rounding are evident (see Section \ref{sec_background-mr}). With rounding, lifting schemes can thus manage integer-to-integer transformations \cite{Calderbank_A_1998_j-acha_wavelet_ttmii}. For synthesis, to reconstruct  resolution $l$ from  resolution $l-1$, we only have to reverse the order and
the sign of the equations:
\begin{align}
 z^{l}[2n] &= z^{l-1}[n] - \left\lfloor \frac{d^{l-1}[n-1] + d^{l-1}[n]}{4} \right\rfloor\,, \label{eq:InverseApproximationFilterUsualBior22}\\
z^{l}[2n+1] &= d^{l-1}[n] + \left\lfloor \frac{z^{l}[2n] + z^{l}[2n+2]}{2} \right\rfloor\,. \label{eq:InverseDetailFilterUsualBior22}
\end{align}

\subsubsection{Managing externalities: borders and boundaries\label{sec-externalities}}

A pertinent  multiresolution on  complex meshes requires to cope with externalities that may hamper their handling: floor and ceil  borders and  outer  boundaries (Figure \ref{fig:fullfield_actnum_0}). First, to keep borders unchanged from the original mesh, throughout all resolutions, the  following constraints must be  met:

\begin{align}
	z^{l-1}[0] &= z^{l}[0]\,, \label{eq:LowerBoundConstraint}\\
z^{l-1}[n_{k}^{l-1}-1] &= z^{l}[n_{k}^{l}-1]\,. \label{eq:UpperBoundConstraint}
\end{align}

Both constraints can be fulfilled if one satisfies the following conditions:
\begin{itemize}
\item Floor border condition to meet  \eqref{eq:LowerBoundConstraint}:
		\begin{equation}
			d^{l-1}[-1] = -d^{l-1}[0]\,, \label{eq:LowerBoundCondition}
		\end{equation}
\item Ceil border condition to meet  \eqref{eq:UpperBoundConstraint}:
\begin{align}
	d^{l-1}[n_{k}^{l-1}-1] = -&d^{l-1}[n_{k}^{l-1}-2]\,,\quad\textrm{($n_{k}^{l}$  odd)} \label{eq:UpperBoundCondition0}\\
  d^{l-1}[n_{k}^{l-1}-1] = -&d^{l-1}[n_{k}^{l-1}-2]\phantom{\,,}\quad\textrm{($n_{k}^{l}$  even)}\nonumber\\
	+&4 z^{l}[n_{k}^{l}-1]-4 z^{l}[n_{k}^{l}-2]\,. \label{eq:UpperBoundCondition1}
\end{align}
\end{itemize}

To complete the MRA of the geometry, the same rounded $1D$ wavelet as in Section \ref{sec_rounded-wavelet} is finally applied to the sets of \emph{floor} and \emph{ceil} nodes of the initial \vm , to get the $x$ and $y$ coordinates of the extremities of the remaining pillars at the lower resolution.

\begin{figure}[ht]
		\centering
    	\subfigure[Boundary artifacts without proper  \actn management.]{
    \includegraphics[width=0.4\linewidth]{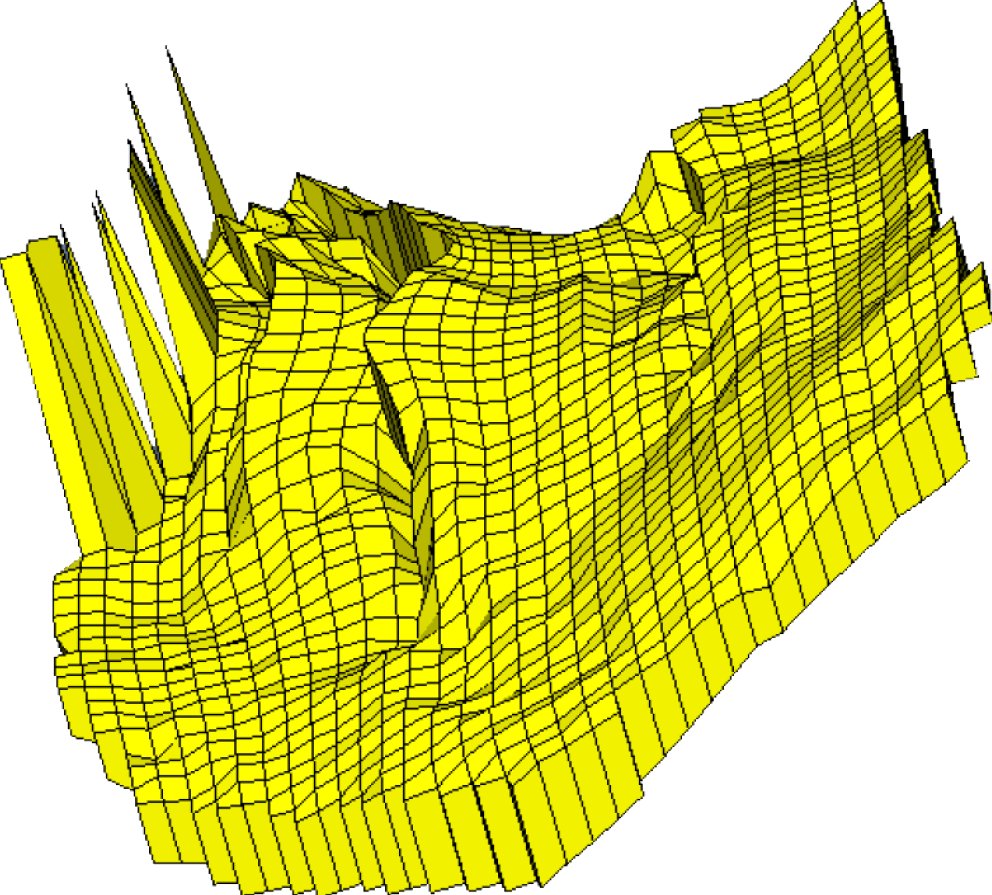}} \quad
        \subfigure[Lower mesh resolution with  efficient  \actn Boolean values care-taking.]{\includegraphics[width=0.4\linewidth]{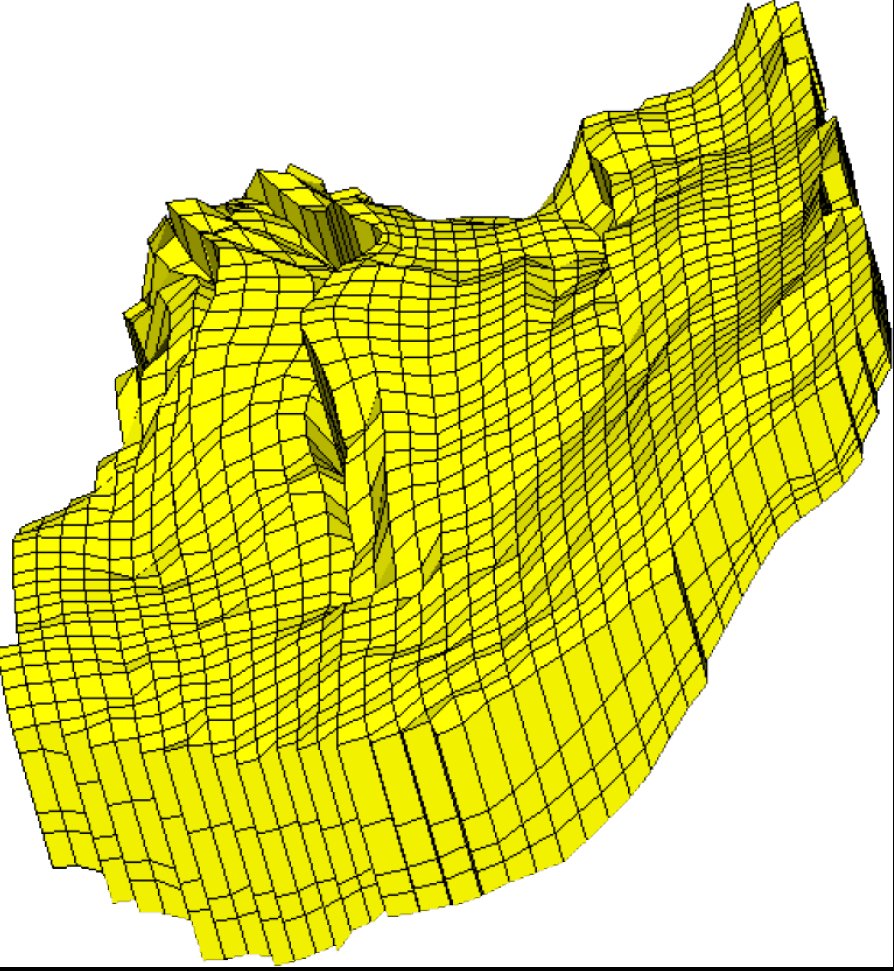}}
	\caption{Inadequate \actn fields management during analysis may lead to severe boundary artifacts (left) that can be dealt with (right) as \emph{examplified} with \ms{5} from Figure \ref{fig:fullfield_actnum_0}. \label{fig:fullfield_bad_good}}
\end{figure}

Second, the  \actn field should also be considered to lessen mesh boundary artifacts. Indeed, severe disturbances may appear at lower resolutions if  not wisely processed during analysis, as shown in Figure \ref{fig:fullfield_bad_good}.

A cell is deemed active if and only if its \num{8} adjacent vertices are active at the resolution $l$. During our study, we find that one vertex at resolution $l-1$ could be considered active if and only if its parent vertices selected  by the morphological wavelet at resolution $l$ (Section \ref{sec_2d-morphological-wavelet}) are active. So, a cell at resolution $l-1$ is considered active if and only if its $8 \times 2$ corresponding parent vertices are active at resolution $l$.

\subsection{Multiresolution scheme for continuous properties\label{sec_continuous_property_compression}}

Once the geometry is coded, one can focus on associated continuous properties. For scalar ones, a value  $p_i \in \mathbb{R}$ is associated to each cell $i$ in the mesh.
Consistently with the handling of cell blocks $\mathcal{C}$ of $2\times2\times 2$ cells throughout scales, we use an adaptation of the well-known \emph{Haar} wavelet. The resolution $l-1$ is a scaled average of cells at resolution $l$. The approximation coefficient $p^{l-1}$ is thus the average value of the related eight property coefficients $\{p_1^l, p_2^l,\ldots,p_8^l\}$.  The seven  details required for synthesis  are  differences with respect to  the approximation coefficient:
$$p^{l-1}  = \frac{1}{8}\sum_{n=1}^8 p_n^l\,; \quad d_n^{l-1} = p_n^l - p^{l-1}\,,\, \forall n \neq 1\,.$$

To deal with real-valued (floating-point) properties, and avoid accuracy imprecision due to the divide operator, we introduce the following modifications. First, reals are mapped into integers up to a user-defined precision, here with a  $10^6$ factor. Second, we disable the division by using a sum. The analysis system thus becomes:
 $$\label{eq:haar_3}p^{l-1}  = \sum_{n=1}^8 p_n^l\,; \quad  d_n^{l-1} = 8p_n^l - p^{l-1}\,,\,\forall n \neq 1\,, $$
 and the synthesis system turns into: 
$$\label{eq:haar_5} p_n^{l} = \frac{1}{8}(d_n^{l-1} + p^{l-1})\,,\,\forall n \neq 1\,; \quad p_1^l = p^{l-1} - \sum_{n=2}^8 p_n^l\,.$$

Approximation and coefficients are stored as is. To recover the accurately scaled values, the division operator should however be applied as a simple linear post-processing.

\subsection{Multiresolution scheme for categorical properties\label{sec_categorial_property_compression}}
We finally complete the global mesh multiresolution decomposition with an original categorical-valued  scheme called \emph{modelet} \cite{Antonini_M_2017_patent_method_ehufmos}. We assume that a mesh cell  category belongs  to a set of classes $\Omega_0 = \{\omega_1, \omega_2, ..., \omega_W \}$, taking discrete values. The cell block $\mathcal{C}^l$
$\{ p_1^l, p_2^l,\ldots,p_8^l \}$ thus contains,  at resolution $l$,  integers indexing   categories from $\Omega_l $. They take values in a subset  of $\Omega$.
The multiresolution scheme is expected to produce, at lower resolutions, discrete values in embedded subsets: $\Omega_0 \supset \Omega_{-1} \supset  \cdots \supset \Omega_l  \supset \cdots $. In other words, a cell category can only belong to an existing category at an upper resolution. We choose here the modal value (mode) \emph{i.e.}, the most frequently represented in $\mathcal{C}^l$. If $|\omega_w|$ denotes the cardinal of this class, then $\sum_{w=1}^W |\omega_w^l| =  |\mathcal{C}^l|=8$. We choose for the modelet:
 $$ p^{l-1} = \arg \max \{ |\omega_w^l|,  \omega_w^l \in \Omega_l \}\,.$$
It may happen that the above definition does not yield a unique maximum. If two or more categories dominate a cell block, a generic approach consists in  taking  into account its first block cell neighborhood (the surrounding \num{26} cells, except at mesh borders and boundaries). We affect  the dominant value in the first neighborhood to $ p^{l-1} $. In case of a draw again, the second-order surrounding can be used, iteratively. In practice for the presented version of \hs, we limit to the first-order  neighborhood, and choose the lowest indexed category when the maximum is not unique. Equipped with this unique lower resolution representative value, we proceed similarly to Section \ref{sec_continuous_property_compression} for details, by using differences between  original categories and the mode. As classes are often indexed by positive integers, a slight motivation allows to get only non-negative indices. By avoiding  negative values, one expects a decrease in  data entropy of around \SI{5}{\percent}, which benefits to compression.

We thus change the sign of a detail coefficient if and only if it generates a value out of the range of $\{\omega_1, \omega_2,\ldots, \omega_W \}$, and then control this condition during reconstruction. So, all details $\{ d_n^{l-1} \}$ for a cell block $\mathcal{C}$ are determined by:
$$ d_n^{l-1} = (-1)^{ (p_n^l - p^{l-1}<0) \land  ((2p^{l-1} - p_n^l) \notin \Omega)  }\times (p_n^l - p^{l-1}).  $$
During synthesis,  coefficients $\{ p_i^l \}$ are obtained thanks to the closed-form equation:
 $$ p_n^l = p^{l-1} + (-1)^{\left( (p^{l-1} + d_n^{l-1}) \notin \Omega \right) } \times d_n^{l-1}.$$

\section{Evaluation methodology, comparative results and discussion\label{sec_results}}
\subsection{Evaluation methodology\label{sec_results_methodo}}
\begin{table*}[t]
	\centering
		\begin{tabular}{|c|c|c|c|c||c|c|c|}
    \hline
    Mesh & \multicolumn{4}{c||}{Characteristics}  & \multicolumn{3}{c|}{Properties} \\
    \cline{2-8}
	index & \# Cells & Dimension &  Faults & Filesize & \actn & Continuous & Categorical \\	
    \hline
	1	&  93,600 &  \num[output-product=\times]{80 x 45 x 26}        & No  &  \SI{4.615}{\mega\byte}  & \SI{100}{\percent}  & Porosity & Rock type  \\
    \hline
	2	& 1,000,000 & \num[output-product=\times]{100 x 100 x 100}  & No  & \SI{42.459}{\mega\byte}  & \SI{100}{\percent}  & --- & ---    \\
    \hline \hline
	3	&  36,816 &  \num[output-product=\times]{59 x 39 x 16}        & Yes  &  \SI{1.458}{\mega\byte}  & \SI{100}{\percent}  & --- & ---  \\
    \hline
4	& 210,000  &  \num[output-product=\times]{100 x 100 x 21}        & Yes  &  \SI{7.881}{\mega\byte}  & \SI{20}{\percent}  & --- & ---  \\
    \hline	
   5	& 450,576  &  \num[output-product=\times]{149 x 189 x 16}        & Yes  &  \SI{22.730}{\mega\byte}  & \SI{46}{\percent}  & Porosity, Permeability  & --- \\
    \hline
6	&  5,577,325 &  \num[output-product=\times]{227 x 95 x 305}        & Yes  &  \SI{274,573}{\mega\byte}  & \SI{97}{\percent}  & Porosity & Rock type  \\
    \hline	
    7	&  13,947,600 &  \num[output-product=\times]{240 x 295 x 197}        & Yes  &  \SI{580.937}{\mega\byte}  & \SI{100}{\percent}  & Porosity & Rock type  \\
    \hline	
\end{tabular}
	\caption{Meshes chosen for evaluation: a compendium of their ontological characteristics and geological properties.\label{Tab-Meshes-Features}}
\end{table*}

Meshes,  hexahedral ones in particular, are complex composite objects. The  ontological  description of their geometry is subject to different options,   ``Block Centered'' or ``Corner Point'' grids for instance. Their generation, cell size and resolution for practical applications  may have undergone more or less complex processing. Mesh complexity can range from simply-layered, homogeneous modes to massively faulted environment with highly varying properties.
Encoded numerical values, albethey cell coordinates, numerical or categorical properties are cast into different possible integer or floating-point precisions. The structure of the raw mesh binary object is itself embedded into enriched formats, for which a few standards exist, as RESQML\texttrademark. The latter also encompasses structural information required to exchange models, generating information overhead. Finally, detail simplification through multiscale decompositions does not possess well-established quality metrics. All of the above hampers exhaustive objective evaluations such as possible in image processing, where metrics and benchmarks have been evaluated for decades.

To evaluate the performance of \hs, we  base our analysis on a  set of seven geological meshes, with geometries  ranging from smooth to fractured, and diverse categorical and continuous properties. Their main characteristics and properties are summarized in Table \ref{Tab-Meshes-Features}.
As will be seen, they appear representative enough to allow one to derive consistent observations and conclusions for different data handling purposes.

They are initially stored in the GRDECL (``GRiD ECLipse'')  file format \cite{Pettersen_O_2006_misc_basics_rsers}. Originally complex geomodels are thus described  with details rendering their geometry explicit and structured, an important feature for geomodelers or flow simulation software (Petrel\texttrademark, SKUA-GOCAD\texttrademark, Eclipse\texttrademark\ldots).

As observed in the state-of-the-art  (Section \ref{sec:mesh_compression}), to our knowledge,  reversible  multiscale representations of geometry and properties --- together with discontinuity preservation --- of hexahedral meshes do not exist. Even if the standardized  $3D$ extension to the JPEG2000  image compression format (termed \jtwoktd) could process the properties as volumetric images, it is not \emph{per se} suited to volume meshes, especially with categorical properties. We thus focus on visualizations,   comparisons with geomodeler upscaling capabilities, and the embedding of our multiscale decompositions into several all-purpose compression algorithms.

The evaluation methodology is twofold.
First, we exemplify  the outcome of \hs on meshes on either their geometry with a continuous and a categorical property at different dyadic scales. This reversible framework is  put into perspective with similar downscaling processes in a reference geomodeler.
Second, a comprehensive evaluation of lossless compression performance is provided, using state-of-the-art coders on either the raw files or their multiscale decomposed counterparts.

\subsection{Reversible multiscale mesh representation\label{sec_results_reversible}}

\begin{figure*}[hp]
\begin{center}
\begin{tabular}{ccc}
Geometry. & Porosity. & Rock type. \\
\includegraphics[width=0.305\textwidth]{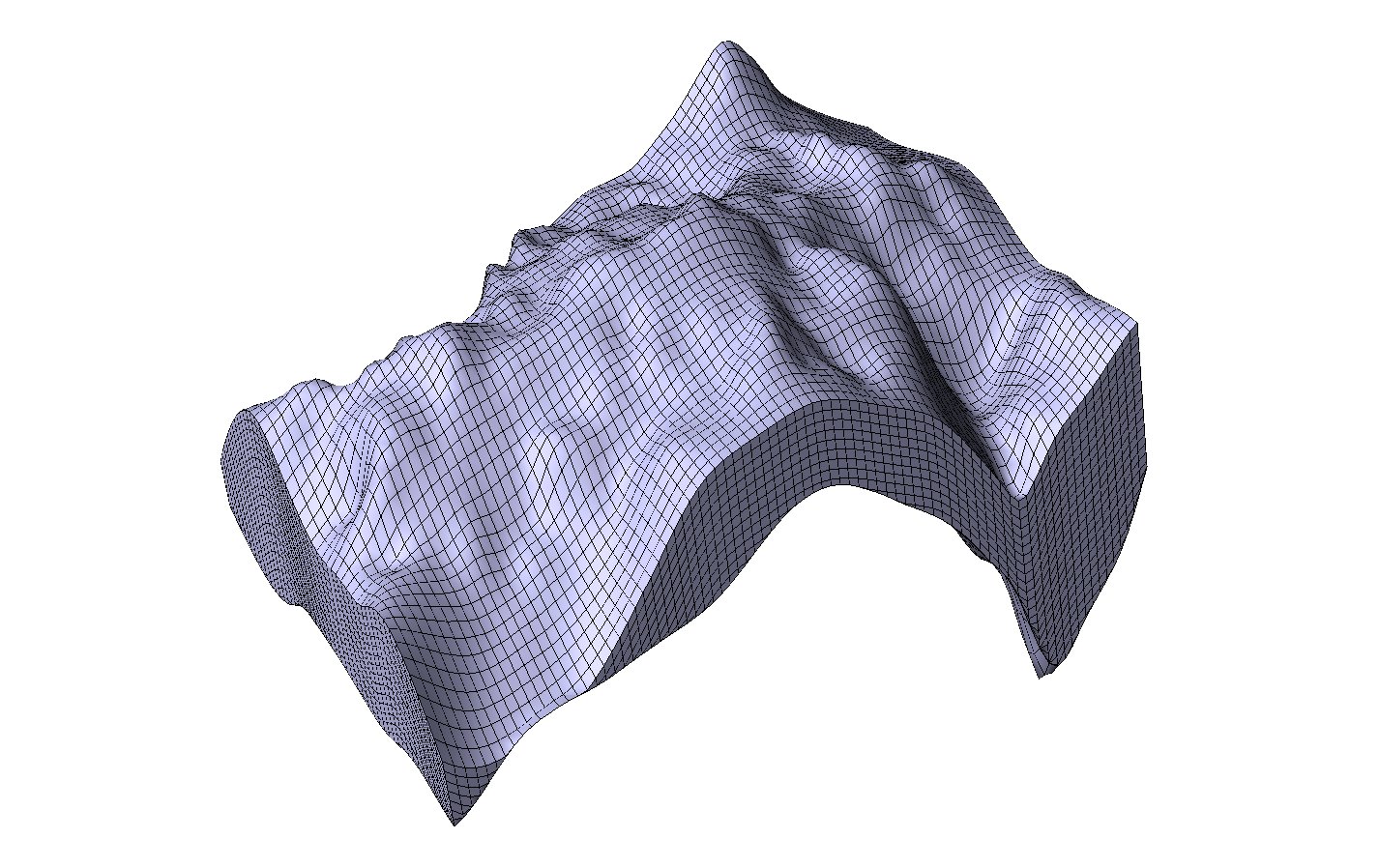}  &
\includegraphics[width=0.305\textwidth]{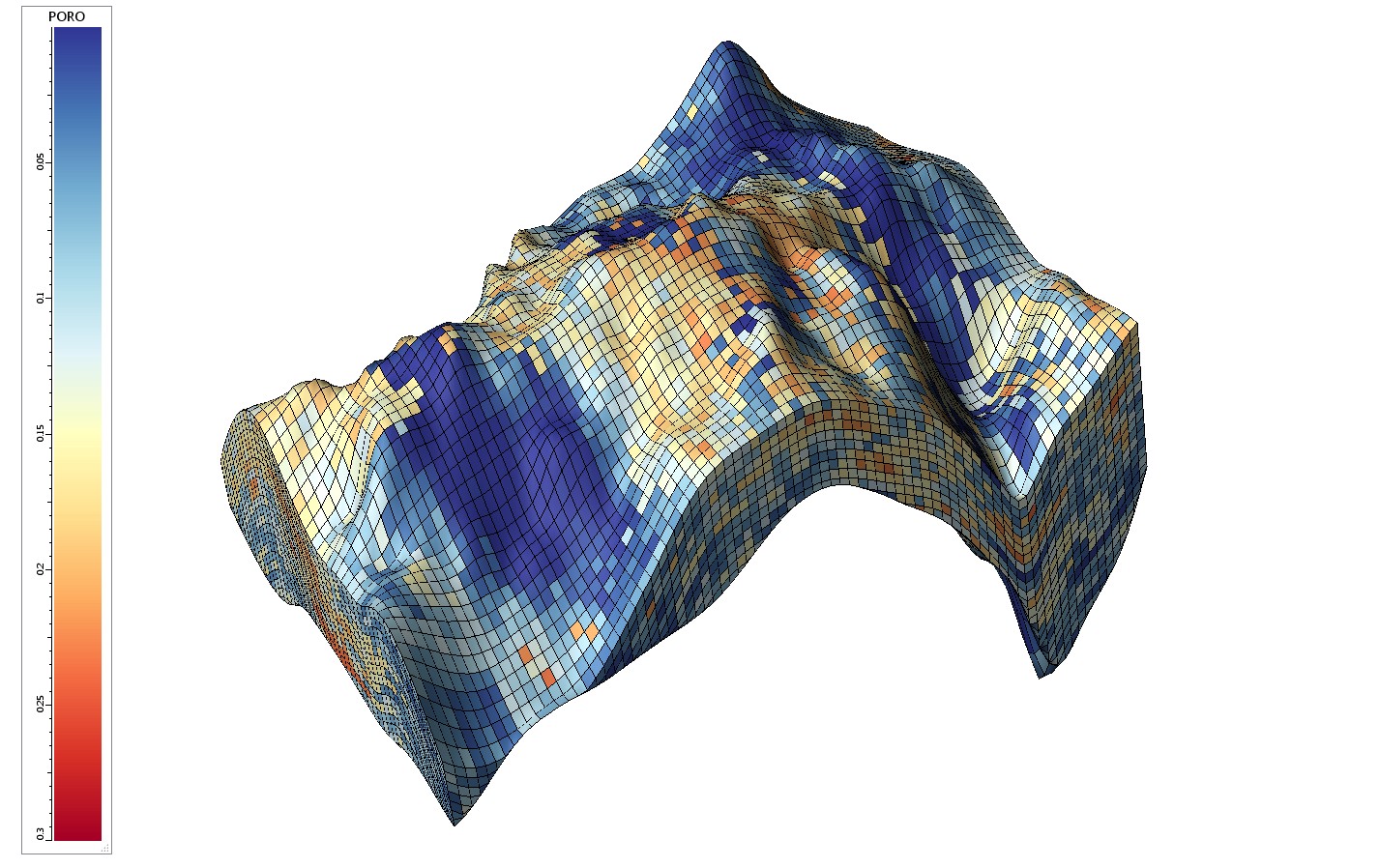} &
\includegraphics[width=0.305\textwidth]{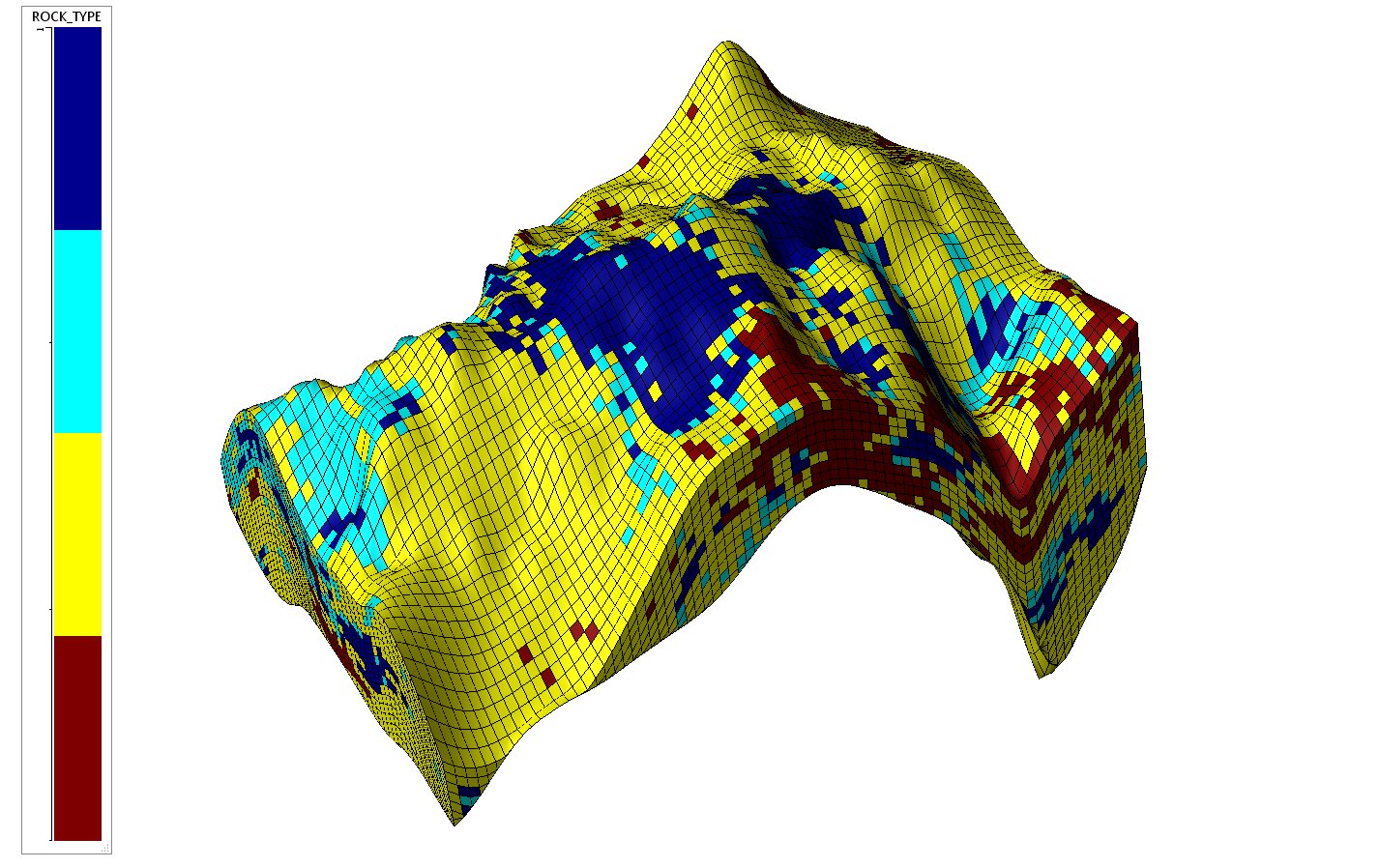} \\
\multicolumn{3}{c}{Original \ms{1}.} \\
\includegraphics[width=0.305\textwidth]{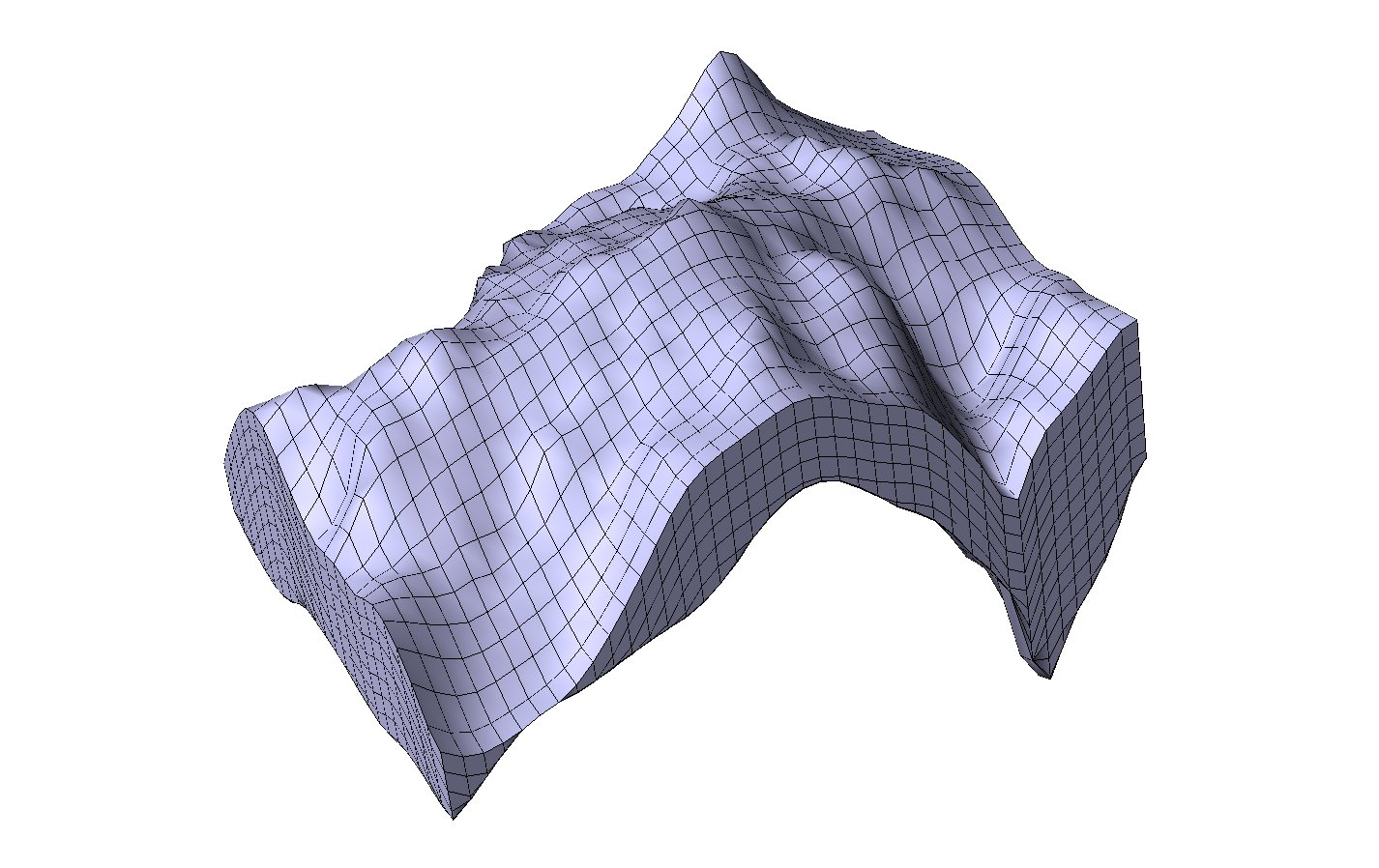}  &
\includegraphics[width=0.305\textwidth]{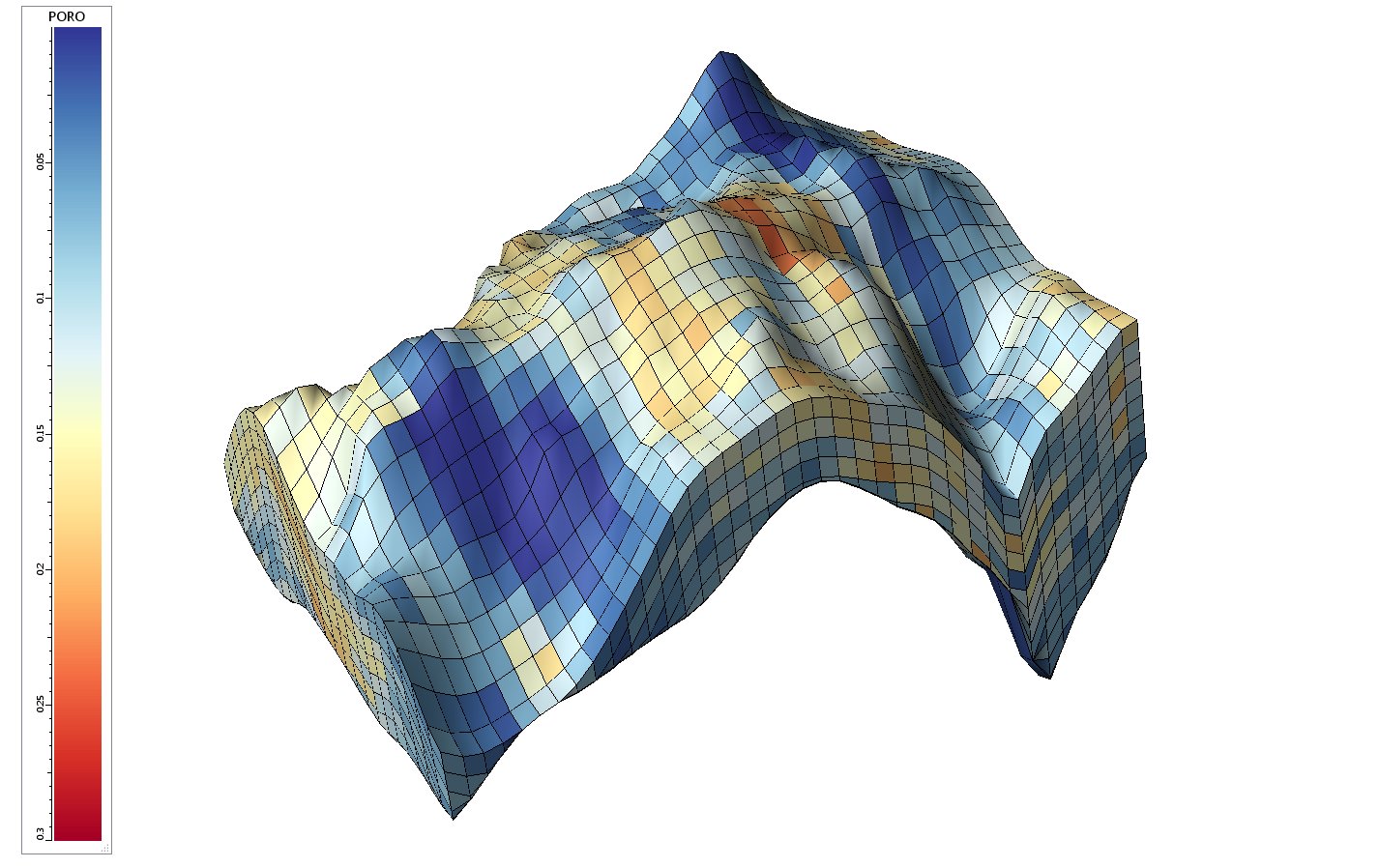} &
\includegraphics[width=0.305\textwidth]{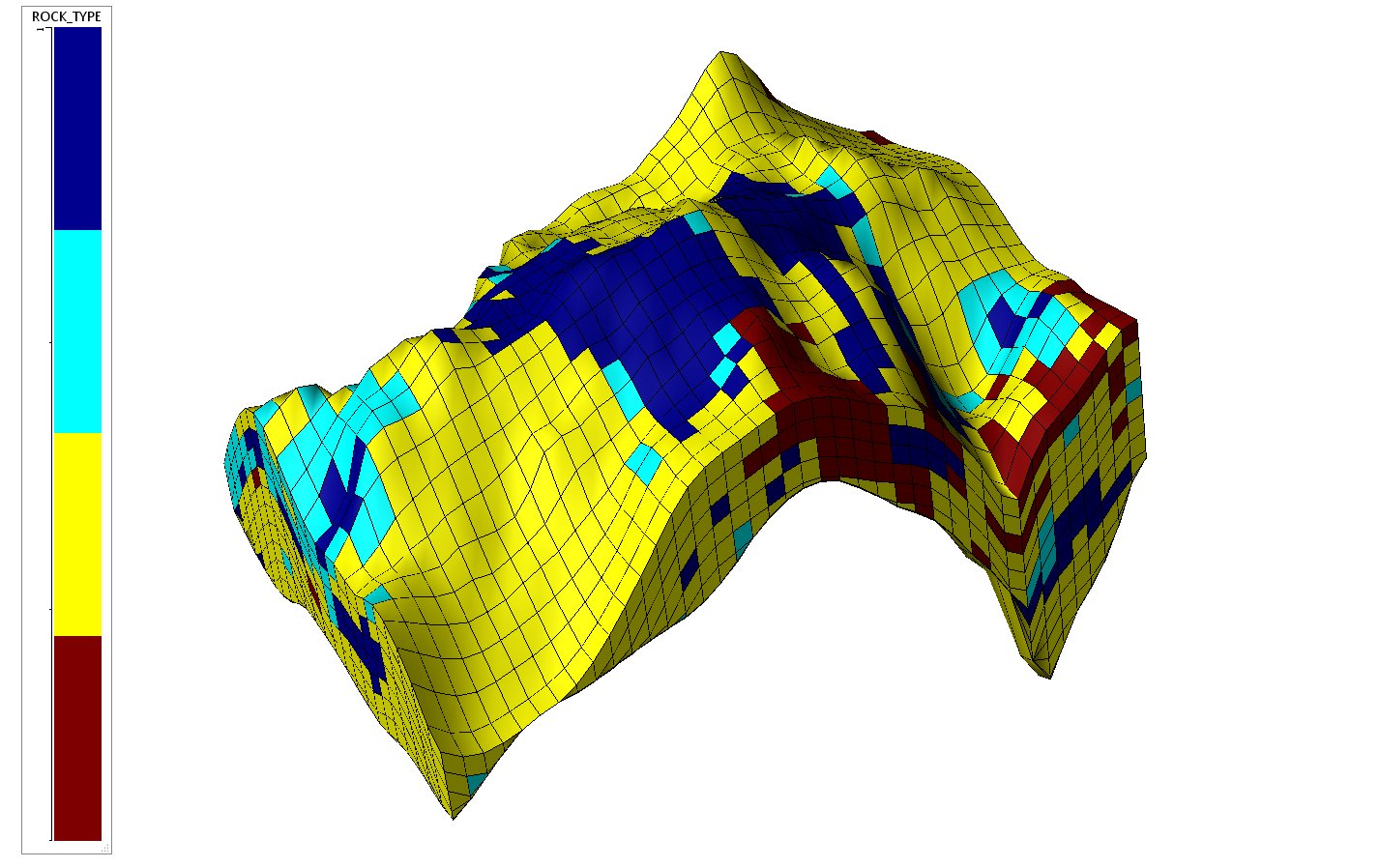} \\
\multicolumn{3}{c}{Resolution $-1$.} \\
\includegraphics[width=0.305\textwidth]{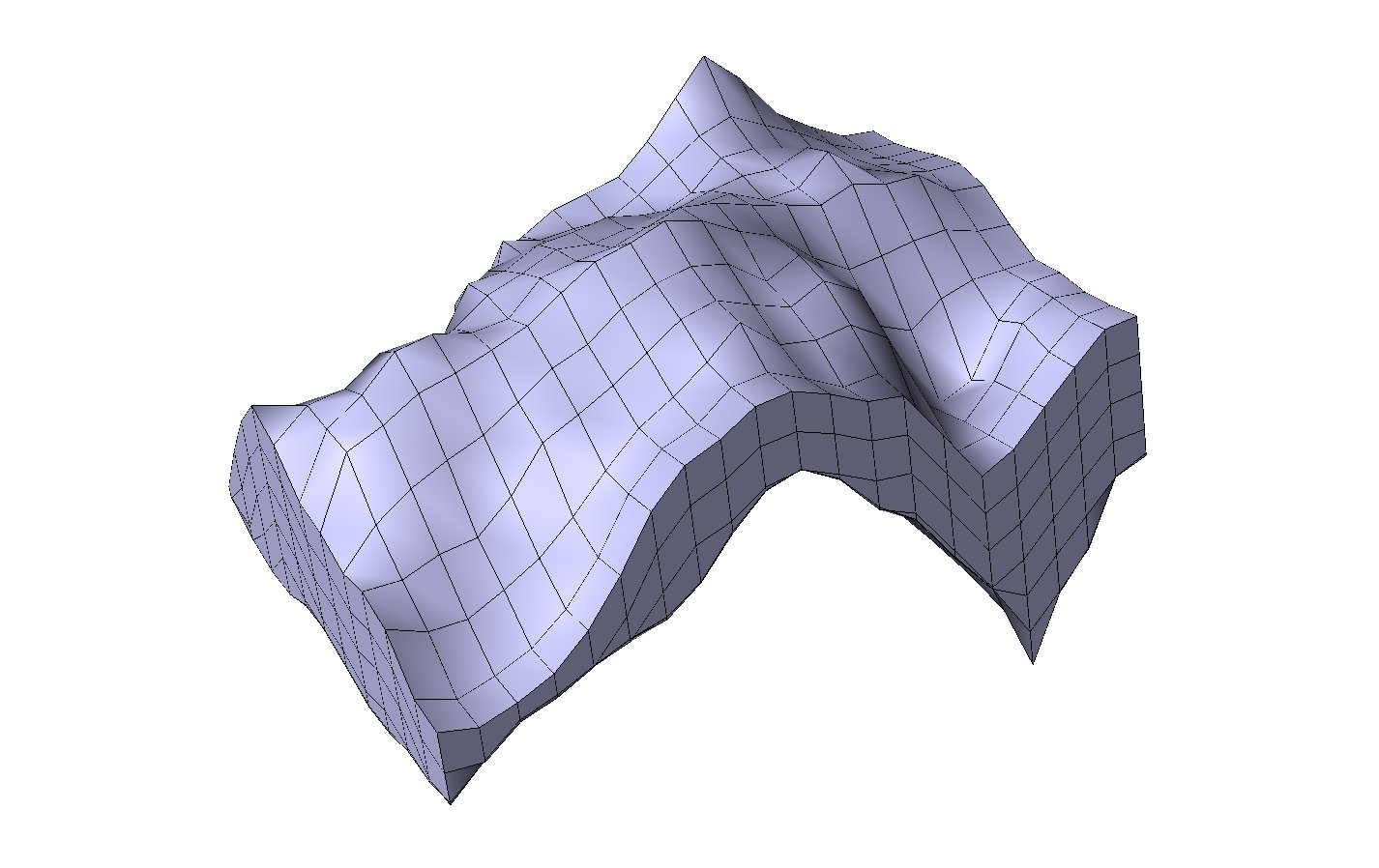}  &
\includegraphics[width=0.305\textwidth]{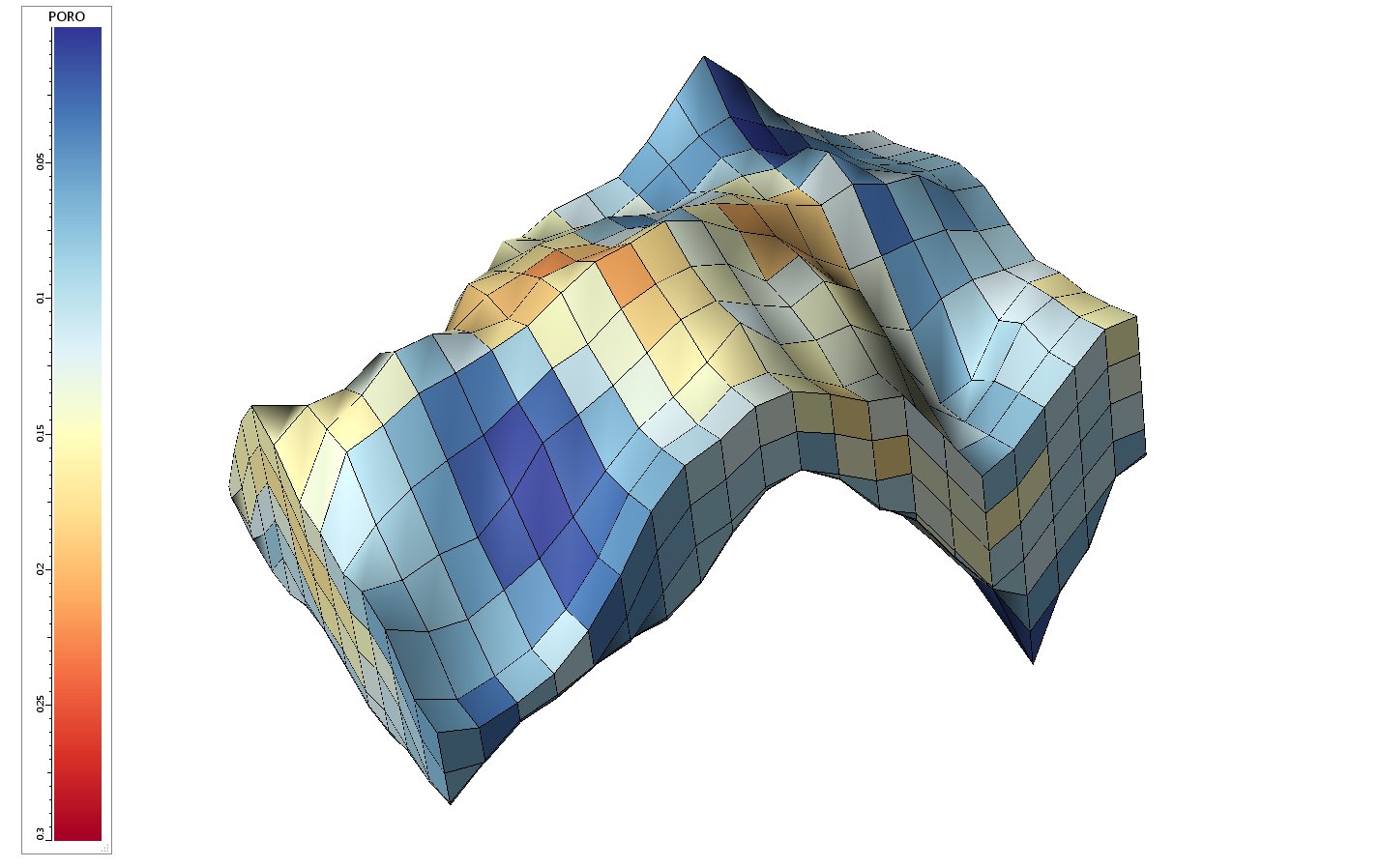} &
\includegraphics[width=0.305\textwidth]{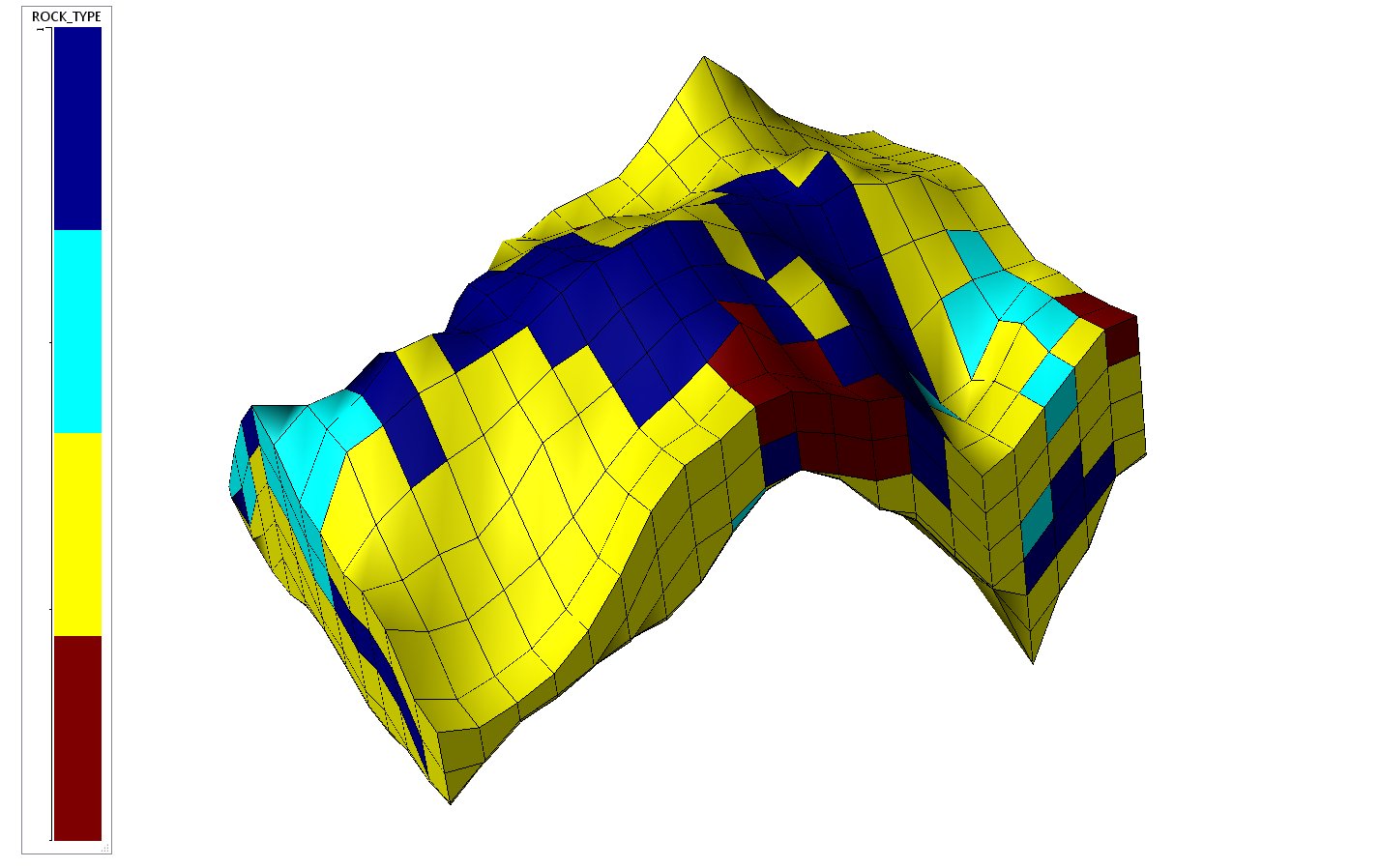} \\
\multicolumn{3}{c}{Resolution $-2$.} \\
\includegraphics[width=0.305\textwidth]{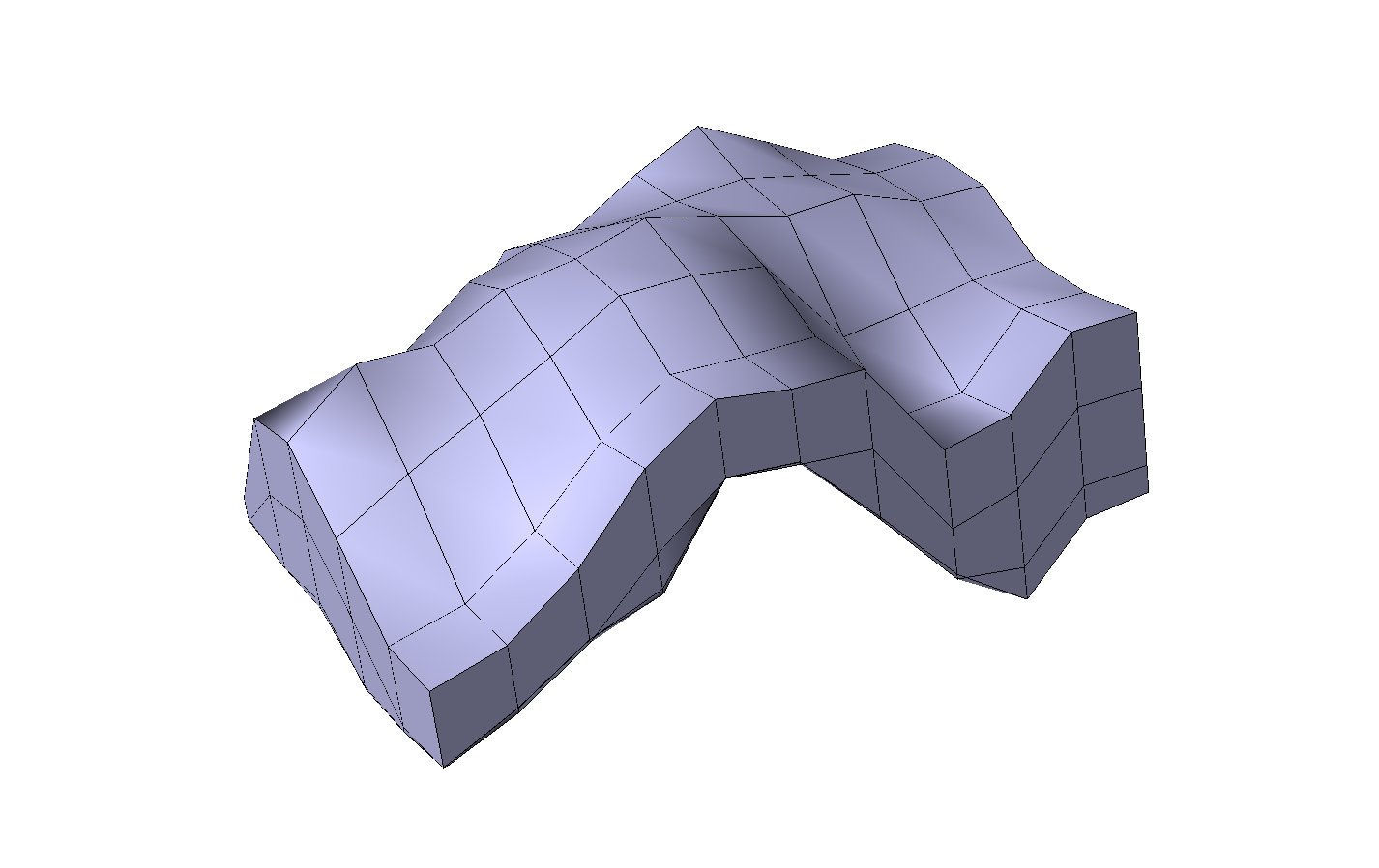}  &
\includegraphics[width=0.305\textwidth]{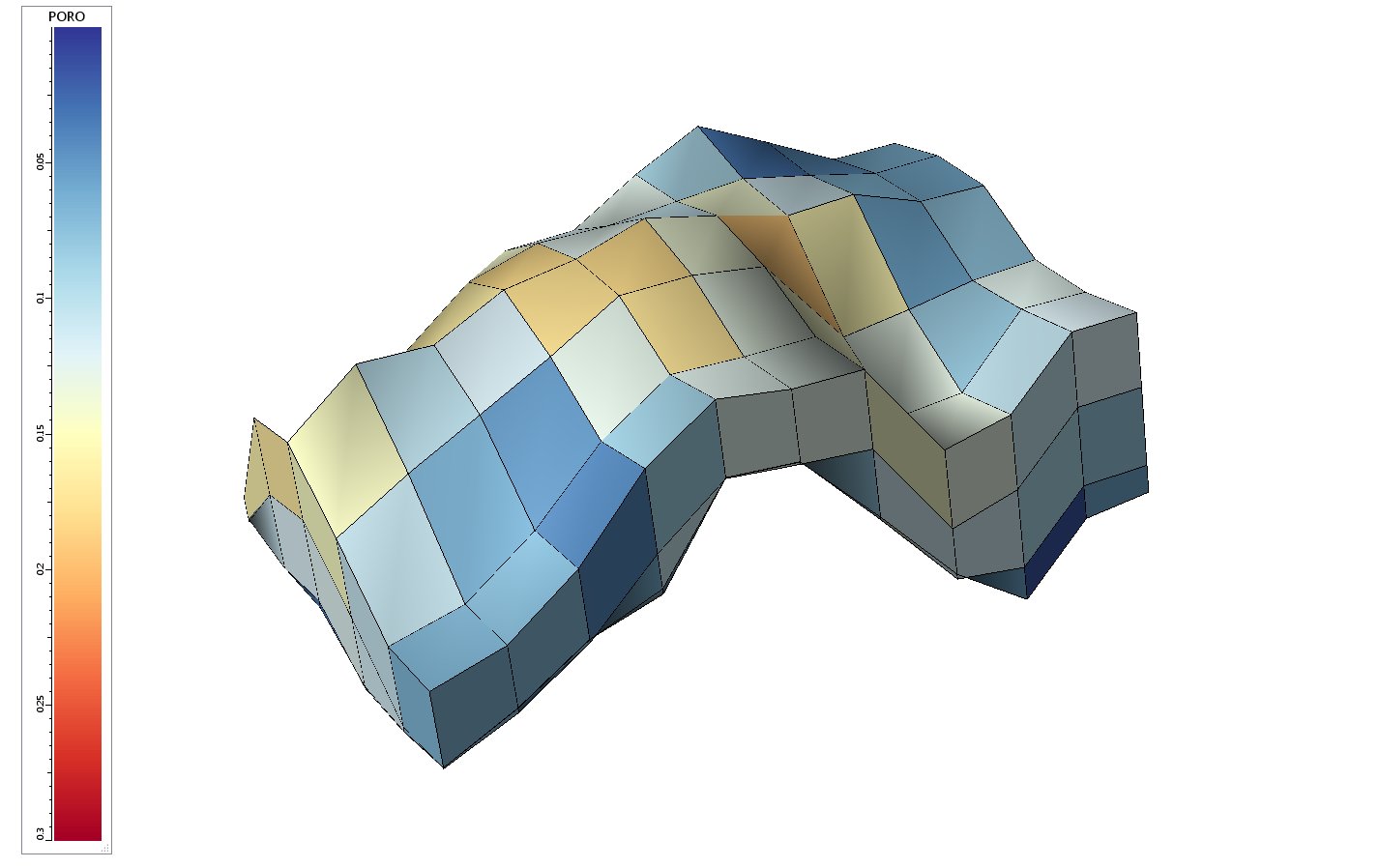} &
\includegraphics[width=0.305\textwidth]{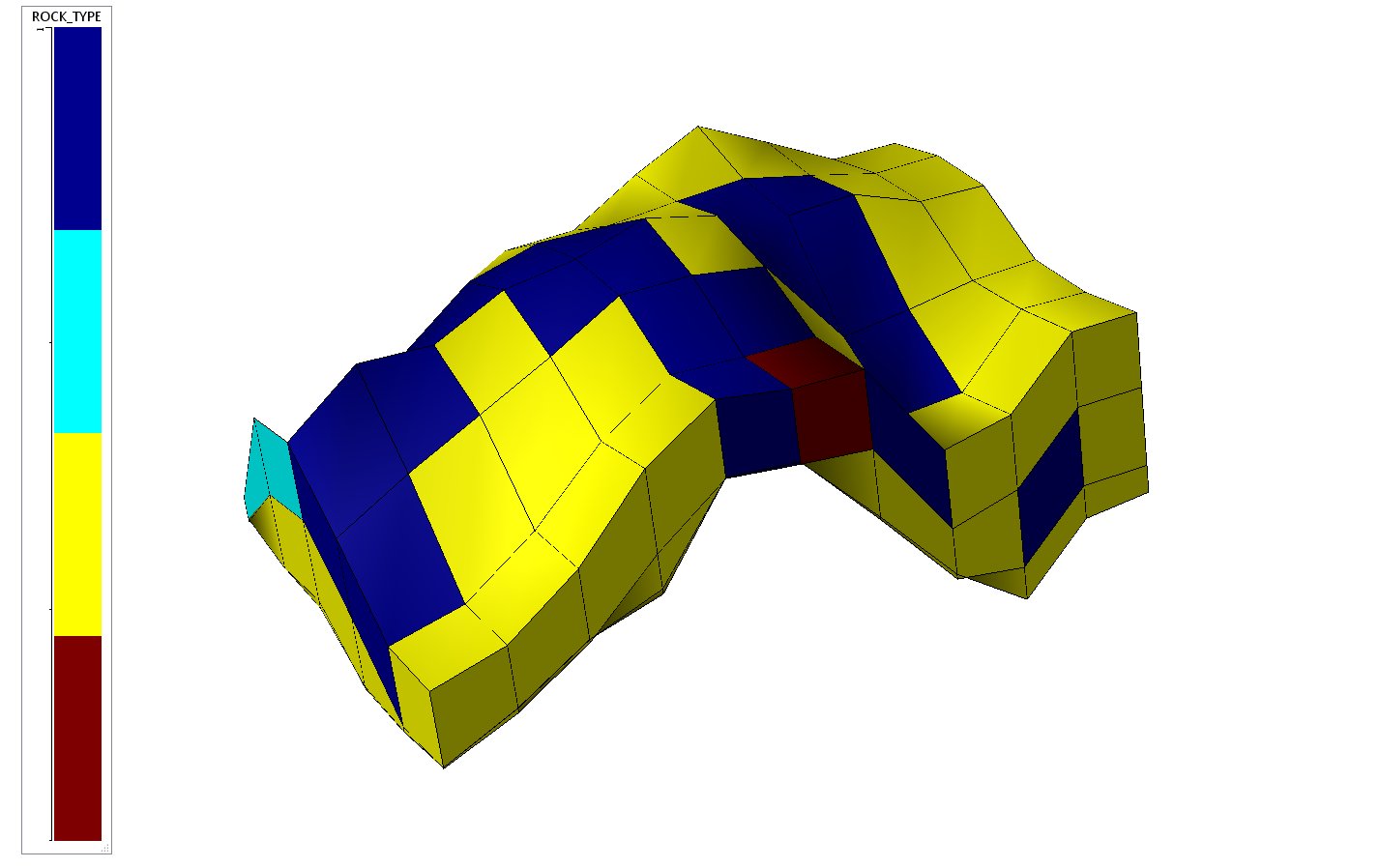} \\
\multicolumn{3}{c}{Resolution $-3$.} \\
\includegraphics[width=0.305\textwidth]{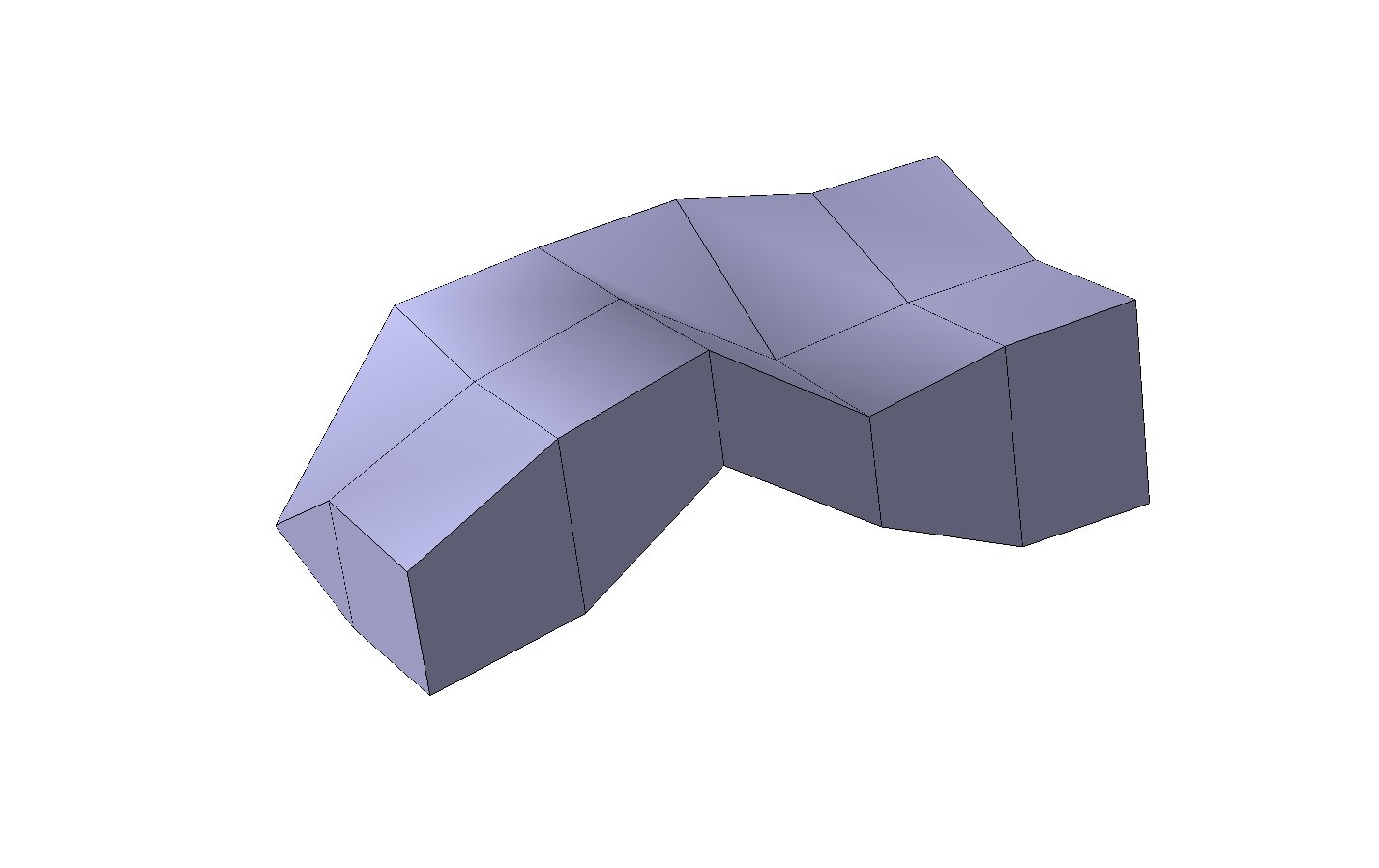}  &
\includegraphics[width=0.305\textwidth]{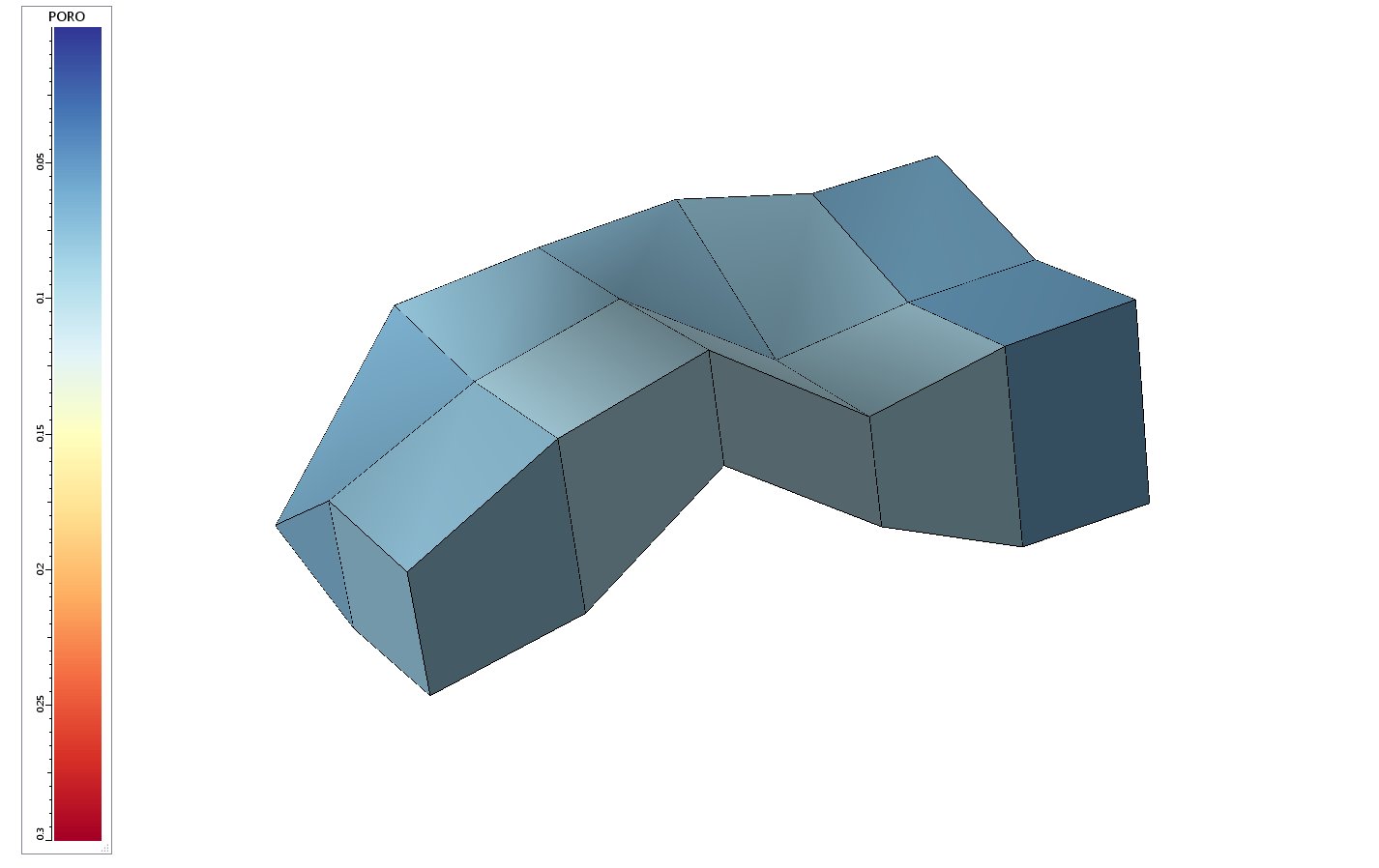} &
\includegraphics[width=0.305\textwidth]{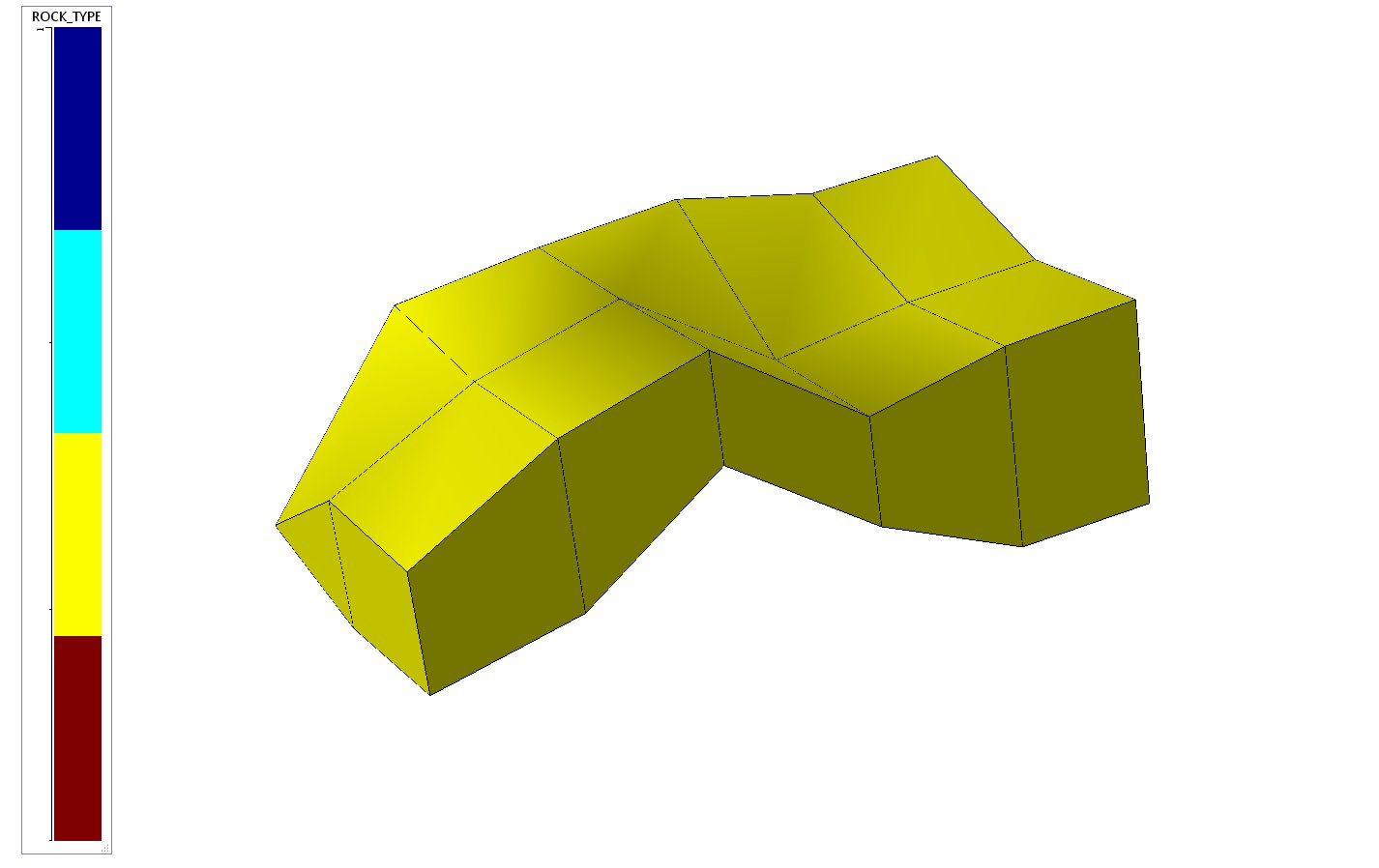} \\
\multicolumn{3}{c}{Resolution $-4$.}
\end{tabular}
\caption{\label{fig:visual_results_coarsePBR} Original \ms{1}, its attributes,   and four levels of resolution generated with \hs.}
\end{center}
\end{figure*}

\begin{figure*}[hp]
\begin{center}
\begin{tabular}{ccc}
Geometry. & Porosity. & Rock type. \\
\includegraphics[width=0.23\textwidth]{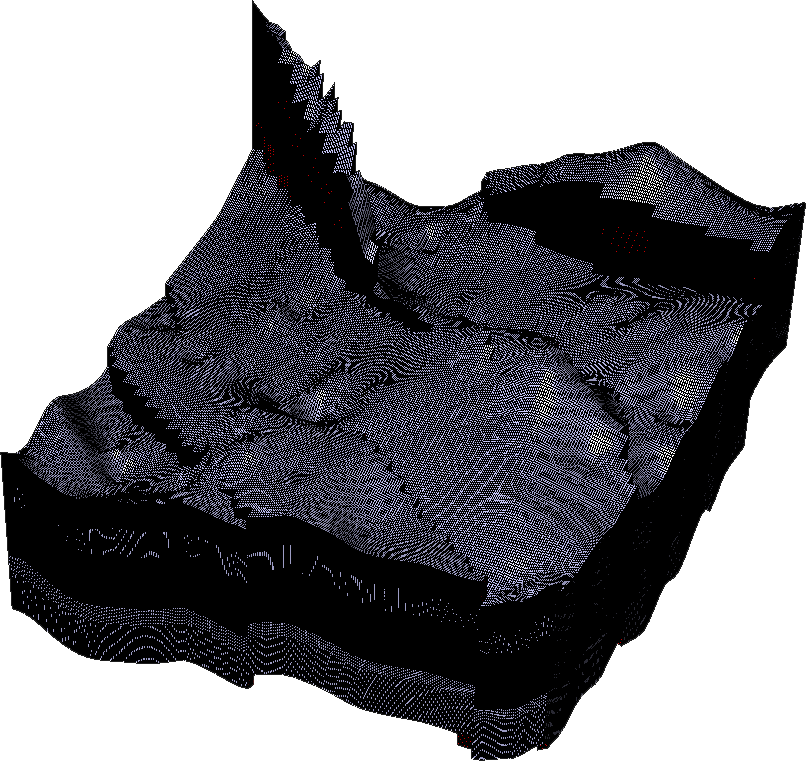}  &
\includegraphics[width=0.295\textwidth]{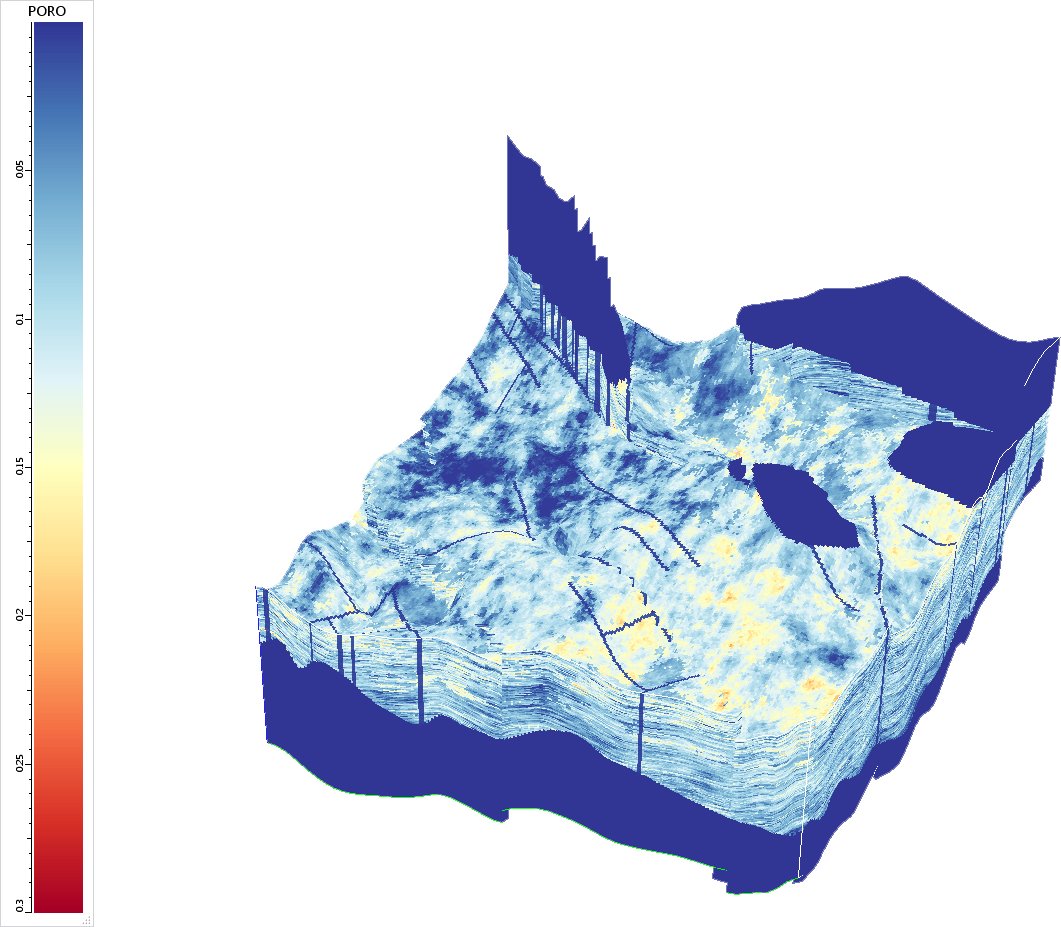} &
\includegraphics[width=0.295\textwidth]{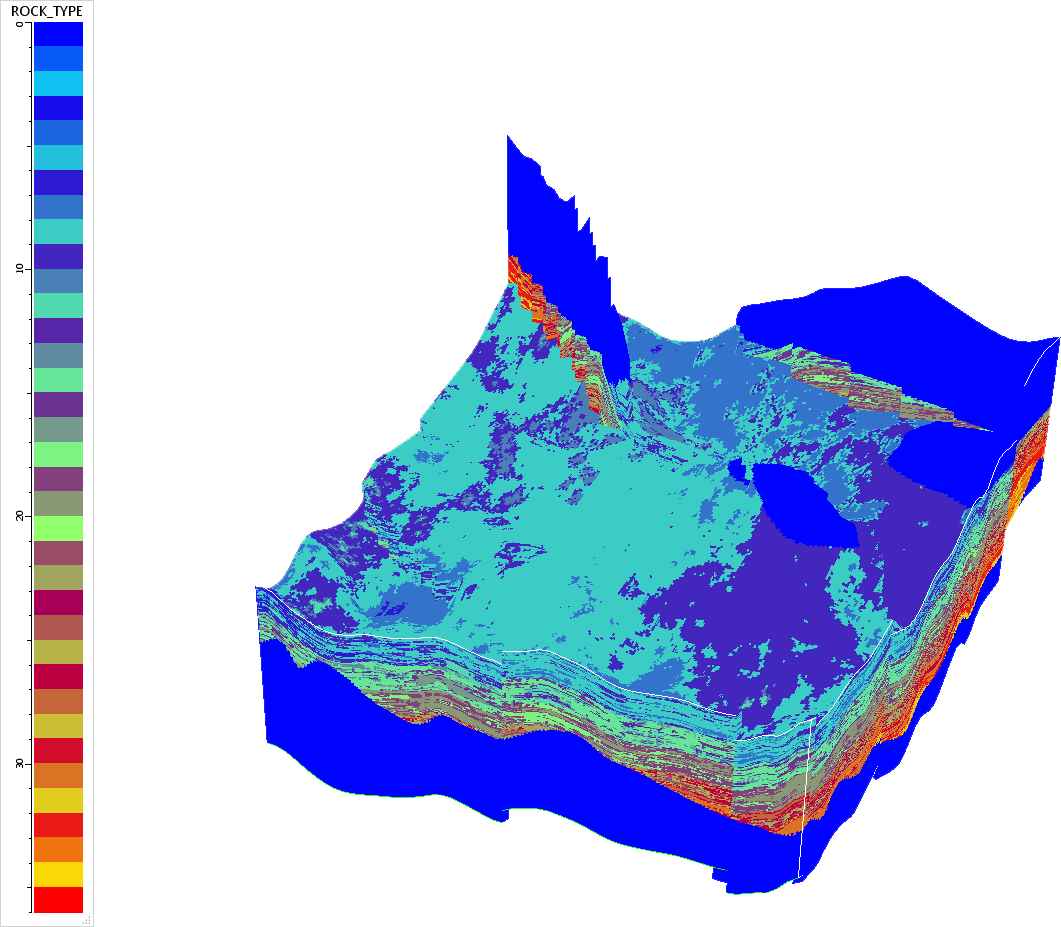} \\
\multicolumn{3}{c}{Original data} \\
\includegraphics[width=0.23\textwidth]{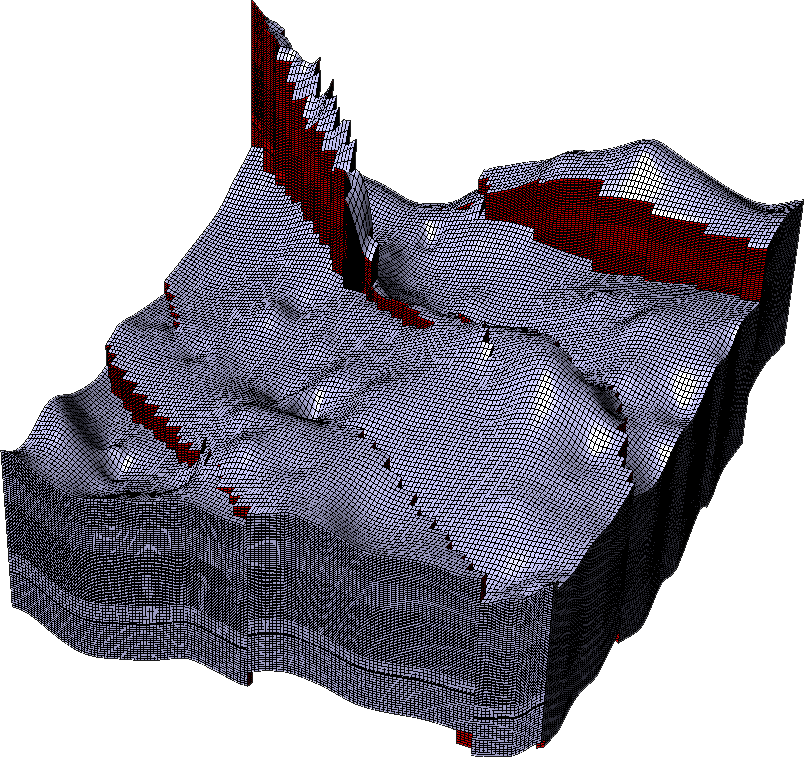}  &
\includegraphics[width=0.295\textwidth]{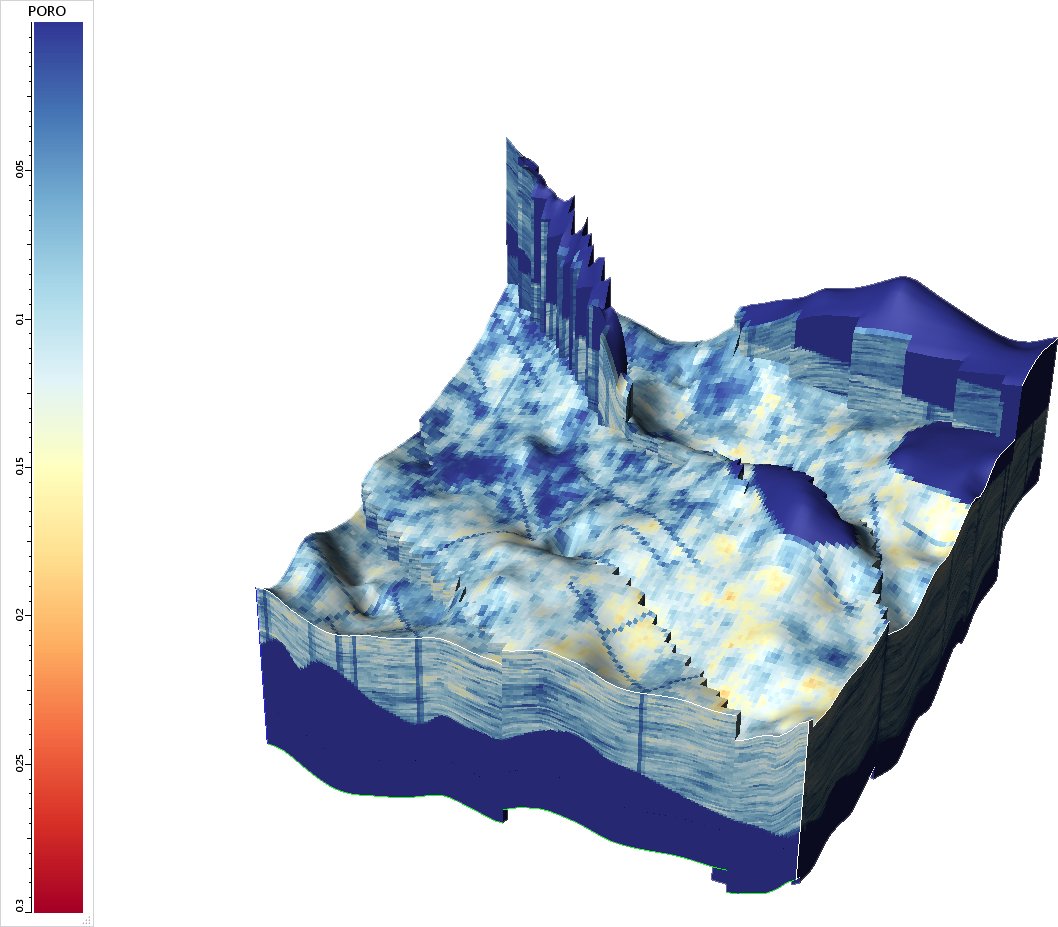} &
\includegraphics[width=0.295\textwidth]{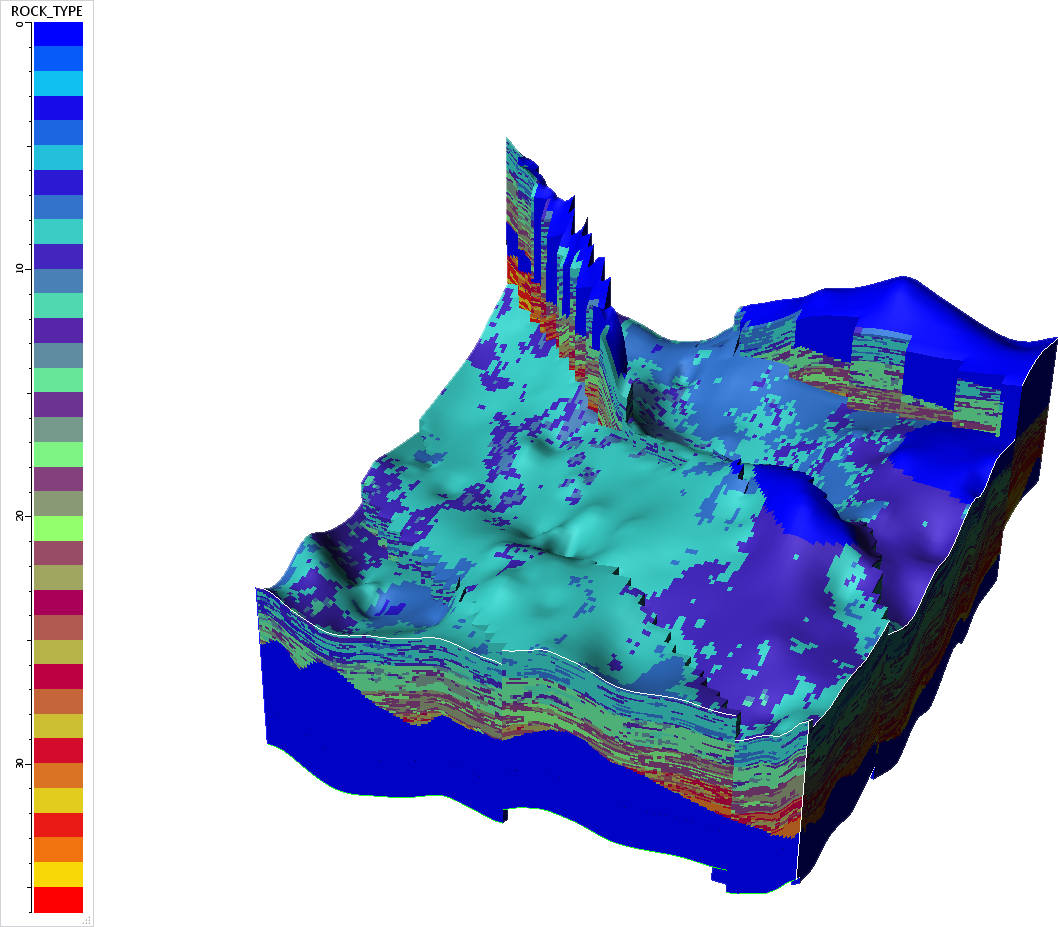} \\
\multicolumn{3}{c}{Resolution $-1$.} \\
\includegraphics[width=0.23\textwidth]{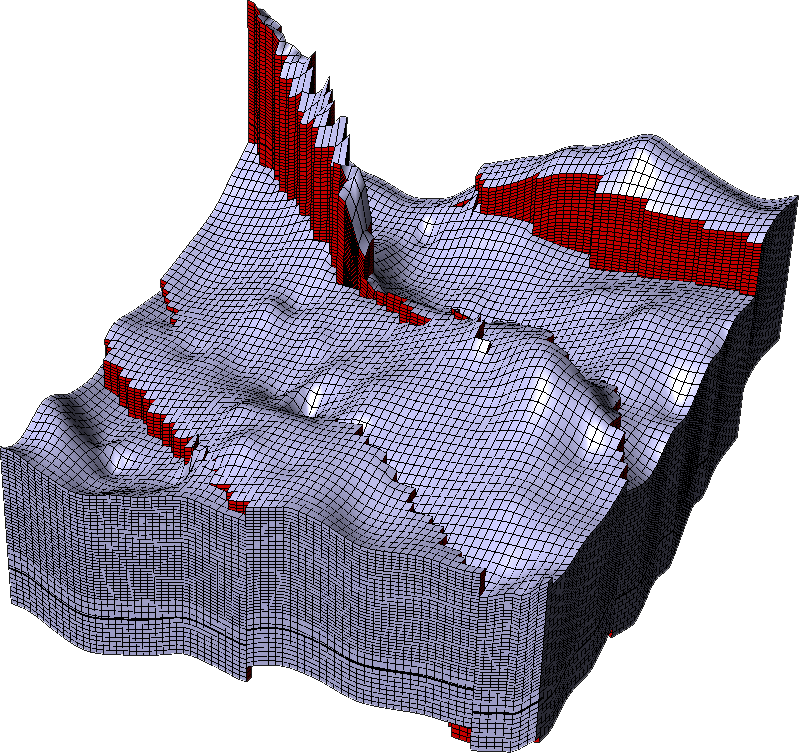}  &
\includegraphics[width=0.295\textwidth]{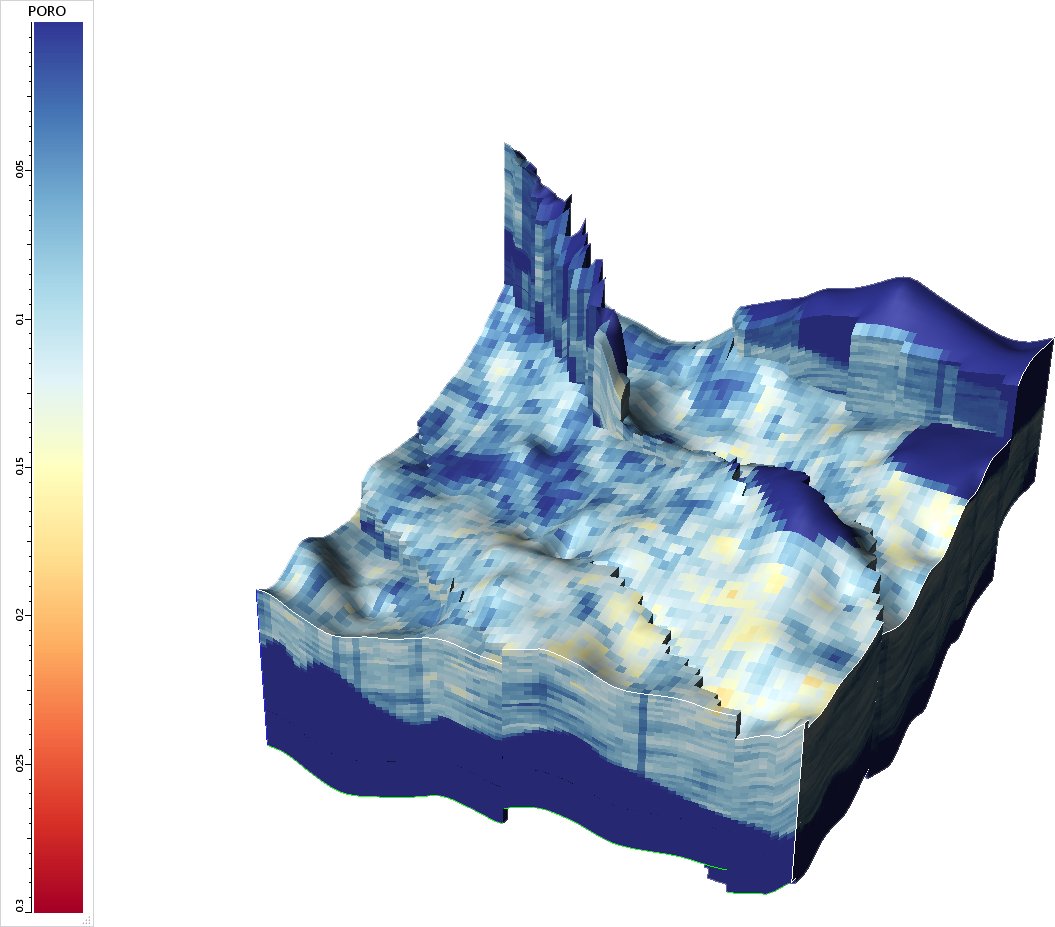} &
\includegraphics[width=0.295\textwidth]{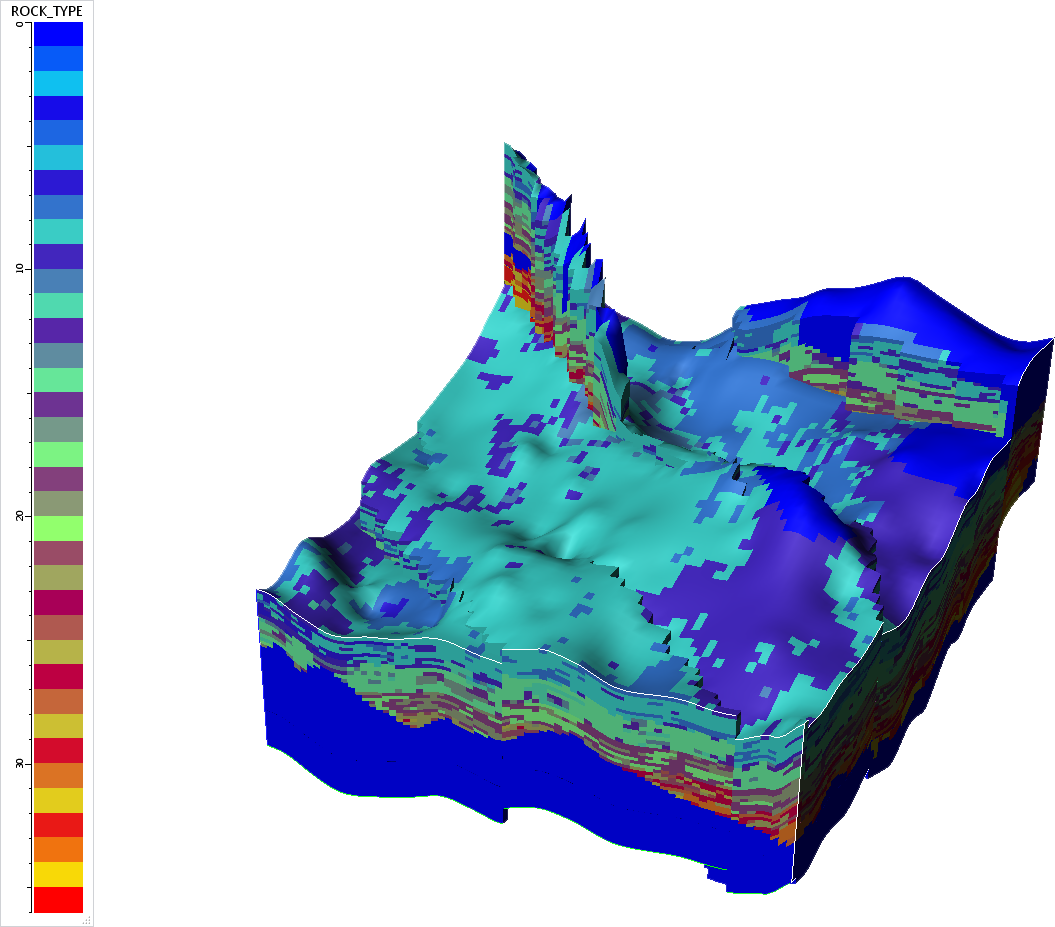} \\
\multicolumn{3}{c}{Resolution $-2$.} \\
\includegraphics[width=0.23\textwidth]{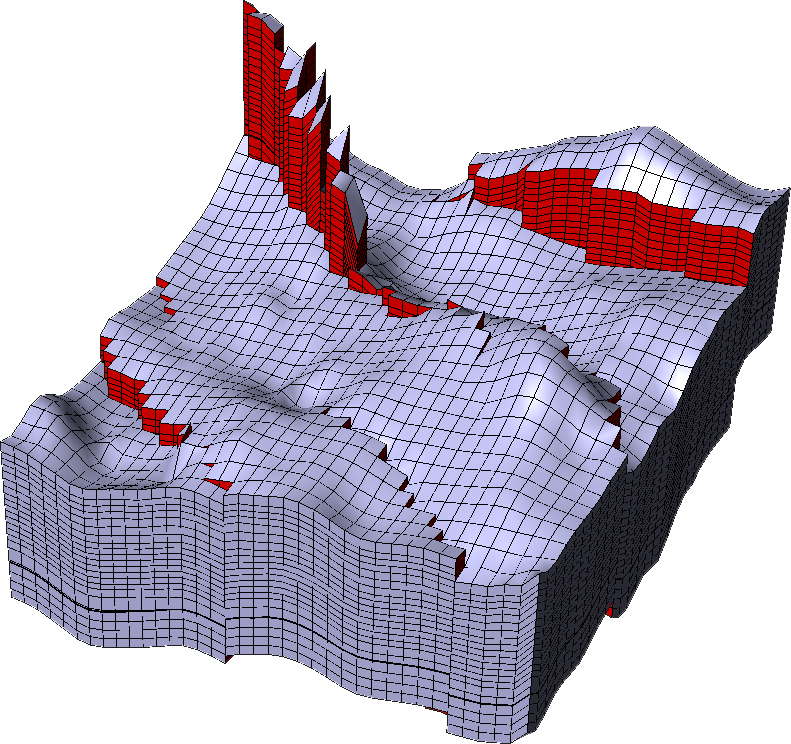}  &
\includegraphics[width=0.295\textwidth]{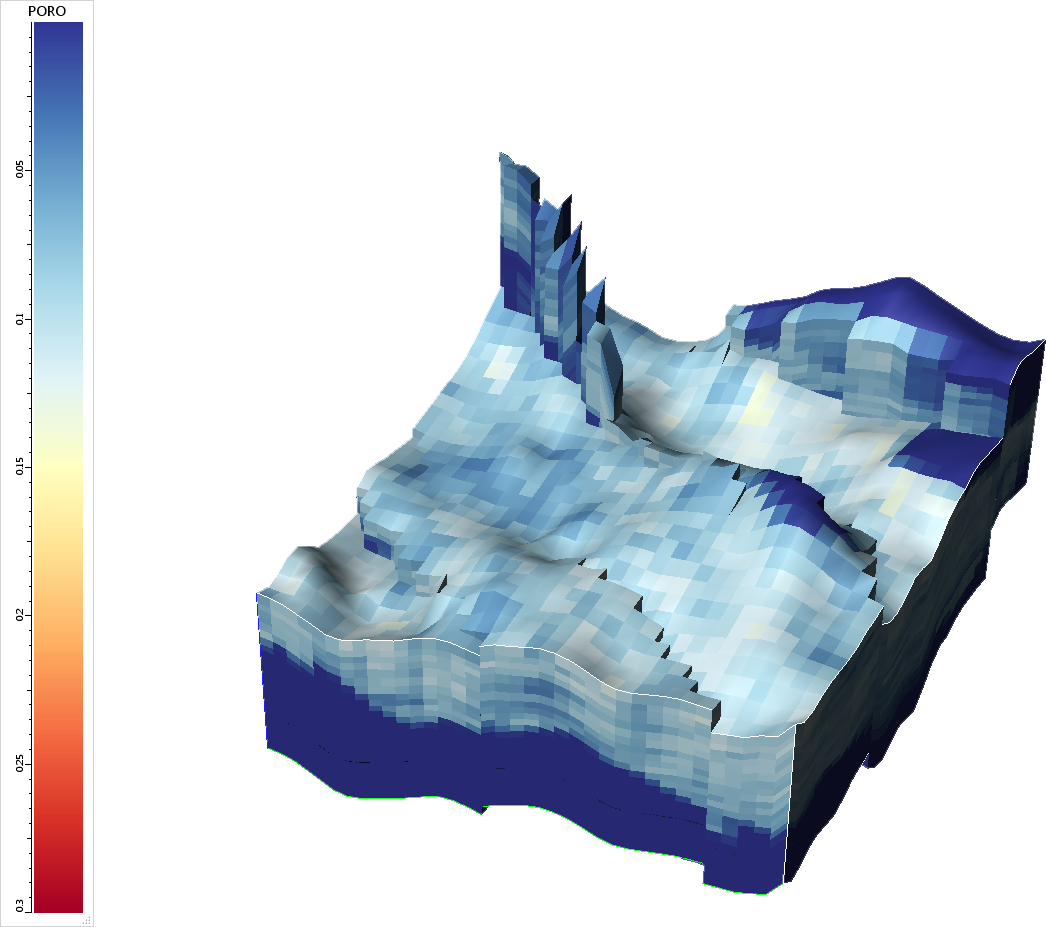} &
\includegraphics[width=0.295\textwidth]{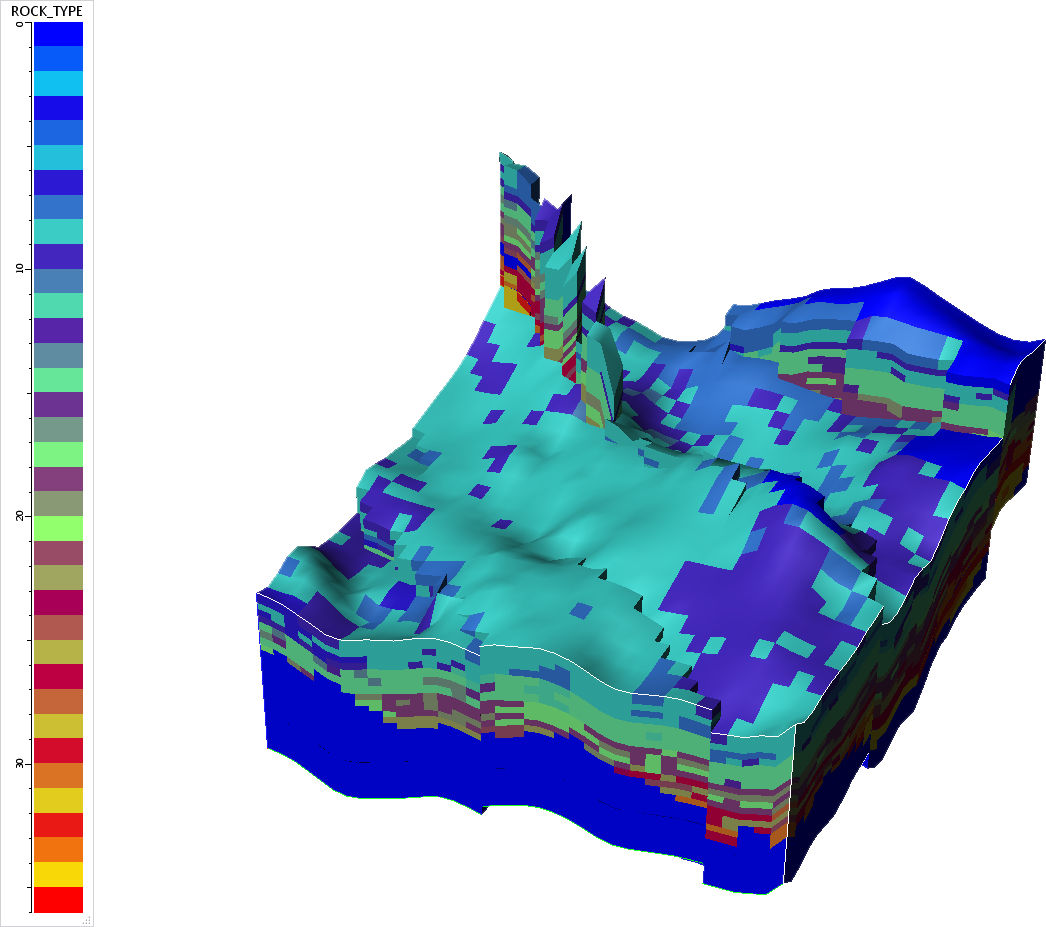} \\
\multicolumn{3}{c}{Resolution $-3$.} \\
\includegraphics[width=0.23\textwidth]{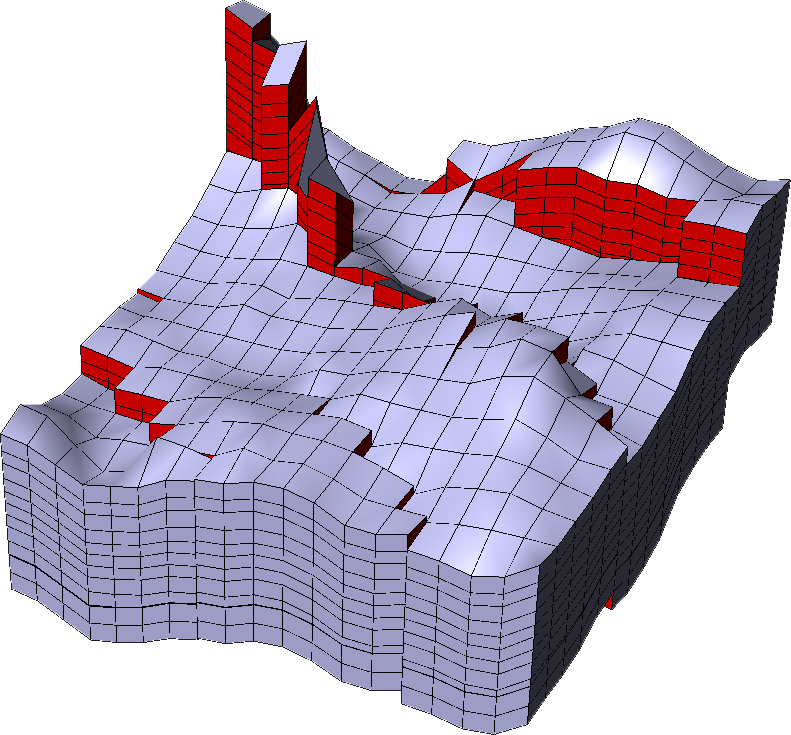}  &
\includegraphics[width=0.295\textwidth]{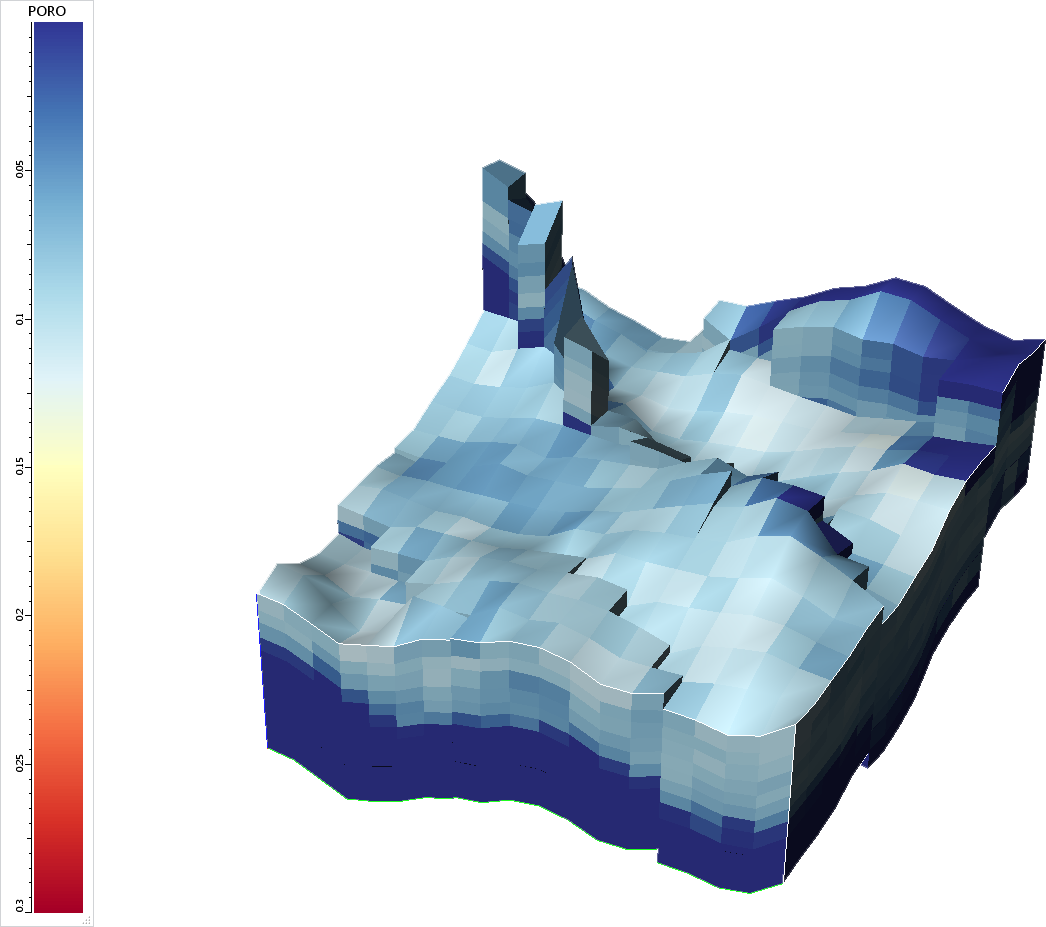} &
\includegraphics[width=0.295\textwidth]{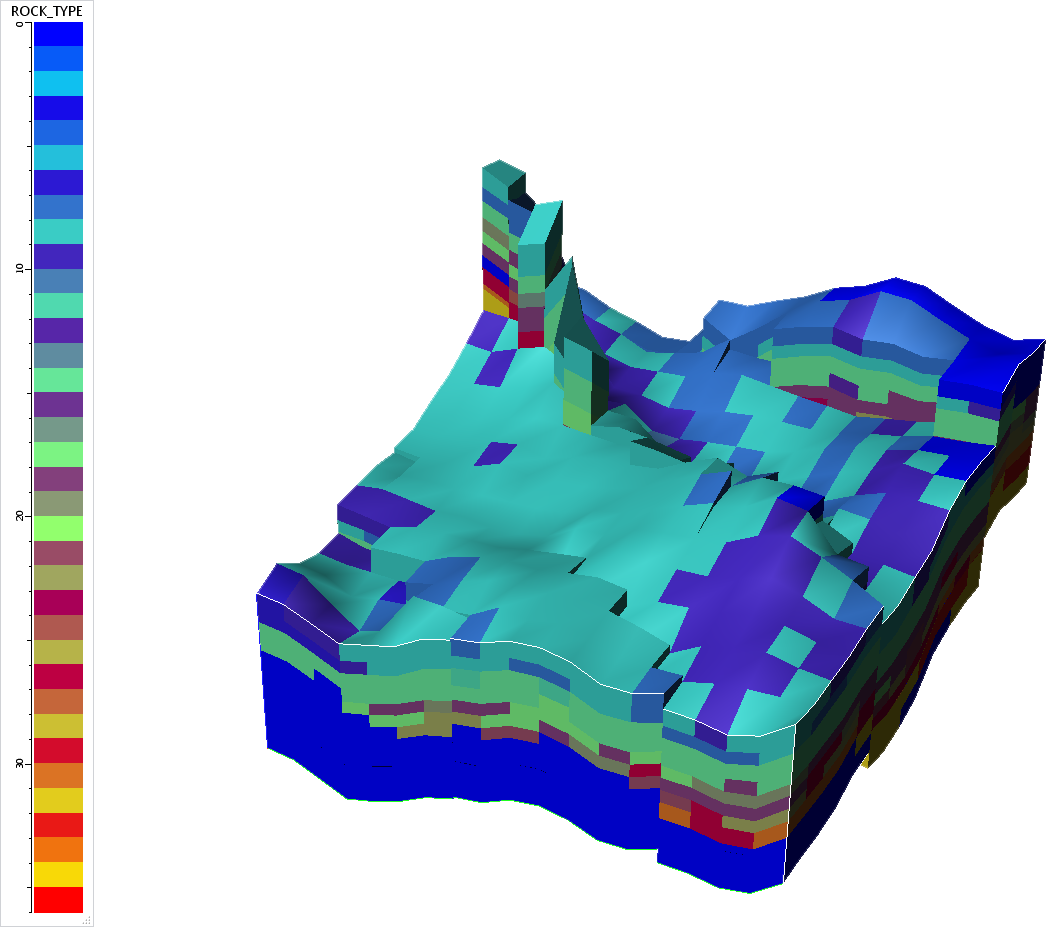} \\
\multicolumn{3}{c}{Resolution $-4$.} \\
\end{tabular}
\caption{\label{fig:visual_results_z13} Original \ms{7},  its attributes, and four levels of resolution generated with \hs.}
\end{center}
\end{figure*}

Figures \ref{fig:visual_results_coarsePBR} and  \ref{fig:visual_results_z13} present the reversible multiscale decompositions generated by \hs for \ms{1} and \ms{7}. The latter contains several faults. Downsampled meshes are arranged in rows by decreasing scale. The first column represents the mesh without any attribute. The second and the third columns represent the same mesh onto which a continuous and a categorical property is mapped, respectively.

\Ms{1} is decomposed to the lowest possible resolution (Figure \ref{fig:visual_results_coarsePBR}, bottom). This is probably not useful from a geologist perspective. However, while all properties are almost constant, the lower arch corresponding to an anticlinal on the mesh at original resolution remains perceptible on the final ``Lego brick'' resolution. Looking at the porosity property (middle column), one observes how the values are progressively homogenized on coarser hexes. Concerning the rock type (last column), one observes that the modelet scheme tends to locally maintain  predominant categories resolution after resolution, which is very satisfactory.

On Figure \ref{fig:visual_results_z13}, the much larger \ms{7} is represented down to a fifth dyadic sub-scale. It contains an isolated fault on the left side (the diagonal crest shape) and a faulty block on the right. Even at the coarsest level, corresponding to a downsampling by $2^4\times2^4\times2^4$, these two structural discontinuities are still present, while keeping a good shape fidelity, globally. Concerning the attributes, the decompositions are also adequate.\\

\begin{figure}[htb]
\begin{center}
\includegraphics[width=\linewidth]{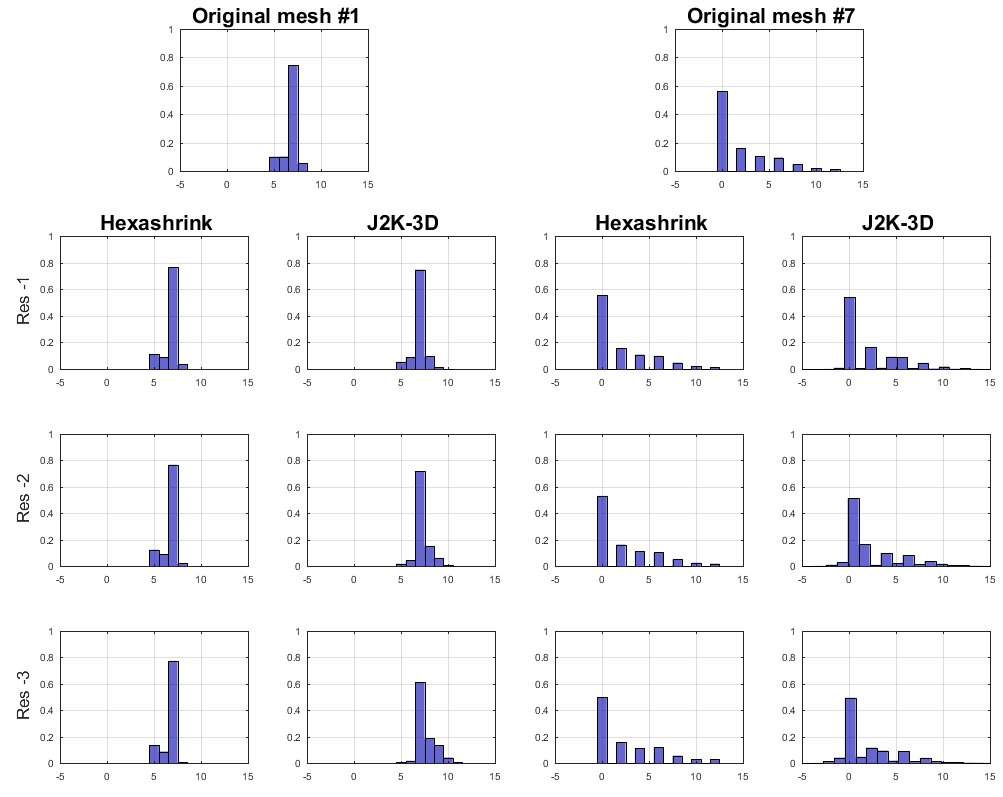}
%
%
\caption{\label{fig:Histogram_categProp}Evolution of the distribution of the \emph{Rock type} categories across resolutions for \ms{1} and \ms{7}, decomposed with either \hs's modelet or the rounded lifting CDF $5/3$ used in the lossless \jtwoktd.}
\end{center}
\end{figure}

To emphasize further the capacity of our scheme to maintain coherency across the resolutions for the properties, Figure \ref{fig:Histogram_categProp} shows the evolution of the \emph{Rock Type} distribution until the third resolution for  \ms{1} and \ms{7}. We observe that \hs preserves the shape of histograms. In other words, the proportion of each category remains consistent across the scales of observation.
Figure \ref{fig:Histogram_categProp} also provides a comparison with the distributions obtained from the reversible  CDF $5/3$  wavelet transform (Section \ref{sec_rounded-wavelet}) used in \jtwoktd \cite{ITU-T_2011_JPEG2000_icsetdd}. One observes that  histograms at lower resolution absolutely do not reflect the original ones, creating interpolated categorical values that do not possess geological meaning. Indeed, a major feature of \hs is to combine, in an overall multiresolution framework, four different downsampling schemes adapted to each property.

Beyond these results in term of geometry and property coherence across resolutions, we recall that our method is deterministic, and exact. The four analysis and synthesis multiresolution schemes allow perfect reconstruction. Contrary to \cite{Chizat_L_2014_msc_multiresolution_scea}, our method is able to manage all the fault configurations. \hs is also scalable, which is indispensable in geosciences, given the steadily growing size of  data volumes and model simulations. This scalability relies on an out-of-core algorithm \cite{Isenburg_M_2003_p-acm-siggraph_out-of-core_cgpm} that splits geometry and property matrices into ``small'' sub-matrices, to process them sequentially. We are thus able to deal with meshes of any size. Lastly, GPU-based parallel computing have been also included, to speed the algorithm up.

As \hs proposes a comprehensive reversible multiscale framework with dyadic downsampling, we compare it to related upscaling  features for geomodels.

\subsection{Geomodel upscaling: SKUA-GOCAD\texttrademark vs \hs }
SKUA-GOCAD\texttrademark or PETREL\texttrademark  are frequently used in geosciences to handle geological objects and to generate meshes  for flow simulation. These specific meshes describe structural discontinuities whose impact is significant on simulation.
To obtain such meshes, the geomodel --- a surface description of horizons or faults --- is fitted into a grid at the desirable resolution.
 Pillar orientation is influenced by fault dip and cell  layer thickness is adapted to the distance between horizons.
Additional properties can then be assigned to  mesh cells: porosity, saturation, rock type\ldots  from well  data or geological interpretation.
Would one wish to lower the resolution, the process described above should be reiterated.

\begin{figure*}[htbp]
\begin{center}
\begin{tabular}{ccc}
\includegraphics[width=0.28\linewidth]{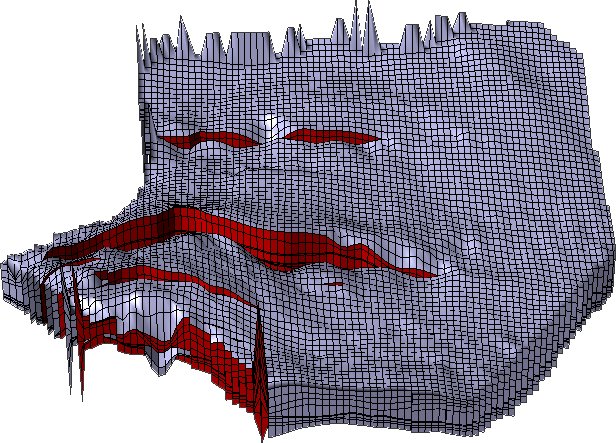}  &
\includegraphics[width=0.28\linewidth]{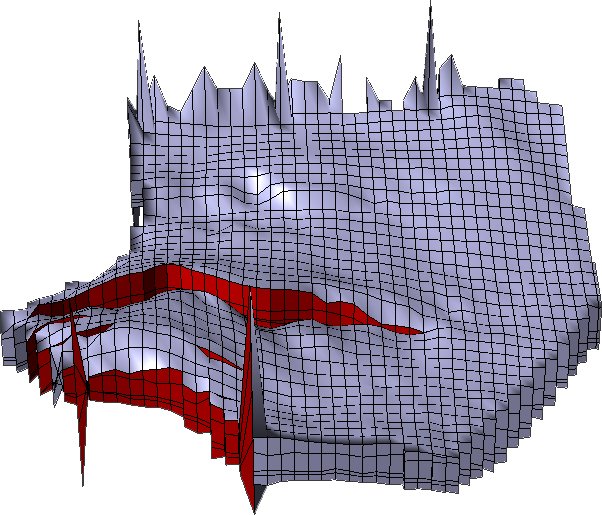} &
\includegraphics[width=0.28\linewidth]{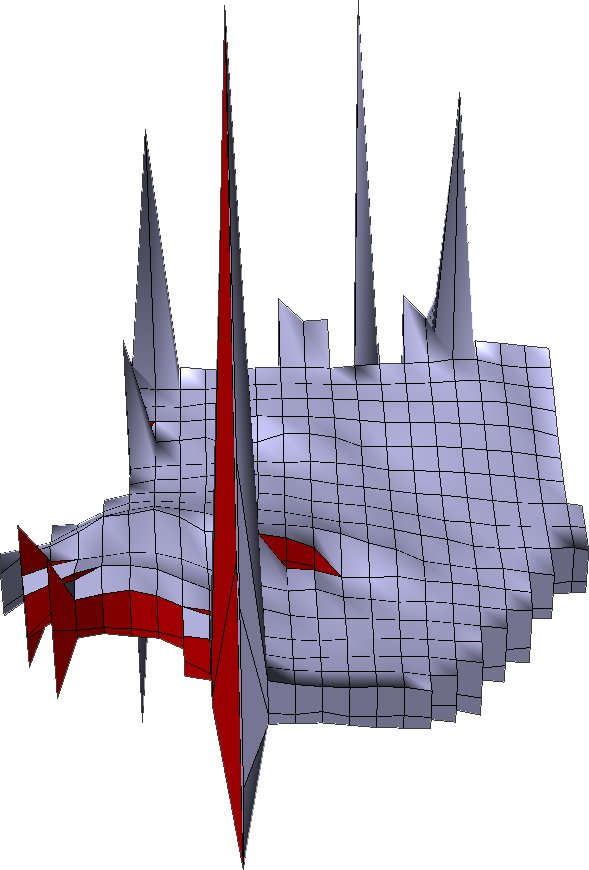} \\
\multicolumn{3}{c}{\Ms{5} with SKUA-GOCAD} \\
\multicolumn{3}{c}{ \ } \\
\includegraphics[width=0.28\linewidth]{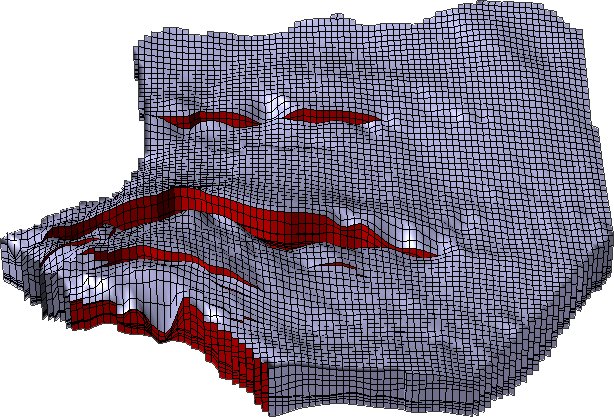}  &
\includegraphics[width=0.28\linewidth]{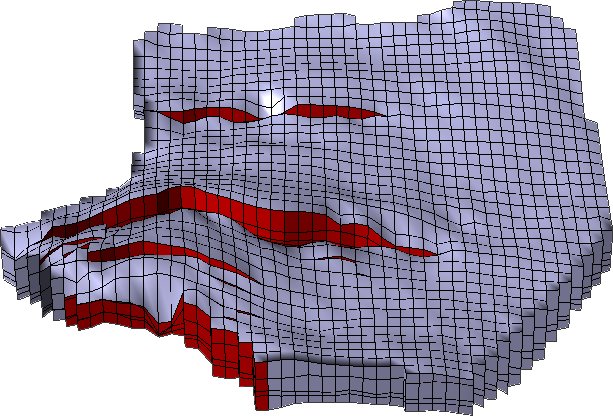} &
\includegraphics[width=0.28\linewidth]{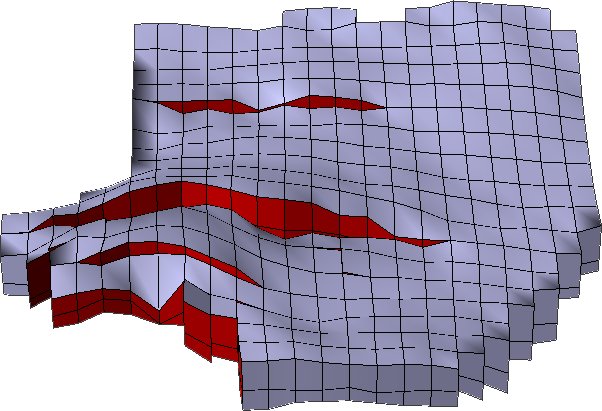} \\
\multicolumn{3}{c}{\Ms{5} with \hs} \\
\multicolumn{3}{c}{ \ } \\ \hline
\multicolumn{3}{c}{ \ } \\
\includegraphics[width=0.305\linewidth]{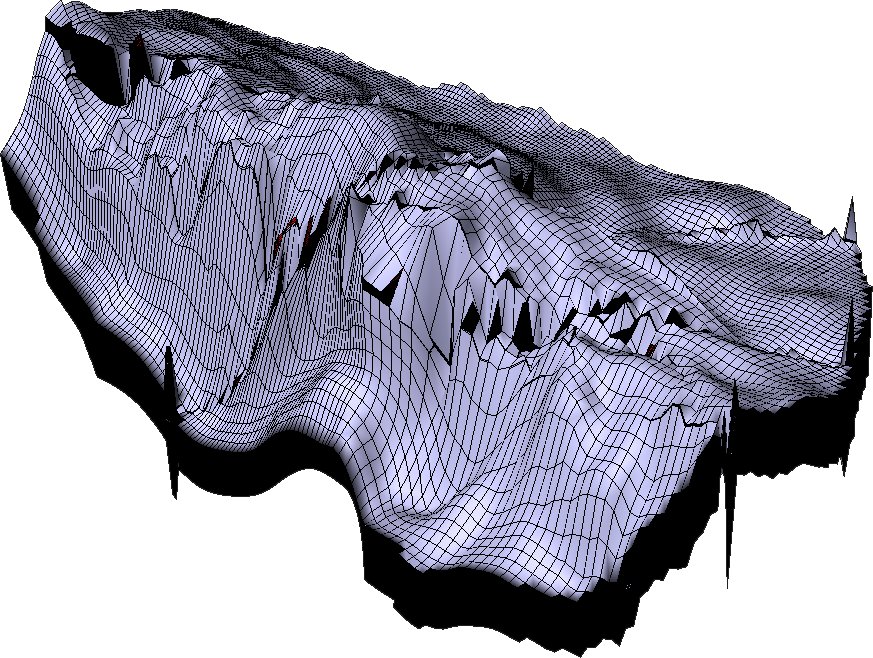}  &
\includegraphics[width=0.305\linewidth]{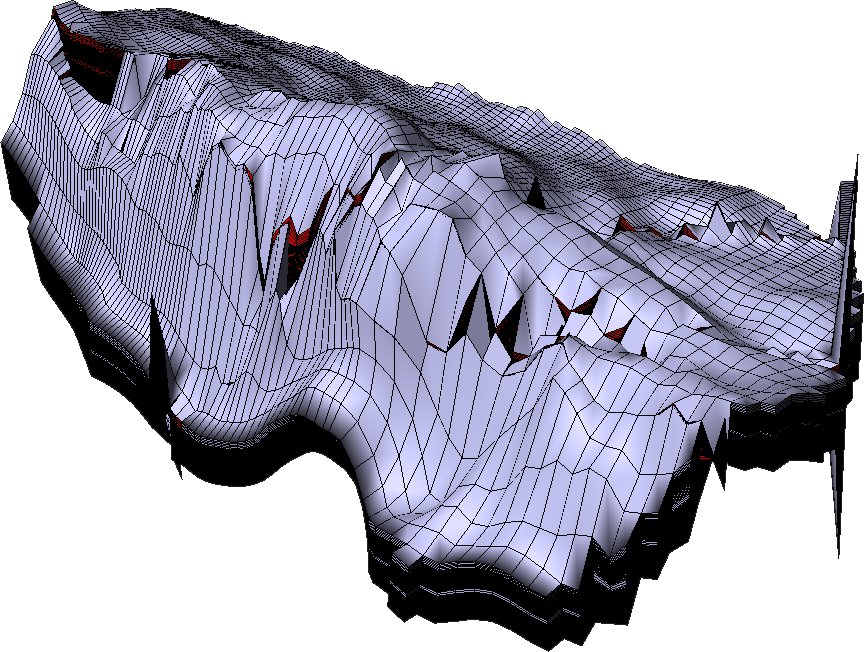} &
\includegraphics[width=0.305\linewidth]{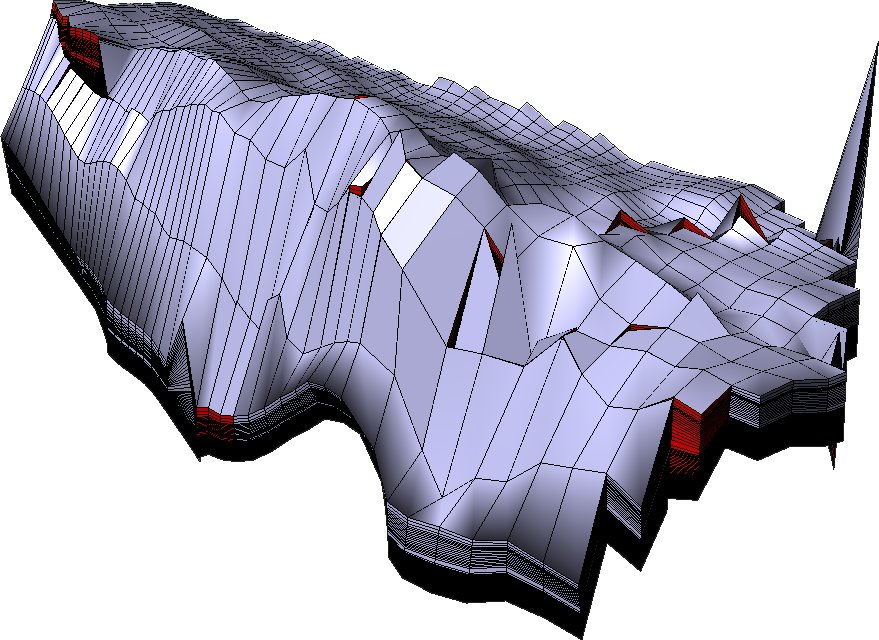} \\
\multicolumn{3}{c}{\Ms{6} with SKUA-GOCAD} \\
\multicolumn{3}{c}{ \ } \\
\includegraphics[width=0.305\linewidth]{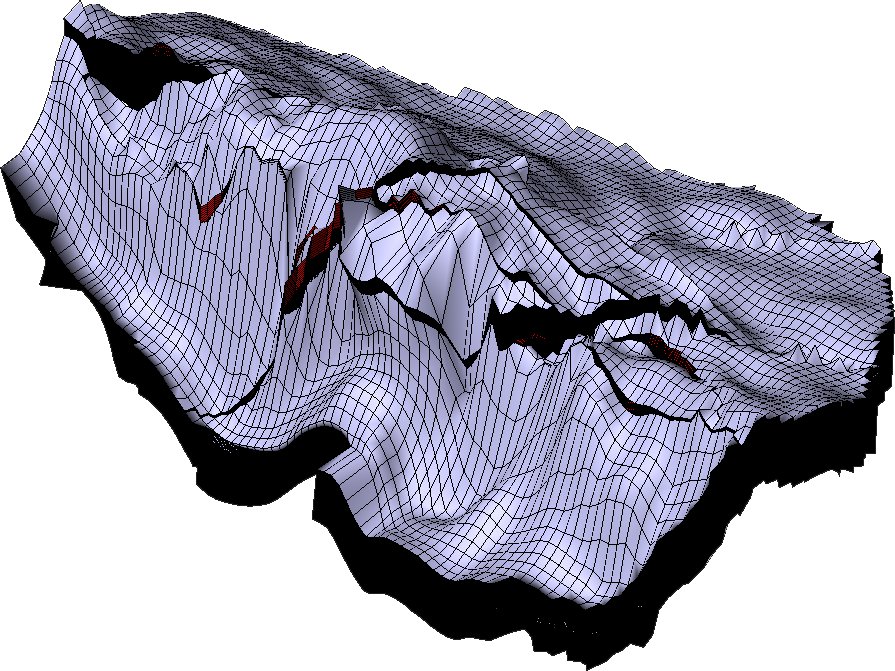}  &
\includegraphics[width=0.305\linewidth]{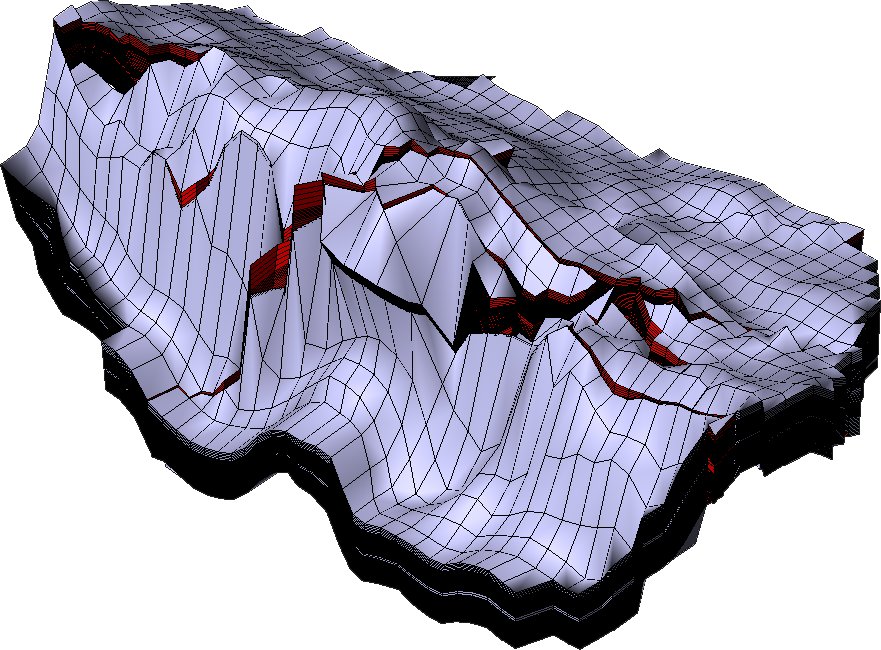} &
\includegraphics[width=0.305\linewidth]{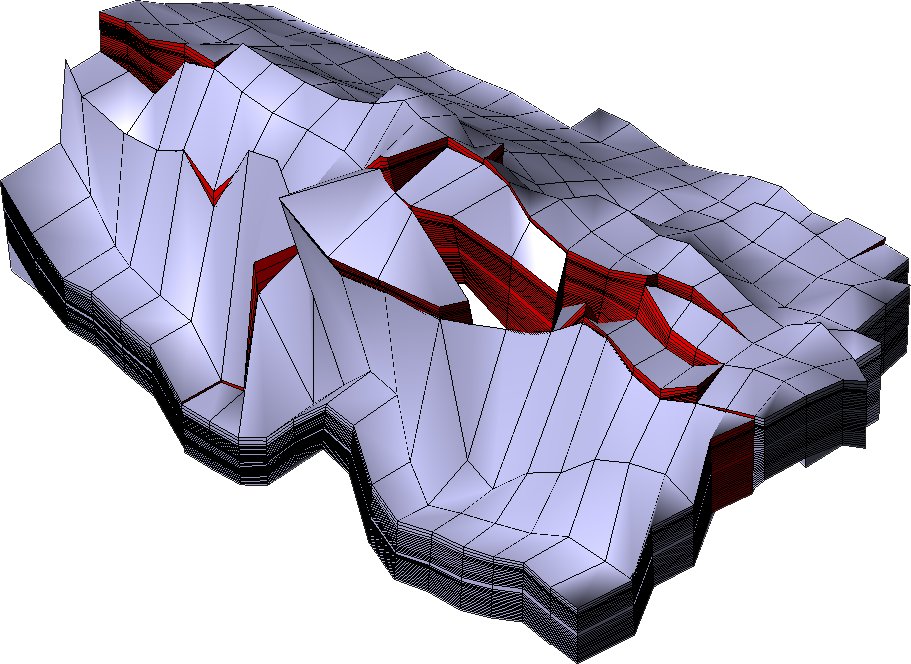} \\
\multicolumn{3}{c}{\Ms{6} with \hs}
\end{tabular}
\caption{\label{fig:actum_preservation} After dyadic downsampling/upscaling, \hs (bottom) better preserves  faults, and manages  non-active cells (\emph{i.e.}, with null \actn values) across scales, yielding nicer borders at each resolution, contrary to GOCAD. From left to right: resolution $-1$, $-2$, and $-3$, respectively.}
\end{center}
\end{figure*}

A simpler alternative proposed by geomodelers consists in \emph{upscaling} meshes. Such methods are usually flexible yet often \emph{ad-hoc}, converting geometry and properties in a non-reversible manner.
Figure \ref{fig:actum_preservation} confronts meshes \#$5$ and \#$6$ downsampled at power-of-two resolutions with SKUA-GOCAD\texttrademark and \hs.
\hs tends to better preserve faults (colored in red), as compared to  SKUA-GOCAD\texttrademark. Figures emphasize an improved preservation of mesh borders, with an efficient management of  \actn  throughout  resolutions. Some  artifacts  may appear with SKUA-GOCAD\texttrademark's  upscaling,  which are automatically averted  by \hs, leading to nicer meshes at low resolution. As a summary, \hs, while being fully reversible at dyadic scales only,  efficiently and automatically manages    structural discontinuities in the \vms. It may provide an interesting complement to  existing irreversible upscaling proposed by several geomodelers.

\subsection{Compression performance comparison\label{sec_results_lossless}}
We now provide an objective evaluation of the \hs multiscale representation for compression purposes. Our main objective is to verify  that binary mesh formats (beyond mere ontological analyses) are indeed compressible, and whether decomposing them in a progressive manner over different scales remains beneficial in size reduction for needs beyond mere visualization (data storage, transfer). Indeed, multiscale representations are thought to enhance the sparsity of locally regular data, often resulting in better predicted properties that can subsequently be compressed.

We first  assess lossless (or perfect) compression. All mesh ontological and geological (cf. Table \ref{Tab-Meshes-Features}, Section \ref{sec_results_methodo}) are thus perfectly restored, whatever the number of decomposition levels (Section \ref{sec_results_reversible}). Since  our mesh objects are heterogeneous, we treat geometry and properties independently, and compress individually their approximations and details.

We compare three generations of lossless all-purpose encoders: gzip (1992), bzip2 (1996), LZMA (1998). They use   Lempel-Ziv, Burrows-Wheeler, Lempel-Ziv-Markov entropic coding, respectively. We refer to \cite{Nelson_M_1995_book_data_cb,Salomon_D_2009_book_handbook_dc} for details regarding these state-of-the-art compression methods.


\begin{table}[htb]

	\centering
		\begin{tabular}{c||cccc}

	Mesh  &  Level & gzip & bzip2 & LZMA\\			
	\hline \hline
		 \multirow{3}{*}{1}  & none &  \num{3.73}  &  \num{4.98}  &  \num{6.43} \\		
		
	  & $1$ & \num{5.62} & \num{6.07} &  \num{7.52} \\	
		  & \numrange{2}{4} & \num{5.67} & \numrange{6.12}{6.13} & \numrange{7.42}{7.44}\\	
			\hline
			
				 \multirow{3}{*}{2}  & none &  \num{3.23}  &  \num{8.41}  &  \num{10.12} \\		
		
	  & $1$ & \num{6.49} & \num{10.82} &  \num{11.81} \\	
		  & \numrange{2}{6} & \numrange{7.48}{7.58} & \numrange{12.75}{13.03} & \num{13.35}\\	
			\hline \hline
			
	 \multirow{3}{*}{3}  & none &  \num{2.67}  &  \num{2.99}  &  \num{3.63} \\		
		
	  & $1$ & \num{3.88} & \num{4.70} &  \num{5.24} \\	
		  & \numrange{2}{4} & \numrange{4.03}{4.05} & \numrange{4.92}{4.93} & \numrange{5.47}{5.48}\\	
			\hline
						
	 \multirow{3}{*}{4}  & none &  \num{1.83}  &  \num{1.89}  &  \num{2.21} \\		
		
	  & $1$ & \num{2.64} & \num{3.06} &  \num{3.48} \\	
		  & \numrange{2}{4} & \num{2.76} & \numrange{3.22}{3.23} & \numrange{3.64}{3.65}\\	
			\hline
						
	 \multirow{3}{*}{5}  & none &  \num{2.46}  &  \num{2.55}  &  \num{3.33} \\		
		
	  & $1$ & \num{3.14} & \num{2.83} &  \num{3.71} \\	
		  & \numrange{2}{4} & \numrange{3.25}{3.26} & \numrange{2.91}{2.92} & \numrange{3.80}{3.81}\\	
			\hline
						
	 \multirow{3}{*}{6}  & none &  \num{2.31}  &  \num{2.25}  &  \num{3.04} \\		
		
	  & $1$ & \num{3.31} & \num{3.53} &  \num{4.44} \\	
		  & \numrange{2}{6} & \numrange{4.14}{4.24} & \numrange{4.48}{4.68} & \numrange{5.54}{5.73}\\	
			\hline
						
	 \multirow{3}{*}{7}  & none &  \num{3.20}  &  \num{5.98}  &  \num{12.52} \\		
		
	  & $1$ & \num{5.42} & \num{7.07} &  \num{8.90} \\	
		  & \numrange{2}{7} & \numrange{5.80}{6.72} & \numrange{7.63}{10.12} & \numrange{9.05}{10.23}\\	
			\hline
		\end{tabular}
	\caption{Comparative lossless coding performances with compression ratios at different \hs resolution levels combining \hs with gzip, bzip2 and LZMA.\label{Tab-Meshes-Lossless}}
\end{table} 

We exhaustively compare compression performances in Table \ref{Tab-Meshes-Lossless}. For the sake of clarity, recall that different computational  methodologies exist: a compression ratio of $4$ is given by the fraction between the sizes of the original file and the compressed one (the larger the ratio, the better the compression). The latter can also be related to its inverse, the smallest file representing \SI{25}{\percent} of the raw data  ($\tfrac{1}{4}\times 100$), or a compression gain due to the reduction in size of \SI{75}{\percent} (corresponding to  $\left(1-\tfrac{1}{4}\right)\times 100$).

As an example, we provide a detailed interpretation of the third row,  corresponding to the mildly complicated and faulty \ms{3}. Without decomposition, \emph{i.e.} by directly compressing the binary formats, gains in file size are already observed, from \SI{62.5}{\percent} ($\left(1-\tfrac{1}{2.67}\right)\times 100$) for gzip to \SI{+72.5}{\percent} for LZMA. We first remark that  improvements sensitively increase as we use more recent entropic coders, with only one exception for \ms{6}, gzip performing slightly better than bzip2. However, the most recent LZMA coder always offers the best performances, with a sufficient gap over the two other methods.

The same trend applies when performing a one-level \hs transformation on \ms{3}, with an additional gain in compression: for instance, the combination of a $1$-level \hs associated with LZMA yields a compressed mesh twice as small (\SI{81.9}{\percent}) as a direct gzip  compaction on the original mesh. This is advantageous, as the proposed method either provides access to a  two-fold downsampled mesh together with a smaller overall size.

One can wish to have access to further levels of approximation. The third line of each table block specifies the range of additional available levels (depending on  mesh size), here from \numrange{2}{4}, with the minimum and maximum  compression ratios attained. While we still observe a marginal improvement over a $1$-level \hs (and again a slightly anomalous behavior for \ms{5}), what is more important is the almost imperceptible variation between the resolutions. Hence, \hs offers in all cases an  interesting compression ratio with the supplementary  interest of getting all intermediate resolutions, as shown earlier in Section \ref{sec_results_methodo}.

Overall, the most basic worse case performance (with gzip) of \hs provides a gain above \SI{60}{\percent} in size, which could be further exploited with hardware acceleration \cite{Abdelfattah_M_2014_p-iwocl_gzip_c}. Or in the best cases, combined with  LZMA, one can expect as much as \numrange{3.64}{13.35} fold compression.

In rare cases, \hs  does not result in clear compression improvement.   For \ms{7} and LZMA, we even observe that the best compression ratio (\num{12.52}) is obtained without \hs decomposition.  With such human or computer-generated objects, by contrast with natural data, this  often stems from the inherent quantization of values. This typically happens in \ms{7} when intermediate coordinates, or properties, are obtained by interpolation, to refine geological layers.    Variables  with apparent high-dynamic range and precision, unbeknown to the user, may  derive  from  easy-to-store indices.
Floating-point depth coordinates may result from a mere affine relationship on a list $\mathcal{I}$ of small integers, with offset $o$ and scale $s$:  $s\times \mathcal{I} +o$.
 LZMA's superior capability owes to its capacity to capture complex models of byte patterns. 
 By contrast, with a wavelet decomposition, the affine  relationship in such a case is poorly captured throughout approximations, due to the rounding in wavelet lifting
 (Section \ref{sec_rounded-wavelet}). Hence, multiscale decompositions may slightly reduce the raw compression performance for meshes presenting initial ``numerical format'' artifacts or illusory floating-point precision. This however does not hamper the usability of \hs for storage and visualization, as the direct  access to a hierarchy of resolutions respecting discontinuities is granted,  while already providing impressive compression rates of about \numrange{8}{9}, superior to most results for the others meshes.

Speed performance is highly dependent on mesh size, discontinuity complexity, levels of details. For a baseline evaluation, a Java implementation was run on a laptop with  Intel Core i7-6820HQ CPU @ \SI{2.70}{\giga\hertz} processor and  \SI{16}{\giga\byte} RAM. Each mesh (from our dataset in Table \ref{Tab-Meshes-Lossless}) was compressed  to the maximum level, and decompressed, twelve times. As the outcomes were relatively stable, they were averaged. Timings  for analysis or synthesis alone, and cumulated with lossless encoding and decoding, are  summarized in Table \ref{Tab-Meshes-Timing}.
\begin{table}[htb]
\setlength{\tabcolsep}{3pt} 
\sisetup{round-mode=figures,round-precision=3}
	\centering
		\begin{tabular}{c||cccc}
	 & [A]nalysis & [A]$+$gzip & [A]$+$bzip2 & [A]$+$LZMA \\			
			\hline
  Min.	  & \num{2.7960} & \num{    3.0610  } & \num{  5.6910 } & \num{   6.3450} \\
 Max.		&\num{319.6890 } & \num{ 353.6980 } & \num{ 374.3520  } & \num{ 760.4050}\\			
	 & [S]ynthesis & [S]$+$gzip & [S]$+$bzip2 & [S]$+$LZMA \\			
			\hline
 Min.	  & \num{ 0.7900} & \num{ 1.0270  } & \num{ 3.3480 } & \num{3.2990} \\
Max.		&\num{264.1510} & \num{ 279.5880 } & \num{  284.4360 } & \num{  277.1470}\\			
		\end{tabular}
	\caption{Cumulative lossless compression and decompression durations of \hs with gzip, bzip2 and LZMA, in seconds.\label{Tab-Meshes-Timing}}
\end{table}

 Analysis is slightly slower than synthesis, both taking from a couple of seconds to a couple of minutes. Concerning the coders, gzip is the fastest, adding little overhead to \hs speed. However,  bzip2 or LZMA durations can reveal more expensive than analysis alone. The largest mesh is compressed in a little less than \SI{13}{\minute}. 
 What is more appealing in applications is that only a few seconds are necessary for small meshes. 
Lossless decompression is very fast, even for the largest mesh (\SIrange{13}{20}{\second}), as synthesis takes most of the time. 
And LZMA is the fastest here, adding only \SI{5}{\percent} overhead to synthesis, for recovering the whole original mesh. Decompressing lower resolutions only is of course even faster. This is interesting, as here we obtain a beneficial asymmetrical compression-decompression scheme, termed ``compress once, decompress many''. Once a model is built, one can afford to compress it only once, whatever the time it takes. Then, after transferring, handling it with the benefit of the smaller sizes, decompressing it many times is less expensive. The above performance could be greatly improved with more involved acceleration techniques.

 As a result, on all tested examples, we demonstrate the possibility of storing independently multiscale mesh properties as approximations and details, while preserving geometry (hence faults). This method additionally has an important benefit in  data handling, visualization  and compression, all at a reasonable computational cost.

\section{Conclusion and perspectives\label{sec_perspectives}}
\hs offers a  comprehensive and efficient  framework for a scalable representation of  hexahedral meshes with continuous and categorical properties, at  dyadic resolutions. It is first dedicated to the visualization of massive structured meshes, as  used in geosciences, that can contain geometrical discontinuities, to describe faults for instance. Four adapted multiresolution representations are matched to the underlying nature of each  data field. They permit to decompose such specific meshes progressively. In particular, this framework includes a morphological transform that takes into account  geometrical discontinuities relative to any fault configuration, and preserve their rendering across  resolutions, while maintaining  structural coherency. 

\hs can process any mesh size, thanks to a  GPU-based out-of-core algorithm. This is crucial, given the constant evolution of  data acquisition density that yields increasingly massive and accurate dataset, associated to more demanding simulations. The \hs decomposition  is consistent with respect to mesh rescaling in geosciences, and provides an option for a reversible upscaling; \cite{Misaghian_N_2018_j-computat-geosci_upscaling_auamruarpa} recently proposed such a wavelet-inspired scheme. Finally, it  lends itself to an efficient lossless compression, which can be used for storage and transfer.

Perspectives can deploy into many directions. Motivated by preliminary progressive lossless compression results,  we aim at  developing a more versatile multiresolution  compression scheme,  to manage the potential evolution of  mesh geometry or properties over time, with a special care for simulation-related quality metrics. As multiresolution analysis and synthesis were computed independently from compression, we also envision a better matched coding of approximations and details, for additional performance, toward lossy compression \cite{Lu_T_2018_p-ieee-ipdps_understanding_mllcshpcsd,Liang_X_2018_p-ieee-ic-big-data_error-controlled_lcohcrsd}.

\section*{Acknowledgements}
The authors would like to thank C. Dawson and  M. F. Wheeler for their help, as well as the  reviewers whose comments helped improve the compression performance assessment and comparison.

\bibliographystyle{spmpsci}

\begin{thebibliography}{10}
\providecommand{\url}[1]{{#1}}
\providecommand{\urlprefix}{URL }
\expandafter\ifx\csname urlstyle\endcsname\relax
  \providecommand{\doi}[1]{DOI~\discretionary{}{}{}#1}\else
  \providecommand{\doi}{DOI~\discretionary{}{}{}\begingroup
  \urlstyle{rm}\Url}\fi

\bibitem{Abdelfattah_M_2014_p-iwocl_gzip_c}
Abdelfattah, M.S., Hagiescu, A., Singh, D.: Gzip on a chip: high performance
  lossless data compression on {FPGA}s using {O}pen{CL}.
\newblock In: Proc. Int. Workshop OpenCL (2014).
\newblock \doi{10.1145/2664666.2664670}

\bibitem{Antonini_M_2017_patent_method_ehufmos}
Antonini, M., Payan, F., Schneider, S., Duval, L., Peyrot, J.-L.: Method of
  exploitation of hydrocarbons of an underground formation by means of
  optimized scaling (2017).
\newblock Patent

\bibitem{Bey_J_1995_j-computing_tetrahedral_gr}
Bey, J.: Tetrahedral grid refinement.
\newblock Computing \textbf{55}(4), 355--378 (1995).
\newblock \doi{10.1007/BF02238487}.

\bibitem{Boscardin_L_2006_j-istec-jcst_wavelets_bdot}
Boscard\'in, L.B., Castro, L.R., Castro, S.M., Giusti, A.D.: Wavelets bases
  defined over tetrahedra.
\newblock INSTEC J. Computer Science and Technology \textbf{6}(1), 46--52
  (2006)

\bibitem{Bruekers_F_1992_j-ieee-sel-areas-com_new_npipr}
Bruekers, F.A.M.L., van~den Enden, A.W.M.: New networks for perfect inversion
  and perfect reconstruction.
\newblock IEEE J. Sel. Areas Comm. \textbf{10}(1), 129--137 (1992).
\newblock \doi{10.1109/49.124464}

\bibitem{Calderbank_A_1998_j-acha_wavelet_ttmii}
Calderbank, A.R., Daubechies, I., Sweldens, W., Yeo, B.L.: Wavelet transforms
  that map integers to integers.
\newblock Appl. Comput. Harmon. Analysis \textbf{5}(3), 332--369 (1998).
\newblock \doi{10.1006/acha.1997.0238}.

\bibitem{Cannon_S_2018_book_reservoir_mpg}
Cannon, S.: Reservoir Modelling: A Practical Guide.
\newblock John Wiley \& Sons (2018)

\bibitem{Caumon_G_2013_j-ieee-tgrs_three-dimensional_ismbrsdtmtarmlpbnem}
Caumon, G., Gray, G., Antoine, C., Titeux, M.O.: Three-dimensional implicit
  stratigraphic model building from remote sensing data on tetrahedral meshes:
  Theory and application to a regional model of {L}a {P}opa basin, {NE}
  {Mexico}.
\newblock IEEE Trans. Geosci. Remote Sens. \textbf{51}(3), 1613--1621 (2013).
\newblock \doi{10.1109/tgrs.2012.2207727}.

\bibitem{Chaux_C_2008_j-ieee-tsp_nonlinear_sbemid}
Chaux, C., Duval, L., Benazza-Benyahia, A., Pesquet, J.C.: A nonlinear {Stein}
  based estimator for multichannel image denoising.
\newblock IEEE Trans. Signal Process. \textbf{56}(8), 3855--3870 (2008).
\newblock \doi{10.1109/TSP.2008.921757}

\bibitem{Chaux_C_2007_j-ieee-tit_noise_cpdtwd}
Chaux, C., Pesquet, J.C., Duval, L.: Noise covariance properties in dual-tree
  wavelet decompositions.
\newblock IEEE Trans. Inform. Theory \textbf{53}(12), 4680--4700 (2007).
\newblock \doi{10.1109/TIT.2007.909104}

\bibitem{Chen_D_2005_p-eusipco_geometry_ctmuop}
Chen, D., Chiang, Y.J., Memon, N., Wu, X.: Geometry compression of tetrahedral
  meshes using optimized prediction.
\newblock In: Proc. Eur. Sig. Image Proc. Conf. (2005).

\bibitem{Chizat_L_2014_msc_multiresolution_scea}
Chizat, L.: Multiresolution signal compression: Exploration and application.
\newblock Master's thesis, ENS Cachan (2014)

\bibitem{Cohen_A_1992_j-comm-acm_biorthogonal_bcsw}
Cohen, A., Daubechies, I., Feauveau, J.C.: Biorthogonal bases of compactly
  supported wavelets.
\newblock Commun. ACM \textbf{45}(5), 485--560 (1992).
\newblock \doi{10.1002/cpa.3160450502}

\bibitem{Courbet_C_2010_j-vis-comput_streaming_chm}
Courbet, C., Isenburg, M.: Streaming compression of hexahedral meshes.
\newblock Vis. Comput. \textbf{26}(6-8), 1113--1122 (2010).
\newblock \doi{10.1007/s00371-010-0481-7}.


\bibitem{Danovaro_E_2002_p-smi_multiresolution_tmac}
Danovaro, E., De~Floriani, L., Lee, M.T., Samet, H.: Multiresolution
  tetrahedral meshes: An analysis and a comparison.
\newblock In: Proc. Shape Modeling International, pp. 83--91 (2002).
\newblock \doi{10.1109/SMI.2002.1003532}.

\bibitem{Dupont_F_2013_incoll_3d_mc}
Dupont, F., Lavou\'e, G., Antonini, M.: 3{D} mesh compression.
\newblock In: L.~Lucas, C.~Loscos, Y.~Remion (eds.) 3{D} Video from Capture to
  Diffusion. Wiley-ISTE (2013)

\bibitem{Foster_I_2017_incoll_computing_jwynodares}
Foster, I., Ainsworth, M., Allen, B., Bessac, J., Cappello, F., Choi, J.Y.,
  Constantinescu, E., Davis, P.E., Di, S., Di, W., Guo, H., Klasky, S., Dam,
  K.K.V., Kurc, T., Liu, Q., Malik, A., Mehta, K., Mueller, K., Munson, T.,
  Ostouchov, G., Parashar, M., Peterka, T., Pouchard, L., Tao, D., Tugluk, O.,
  Wild, S., Wolf, M., Wozniak, J.M., Xu, W., Yoo, S.: Computing just what you
  need: Online data analysis and reduction at extreme scales.
\newblock In: Lecture Notes in Computer Science, pp. 3--19. Springer
  International Publishing (2017).


\bibitem{Gumhold_S_1999_p-visualization_tetrahedral_mccbm}
Gumhold, S., Guthe, S., Stra{\ss}er, W.: Tetrahedral mesh compression with the
  cut-border machine.
\newblock In: Proc. IEEE Visualization Conf., pp. 51--58 (1999).

\bibitem{Gumhold_S_1998_p-acm-siggraph-comput-graph_real_tctmc}
Gumhold, S., Stra{\ss}er, W.: Real time compression of triangle mesh
  connectivity.
\newblock In: Proc. ACM SIGGRAPH Comput. Graph., pp. 133--140 (1998).
\newblock \doi{10.1145/280814.280836}.

\bibitem{Hoppe_H_1993_p-acm-siggraph-comput-graph_mesh_o}
Hoppe, H., DeRose, T., Duchamp, T., McDonald, J., Stuetzle, W.: Mesh
  optimization.
\newblock In: Proc. ACM SIGGRAPH Comput. Graph., pp. 19--26 (1993).
\newblock \doi{10.1145/166117.166119}.

\bibitem{Ibarria_L_2007_p-dcc_spectral_p}
Ibarria, L., Lindstrom, P., Rossignac, J.: Spectral predictors.
\newblock In: Proc. Data Compression Conf., pp. 163--172 (2007).
\newblock \doi{10.1109/DCC.2007.72}.

\bibitem{Isenburg_M_2003_j-graph-model_compressing_hvm}
Isenburg, M., Alliez, P.: Compressing hexahedral volume meshes.
\newblock Graph. Model. \textbf{65}(4), 239--257 (2003).
\newblock \doi{10.1016/s1524-0703(03)00044-4}.

\bibitem{Isenburg_M_2003_p-acm-siggraph_out-of-core_cgpm}
Isenburg, M., Gumhold, S.: Out-of-core compression for gigantic polygon meshes.
\newblock In: Proc. SIGGRAPH Int. Conf. Comput. Graph. Interactive Tech., pp.
  935--942 (2003).
\newblock \doi{10.1145/1201775.882366}.

\bibitem{Isenburg_M_2006_p-graphics-interface_streaming_ctvm}
Isenburg, M., Lindstrom, P., Gumhold, S., Shewchuk, J.: Streaming compression
  of tetrahedral volume meshes.
\newblock In: Proc. Graphics Interface, pp. 115--121 (2006).

\bibitem{Isenburg_M_2005_p-eurographics-symp-geom-process_streaming_ctm}
Isenburg, M., Lindstrom, P., Snoeyink, J.: Streaming compression of triangle
  meshes.
\newblock In: Proc. Eurographics Symp. Geom. Process., vol. 255, pp. 111--118
  (2005).


\bibitem{ITU-T_2011_JPEG2000_icsetdd}
{ITU-T T.809}: {JPEG}2000 image coding system: Extensions for three-dimensional
  data (2011).
\newblock ISO/IEC 15444-10:2011

\bibitem{Jacques_L_2011_j-sp_panorama_mgrisdfs}
Jacques, L., Duval, L., Chaux, C., Peyr{\'e}, G.: A panorama on multiscale
  geometric representations, intertwining spatial, directional and frequency
  selectivity.
\newblock Signal Process. \textbf{91}(12), 2699--2730 (2011).
\newblock \doi{10.1016/j.sigpro.2011.04.025}.

\bibitem{Kober_C_2001_j-eng-comput_case_shmgshm}
Kober, C., M\"uller-Hannemann, M.: A case study in hexahedral mesh generation:
  Simulation of the human mandible.
\newblock Eng. Comput. \textbf{17}(3), 249--260 (2001).
\newblock \doi{10.1007/pl00013389}

\bibitem{Kovacevic_J_2012_book_signal_pfwr}
Kova\v{c}evi\'{c}, J., Goyal, V., Vetterli, M.: Signal Processing Fourier and
  Wavelet Representations (2012)

\bibitem{Krivograd_S_2008_j-comput-aided-des_hexahedral_mccvd}
Krivograd, S., Trlep, M., \v{Z}alik, B.: A hexahedral mesh connectivity
  compression with vertex degrees.
\newblock Comput. Aided Des. \textbf{40}(12), 1105--1112 (2008).
\newblock \doi{10.1016/j.cad.2008.10.013}.


\bibitem{LeGall_D_1988_p-icassp_sub-band_cdiusskfact}
Le~Gall, D., Tabatabai, A.: Sub-band coding of digital images using symmetric
  short kernel filters and arithmetic coding techniques.
\newblock In: Proc. Int. Conf. Acoust. Speech Signal Process. (1988).
\newblock \doi{10.1109/icassp.1988.196696}

\bibitem{Liang_X_2018_p-ieee-ic-big-data_error-controlled_lcohcrsd}
Liang, X., Di, S., Tao, D., Li, S., Li, S., Guo, H., Chen, Z., Cappello, F.:
  Error-controlled lossy compression optimized for high compression ratios of
  scientific datasets.
\newblock In: {IEEE} Int. Conf. Big Data (2018).
\newblock \doi{10.1109/bigdata.2018.8622520}

\bibitem{Lie_K_2016_book_introduction_rsumugmrstmrst}
Lie, K.A.: An Introduction to Reservoir Simulation Using {MATLAB}. User Guide
  for the Matlab Reservoir Simulation Toolbox ({MRST}) (2016).
\newblock SINTEF ICT, Departement of Applied Mathematics

\bibitem{Lie_K_2017_j-computat-geosci_successful_ammrrse}
Lie, K.A., M{\o}yner, O., Natvig, J.R., Kozlova, A., Bratvedt, K., Watanabe,
  S., Li, Z.: Successful application of multiscale methods in a real reservoir
  simulator environment.
\newblock Computat. Geosci. \textbf{21}(5-6), 981--998 (2017).
\newblock \doi{10.1007/s10596-017-9627-2}

\bibitem{Lindstrom_P_2008_p-dcc_lossless_chm}
Lindstrom, P., Isenburg, M.: Lossless compression of hexahedral meshes.
\newblock In: Proc. Data Compression Conf., pp. 192--201 (2008).
\newblock \doi{10.1109/dcc.2008.12}.

\bibitem{Lu_T_2018_p-ieee-ipdps_understanding_mllcshpcsd}
Lu, T., Liu, Q., He, X., Luo, H., Suchyta, E., Choi, J., Podhorszki, N.,
  Klasky, S., Wolf, M., Liu, T., Qiao, Z.: Understanding and modeling lossy
  compression schemes on {HPC} scientific data.
\newblock In: IEEE International Parallel and Distributed Processing Symposium
  (2018).
\newblock \doi{10.1109/ipdps.2018.00044}.

\bibitem{Misaghian_N_2018_j-computat-geosci_upscaling_auamruarpa}
Misaghian, N., Assareh, M., Sadeghi, M.: An upscaling approach using adaptive
  multi-resolution upgridding and automated relative permeability adjustment.
\newblock Computat. Geosci. \textbf{22}(1), 261--282 (2018).
\newblock \doi{10.1007/s10596-017-9688-2}

\bibitem{Nelson_M_1995_book_data_cb}
Nelson, M., Gailly, J.L.: The Data Compression Book.
\newblock Wiley (1995)

\bibitem{Owen_S_2017_j-procedia-eng_hexahedral_mgcmm}
Owen, S.J., Brown, J.A., Ernst, C.D., Lim, H., Long, K.N.: Hexahedral mesh
  generation for computational materials modeling.
\newblock Procedia Eng. \textbf{203}, 167--179 (2017).
\newblock \doi{10.1016/j.proeng.2017.09.803}

\bibitem{Pajarola_R_1999_p-visualization_implant_scptmc}
Pajarola, R., Rossignac, J., Szymczak, A.: Implant sprays: Compression of
  progressive tetrahedral mesh connectivity.
\newblock In: Proc. IEEE Visualization Conf., pp. 299--305 (1999).
\newblock \doi{10.1109/VISUAL.1999.809901}.

\bibitem{Perrons_R_2015_j-energy-pol_data_awogsloiabd}
Perrons, R.K., Jensen, J.W.: Data as an asset: What the oil and gas sector can
  learn from other industries about "big data".
\newblock Energy Pol. \textbf{81}, 117--121 (2015).
\newblock \doi{10.1016/j.enpol.2015.02.020}.

\bibitem{Pettersen_O_2006_misc_basics_rsers}
Pettersen, {\O}.: Basics of reservoir simulation with the {E}clipse reservoir
  simulator.
\newblock Department of Mathematics, University of Bergen, Norway (2006).
\newblock Lecture Notes

\bibitem{Peyrot_J_2016_p-icip_hexashrink_mclshmdg}
Peyrot, J.-L., Duval, L., Schneider, S., Payan, F., Antonini, M.:
  ({H})exashrink: Multiresolution compression of large structured hexahedral
  meshes with discontinuities in geosciences.
\newblock In: Proc. Int. Conf. Image Process., pp. 1101--1105 (2016).
\newblock \doi{10.1109/ICIP.2016.7532528}

\bibitem{Rao_R_1998_book_wavelet_tita}
Rao, R.M., Bopardikar, A.S.: Wavelet Transforms: Introduction to Theory and
  Applications.
\newblock Prentice Hall (1998)

\bibitem{Rezapour_A_2019_j-transp-porous-med_upscaling_gmorugulbgwt}
Rezapour, A., Ortega, A., Sahimi, M.: Upscaling of geological models of oil
  reservoirs with unstructured grids using lifting-based graph wavelet
  transforms.
\newblock Transp. Porous Med. \textbf{127}, 661--684 (2019).
\newblock \doi{10.1007/s11242-018-1219-7}

\bibitem{Roe_P_2016_j-computat-geosci_volume-conserving_rcfcpg}
R{\o}e, P., Hauge, R.: A volume-conserving representation of cell faces in
  corner point grids.
\newblock Computat. Geosci. \textbf{20}(3), 453--460 (2016).
\newblock \doi{10.1007/s10596-015-9500-0}.

\bibitem{Rossignac_J_1999_j-ieee-trans-visual-comput-graph_edgebreaker_cctm}
Rossignac, J.: Edgebreaker: Connectivity compression for triangle meshes.
\newblock IEEE Trans. Visual Comput. Graph. \textbf{5}(1), 47--61 (1999).
\newblock \doi{10.1109/2945.764870}.

\bibitem{Salomon_D_2009_book_handbook_dc}
Salomon, D., Motta, G.: Handbook of Data Compression.
\newblock Springer (2009)

\bibitem{Staadt_O_1998_p-visualization_progressive_t}
Staadt, O.G., Gross, M.H.: Progressive tetrahedralizations.
\newblock In: Proc. IEEE Visualization Conf., pp. 397--402 (1998).
\newblock \doi{10.1109/VISUAL.1998.745329}.


\bibitem{Sweldens_W_1996_j-acha_lifing_scdcbw}
Sweldens, W.: The lifting scheme: a custom-design construction of biorthogonal
  wavelets.
\newblock Appl. Comput. Harmon. Analysis \textbf{3}(2), 186--200 (1996).
\newblock \doi{10.1006/acha.1996.0015}.

\bibitem{Szymczak_A_2000_j-comput-aided-des_grow_fcctm}
Szymczak, A., Rossignac, J.: Grow \& fold: Compressing the connectivity of
  tetrahedral meshes.
\newblock Comput. Aided Des. \textbf{32}(8-9), 527--537 (2000).
\newblock \doi{10.1016/S0010-4485(00)00040-3}.

\bibitem{Touma_C_1998_p-gi_triangle_mc}
Touma, C., Gotsman, C.: Triangle mesh compression.
\newblock In: Proc. Graphics Interface Conf., pp. 26--34. Vancouver, Canada
  (1998).
\newblock \doi{10.20380/GI1998.04}

\bibitem{Witten_I_1987_j-comm-acm_arithmetic_cdc}
Witten, I.H., Neal, R.M., Cleary, J.G.: Arithmetic coding for data compression.
\newblock Commun. ACM \textbf{30}(6), 520--540 (1987).
\newblock \doi{10.1145/214762.214771}.

\end{thebibliography}

\end{document}